\pdfoutput=1
\documentclass[11pt]{article}

\usepackage{etex}
\reserveinserts{28}

\usepackage{amsmath,amssymb,amsfonts,bbm,verbatim}
\usepackage[multiple,stable]{footmisc}
\usepackage[amsmath,amsthm,thmmarks]{ntheorem}
\allowdisplaybreaks[4]
\usepackage[noconfig]{refstyle}
\usepackage[dvipsnames]{xcolor}
\usepackage[hyperfootnotes=false, linktocpage=true, colorlinks, citecolor=blue, linkcolor=blue, urlcolor=Maroon]{hyperref}
\usepackage{array,tabularx,booktabs,geometry,subfigure,graphicx,longtable}
\usepackage[bf]{caption}
\usepackage{tikz}
\usetikzlibrary{shapes,arrows}
\tikzstyle{block} = [rectangle, draw, text width=7em, text centered, rounded corners, minimum height=3em]
\usepackage{graphicx,color}
\usetikzlibrary{positioning,arrows}

\usepackage{cite}
\usepackage{arydshln}
\usepackage{tikz}
\usetikzlibrary{positioning,automata,decorations.pathreplacing,shapes}
\usepackage[english]{babel}
\usepackage{microtype}

\usepackage[all,cmtip]{xy}
\usetikzlibrary{matrix, arrows}

\newtheorem{theorem}{Theorem}[section]
\newtheorem{lemma}[theorem]{Lemma}

\usepackage{hyperref}
\usepackage{extarrows}

\newcommand{\eref}[1]{(\ref{#1})}
\newcommand{\ra}{\rightarrow}

\newcommand{\eeq}{\end{equation}}
\newcommand{\beq}{\begin{equation}}
\newcommand{\ba}{\begin{array}}
\newcommand{\ea}{\end{array}}
\newcommand{\cL}{{\cal L}}

\newcommand{\nn}{\nonumber}

\newcommand{\cK}{{\cal K}}
\newcommand{\cF}{{\cal F}}

\newcommand{\cO}{{\cal O}}
\newcommand{\cA}{{\cal A}}
\newcommand{\IP}{\mathbb P}
\newcommand{\IC}{\mathbb C}

\def\IP{\mathbb{P}}

\def\IZ{\mathbb{Z}}

\def\IC{\mathbb{C}}

\newcommand{\id}{\mathbf{1}}

\def\cA{\mathcal{A}}
\def\cC{\mathcal{C}}
\def\cD{\mathcal{D}}

\def\cF{\mathcal{F}}

\def\cL{\mathcal{L}}

\def\cN{\mathcal{N}}
\def\cO{\mathcal{O}}

\def\ra{\rightarrow}

\def\fto{\longrightarrow}

\def\clap#1{\hbox to 0pt{\hss#1\hss}}

\def\fto{\longrightarrow}

\newcommand{\be}{\begin{equation}}
\newcommand{\ee}{\end{equation}}
\newcommand{\eea}{\end{eqnarray}}
\newcommand{\bea}{\begin{eqnarray}}

\newcommand{\iddots}{\mathinner{\mkern2mu\raise1pt\hbox{.}\mkern2mu \raise4pt\hbox{.}\mkern2mu\raise7pt\hbox{.}\mkern1mu}}

\providecommand{\id}{\leavevmode\hbox{\small$\mathrm{1}$\kern-3.8pt\normalsize$\mathrm{1}$}}
\def\fnote#1#2{\begingroup\def\thefootnote{#1}\footnote{#2}
     \addtocounter{footnote}{-1}\endgroup}
 
 \newcommand{\bi}{\begin{itemize}}
\newcommand{\ei}{\end{itemize}}

\begin{document}

\vspace{1cm}

\title{
       \vskip 40pt
       {\huge  Fibrations in CICY Threefolds}
       }
       \date{}
\maketitle
%\vspace{2cm}

\begin{center}
\author{Lara B. Anderson${}{}$, Xin Gao${}{}$, James Gray${}{}$, and Seung-Joo Lee${}{}$}
\end{center}

\begin{center} {\small ${}${\it Physics Department, Robeson Hall, Virginia Tech, Blacksburg, VA 24061, USA}}\\
\fnote{}{lara.anderson@vt.edu,  xingao@vt.edu, jamesgray@vt.edu, seungsm@vt.edu}
\end{center}

\begin{abstract}
\noindent In this work we systematically enumerate genus one fibrations in the class of $7,890$ Calabi-Yau manifolds defined as complete intersections in products of projective spaces, the so-called CICY threefolds. This survey is independent of the description of the manifolds and improves upon past approaches that probed only a particular algebraic form of the threefolds (i.e. searches for ``obvious" genus one fibrations as in \cite{Gray:2014fla,Anderson:2016cdu}). We also study K3-fibrations and nested fibration structures. That is, K3 fibrations with potentially many distinct elliptic fibrations. 
To accomplish this survey a number of new geometric tools are developed including a determination of the full topology of all CICY threefolds, including triple intersection numbers. In $2,946$ cases this involves finding a new ``favorable" description of the manifold in which all divisors descend from a simple ambient space. Our results consist of a survey of obvious fibrations for all CICY threefolds and a complete classification of all genus one fibrations for $4,957$ ``K\"ahler favorable" CICYs whose K\"ahler cones descend from a simple ambient space. Within the CICY dataset, we find $139,597$ obvious genus one fibrations, $30,974$ obvious K3 fibrations and $208,987$ nested combinations. For the K\"ahler favorable geometries we find a complete classification of $377,559$ genus one fibrations. For one manifold with Hodge numbers $(19,19)$ we find an explicit description of an infinite number of distinct genus-one fibrations extending previous results for this particular geometry that have appeared in the literature. The data associated to this scan is available here \cite{website}.
\end{abstract}

\thispagestyle{empty}
\setcounter{page}{0}
\newpage

\tableofcontents

\section{Introduction: fibrations in Calabi-Yau threefolds}\label{sec:Intro}

Calabi-Yau manifolds admitting fibrations have long played a central role in the study of string compactifications. This has included bringing to light remarkable string dualities including heterotic/Type IIA duality, heterotic/F-theory duality and F-/M-theory duality, among others. Crucially, because F-theory arises from a ``geometrization" of the axio-dilaton of Type IIB string theory \cite{Vafa:1996xn}, the structure of effective theories in this context is intrinsically linked to the geometry of elliptically (or more generally, genus one) fibered Calabi-Yau (CY) manifolds. In addition, genus one fibered CY geometries are significant because they provide an important foothold into attempts to classify \emph{all compactification geometries} since the set of all genus one fibered CY $3$-folds has been proven to be finite \cite{gross_finite}. Recent progress \cite{2016arXiv160802997D} has given evidence of finiteness for genus one fibered CY $4$- and $5$-folds as well. From a mathematical perspective, these classifications \cite{grassi,gross_finite} were motivated by the hope that they could provide tools which might be used to establish the finiteness of the set of all CY $n$-folds. However, despite these hopes, and the manifest utility of CY fibrations for string dualities, for many years it was generally thought that CY manifolds which admit fibrations (i.e. genus-one, $K3$, or abelian surface fibrations) would likely be rare within the set of all CY geometries. 

Recent work has made clear that, in fact, the vast majority of all known Calabi-Yau manifolds are genus-one fibered \cite{Rohsiepe:2005qg,Johnson:2014xpa,Gray:2014fla,Johnson:2016qar,Candelas:2012uu,Anderson:2016cdu}. Further, these manifolds also appear to be generically multiply fibered, that is that they can be written in more than one way as a genus-one fibration, over topologically distinct bases \cite{Rohsiepe:2005qg,Gray:2014fla,Anderson:2016cdu,Morrison:2016lix}. More explicitly, a \emph{multiply elliptically fibered} (or genus one fibered in the case without section) CY $n$-fold admits multiple descriptions of the form $\pi_i: X_{n} \longrightarrow B^{(i)}_{n-1}$ with elliptic fiber $\mathbb{E}_{(i)b}=\pi^{-1}(b\in B^{(i)}_{n-1})$ (denoted succinctly by $\pi_i: X_{n} \stackrel{\mathbb{E}_{(i)}}{\longrightarrow} B^{(i)}_{n-1}$).  That is,
\beq\label{manyfibs}
\xymatrix{
& X_{n} \ar[ld]_{\mathbb{E}_{(1)}} \ar[d]^{\mathbb{E}_{(2)}}  \ar[rd]^{\mathbb{E}_{(i)}} &\\
B^{(1)}_{n-1} & B^{(2)}_{n-1} \ldots & B^{(i)}_{n-1}}
\eeq
For each fibration, $\pi_i$, the form of the associated Weierstrass model \cite{nakayama}, the structure of the singular fibers, discriminant locus, fibral divisors and Mordell-Weil group can all be different, as can the topology of the base manifolds $B^{(i)}_{n-1}$. Initial steps to explore such prolific fibration structures were taken for CICY four-folds in \cite{Gray:2014fla} and some examples were studied for three-folds in \cite{Anderson:2016cdu}. 

In this work, we will be focused on \emph{systematically enumerating such fibration structures} for a simple dataset of CY threefolds. To begin, we will consider a dataset that is sufficiently large in scope to be interesting, but small enough to be tractable -- the set of $7890$ CY manifolds constructed as complete intersections in products of projective spaces (CICYs) \cite{Candelas:1987kf,Gray:2013mja,Gray:2014kda,Gray:2014fla}. However many of the tools and observations could equally well be applied to complete intersections in toric varieties \cite{Kreuzer:2000xy,Rohsiepe:2005qg,Braun:2011ux} or the recently constructed gCICY manifolds \cite{Anderson:2015iia,Anderson:2015yzz,Berglund:2016yqo,Berglund:2016nvh,Garbagnati:2017rtb}.

A CICY manifold can be described by a so-called ``configuration matrix" which encodes the data essential to the algebraic definition of the manifold. In general, a three-fold $X$ can be defined as the complete intersection of $K$ polynomials, $p_{\alpha}$ where $\alpha=1, \cdots, K$, in an ambient space, $\cA= \IP^{n_1} \times \cdots \times \IP^{n_m}$.   The polynomials $p_\alpha$ are sections of appropriate line bundle $\cO_\cA(a^{1}_\alpha, \cdots, a^{m}_\alpha)$, with $a^{r}_{\alpha} \geq 0$ specifying the non-negative homogeneous degree of $p_\alpha$ in the $r$-th projective piece. Here the indices $r, s, \cdots  = 1, \cdots, m$ are used to label the projective ambient space factors $\IP^{n_r}$, and the indices $\alpha, \beta, \cdots = 1, \cdots, K$, to label the polynomials, $p_\alpha$. A family of such geometries can be characterized by a configuration matrix of the form, 
 
 \bea\label{confForM}
X=\left[ \,\cA\, ||\, \{\bf a_\alpha\} \,\right] = \def\arraystretch{1.2}\left[\ba{c|ccc} 
\IP^{n_1}&  a^{1}_{1} & \cdots &a^{1}_{K} \\
\IP^{n_2} &  a^{2}_{1} & \cdots & a^{2}_{K}\\
\vdots &  \vdots &\ddots&\vdots\\
\IP^{n_m} & a^{m}_{1} & \cdots& a^{m}_{K}\\ 
\ea\right]  
\eea
where
\beq
{\rm dim}_\IC \; X = \sum\limits_{r=1}^m n_r - K=3  \ ,
\eeq  
and the Calabi-Yau condition leads to the degree constraints,
\beq\label{deg-constraint}
n_r +1 = \sum\limits_{\alpha=1}^K a_\alpha^r \ , 
\eeq
for each $r=1, \cdots, m$.

Within this dataset, \emph{many fibration structures are ``obvious" from the form of the configuration matrix above}. It should be noted that it is possible to perform arbitrary row and column permutations on a configuration matrix without
changing the geometry that is described. These operations simply correspond to reordering the $\IP^{n_r}$
ambient factors and the hypersurface equations, respectively. Thus, we can ask whether the configuration matrix can be put in the following form by row and column permutations:
\begin{equation}\label{mrfib}
{ X}= \left[\begin{array}{c|c:c}  {\cal A}_1 & 0 & {\cal F} \\ \hdashline
{\cal A}_2 & {\cal B} & {\cal T} \end{array}\right] .
\end{equation}
where  ${\cal A}_1$ and ${\cal A}_2$ are both products of projective spaces, while ${\cal F}, {\cal B}$ and ${\cal T}$ are block sub-matrices. Such a configuration matrix describes a fibration of the manifold described by $\left[ {\cal A}_1 | {\cal F}\right]$ over the base $\left[ {\cal A}_2 | {\cal B}\right]$ where the ``twisting" of the fibre over the base is determined by the matrix ${\cal T}$. Therefore, as long as the number of columns of $\cF$ and the dimension of  $\cA_1$ are such that
$\cF$ is of complex dimension $1$, \eref{deg-constraint} guarantees that the fibers will be Calabi-Yau one-folds: that is genus-one curves. It follows that the base of the fibration will then be of complex dimension $n-1$.

As a simple example, consider the following configuration matrix defining the tetra-quadric threefold:
 \begin{eqnarray} \label{thisone}
 \quad X_{\{4,68\}} =\def\arraystretch{1.2}\left[\ba{c|c} 
\IP^1 &  2 \\
\IP^1 &  2 \\ 	
\IP^1 &  2  \\
\IP^1 & 2  \\ 
\ea\right]
 \end{eqnarray}
as a single hypersurface of multi-degree $\{2,2,2,2\}$ in a product of four $\mathbb{P}^1$ factors. By choosing a point in a surface defined by any two ambient $\mathbb{P}^1$ factors, it is clear that the defining equation takes the form of a genus one curve defined via a $\{2,2\}$ hypersurface in the remaining $\mathbb{P}^1 \times \mathbb{P}^1$ factors. Thus, this manifold can be described as a genus one fibration $\pi: X \to \mathbb{P}^1 \times \mathbb{P}^1$. There are $6$ distinct (but equivalent) fibrations of this type. Likewise, there are $4$ manifest $K3$ fibrations, $\rho: X \to \mathbb{P}^1$, in which the $K3$ fiber is itself genus one fibered and is described as a $\{2,2,2\}$ hypersurface in a product of three $\mathbb{P}^1$ factors.

Fibers of the type described above -- evident from the algebraic description of the manifold -- have been referred to as Obvious Genus-One Fibrations (OGFs). As noted above, nearly all CICYs admit multiple fibrations of this kind. Of the 7,890 CICY three-fold configuration matrices it was noted in \cite{Anderson:2016cdu} that 7,837 admit at least one such fibration, with the average number of inequivalent obvious fibrations per manifold being 9.85. For the CICY four-folds, the percentage of obviously fibered manifolds is even higher with 921,420 out of 921,497 cases admitting such a fibration (here the average manifold can be described as OGF in 54.6 different ways \cite{Gray:2014fla}).

It is important to note however, that the existence of such obvious fibration structures can be dependent on the algebraic form of the manifold and hence, potentially incomplete. For example, consider the following CY threefold with Hodge numbers $(h^{1,1}, h^{1,2}) = (6, 51)$.
\bea\label{firstguy}
X_{\{6,51\}}= \left[
\begin{array}{c|ccc}
\IP^2 & 2 & 0 & 1  \\
\IP^1 & 1 & 1 & 0  \\
\IP^3 & 0 & 1 & 3 
\end{array}
\right]
\eea
By inspection, this manifold admits two obvious genus one fibrations of the form described in \eref{mrfib}, $\pi_1: X_{\{6,51\}} \to \mathbb{P}^2$ and $\pi_2:X_{\{6,51\}} \to dP_4$ where $dP_4$ denotes the fourth del Pezzo surface ($\mathbb{P}^2$ blown up at four generic points). These can be seen by splitting the configuration matrix up into two pieces, one describing the base and the other the fiber.
\beq \label{eg1aa}
\def\arraystretch{1.2}\left[\ba{c|:ccc} 
\IP^1 &  1 & 1 & 0 \\
\IP^3 &  0 & 1 & 3 \\ \hdashline
\IP^2 &  2 & 0 &1 \\
\ea\right]   \;\;\ , \;\;\;\;\;\;\; 
\def\arraystretch{1.2}\left[\ba{c|c:cc} 
\IP^3 &  0 & 1 & 3 \\ \hdashline
\IP^1 & 1 & 1 & 0 \\ 
\IP^2 &  2 & 0 & 1 \\
\ea\right]  \ .
\eeq
In the first case, the rows of the configuration matrix have been reordered to separate the $\mathbb{P}^2$ base from the fiber and in the second case, the base surface
\bea
dP_4=\left[\ba{c|c} 
\IP^1 & 1  \\ 
\IP^2 & 2  \\
\ea\right]  \ 
\eea
has been made clear. In each case if any point is selected on the base manifold, substituting the coordinates of this point into the remaining defining relations leads to (a specific complex structure and) equations which now depend only upon the coordinates in the first projective space factors (given above the dotted horizontal line in the two cases above). The degrees of the equations in the remaining variables satisfy \eref{deg-constraint} thus, these equations describe a Calabi-Yau one-fold -- a torus. If the choice of point in the base is varied, the complex structure describing the associated torus fiber will change, and so it is clear that each of the configuration matrices in \eref{eg1aa} is a non-trivial fibration of a genus-one curve over that base.  However, it must be noted that the description given in \eref{firstguy} is not unique. The \emph{same CY manifold} can also be described by the configuration matrix:
\bea\label{secondguy}
\tilde{X}_{\{6,51\}}= \left[
\begin{array}{c|cc:ccc}
\IP^3 & 0 & 0 & 0 &1 &3  \\
\IP^1 & 0 & 0 & 1& 1 & 0  \\  \hdashline
\IP^2 & 1 &1 &1 & 0 &0 \\
\IP^2 & 1 & 1 & 0 & 0 &1 
\end{array}
\right].
\eea
This description makes evident yet another fibration $\pi_3: \tilde{X}_{\{6,51\}} \to dP_3$ given by
\bea\label{newbase}
dP_3= \left[
\begin{array}{c|cc}
\IP^2 & 1 &1 \\
\IP^2 & 1 & 1  
\end{array}
\right].
\eea
The existence of OGF structures has also been observed in other constructions of CY manifolds (e.g. toric \cite{Kreuzer:2000xy}, gCICY constructions \cite{Anderson:2015iia} and CY quotient geometries \cite{Braun:2010vc,Constantin:2016xlj}) and its ubiquitous nature is suggestive of the fact that most CY manifolds with large enough topology may admit a genus one fibration. However, the above example illustrates that any characterization of fibrations that relies on one algebraic description of a given CY manifold is destined to be incomplete and that a \emph{full classification can only be possible via criteria that rely only on the fundamental topology of the CY manifold}. Fortunately, just such a tool exists for CY 3-folds and we will employ it in this work.

\subsection{Criteria for the existence of a genus one fibration}\label{oguiso-wilson}
Throughout this work we will refer to a fibration in which the generic fiber is a complex curve of genus one as a \emph{genus one fibration}\footnote{Sometimes in the literature the terms ``elliptic fibration" and ``genus one fibration" are used interchangeably. Here we will follow the convention that an elliptic fibrations is a genus one fibration with section. In the present work we do not attempt to identify sections in our classification of genus one fibrations (although tools to do this for CICYs have recently been developed in \cite{Anderson:2016ler}).}. The existence of a genus-one fibration in a Calabi-Yau $n$-fold has been conjectured by Koll\'ar \cite{kollar-criteria} to be determined by the following criteria:

\vspace{0.3cm}

\noindent \emph{Conjecture \cite{kollar-criteria}: Let $X$ be a Calabi-Yau $n$-fold. Then $X$ is genus-one fibered iff there exists a $(1,1)$-class $D$ in $H^2(X, \mathbb{Q})$ such that $D \cdot C \geq 0$ for every algebraic curve $C \subset X$, $D^{{\rm dim}(X)}=0$ and $D^{{\rm dim}(X)-1} \neq 0$.}

\vspace{0.2cm}

\noindent In the case that $X$ is a Calabi-Yau threefold this conjecture has been proven by Oguiso and Wilson subject to the additional constraints that $D$ is effective or $D \cdot c_2(X) \neq 0$ \cite{ogu,wil}. Phrased simply these criteria are characterizing the existence of a fibration by \emph{characterizing a particular divisor in the base manifold of that fibration}. In particular, the role of the divisor $D$ above is that of a pull-back of an ample divisor in the base, $B$, where the fibration of $X$ is written $\pi: X \to B$. Such a divisor in $X$ is sometimes referred to as \emph{semi-ample} \cite{kollar-criteria}. The existence of $D=\pi^*(D_{base})$ makes it possible to define the form dual to points on the base (i.e. $D^{{\rm dim}(X)-1}$) which in turn determines the class of the genus-one fiber itself\footnote{It should be noted that the existence of a fibration structure within a smooth Calabi-Yau $n$-fold with $n>2$ is a deformation invariant quantity ({i.e.} given a fibered manifold, every small deformation is also fibered)\cite{kollar-criteria,wil,wilson2}. Indeed this must clearly be the case if the above conjecture is to make sense.}.  While Koll\'ar's conjecture has yet to be proven for CY manifolds in arbitrary dimensions, for threefolds, this is a well-established if and only if condition that can be used to determine whether or not fibrations exist. In this paper we will employ the criteria above to enumerate all genus one fibrations in a set of CICY 3-folds ($K3$ fibrations will also be enumerated using different means in Section \ref{ogf_sec}). Throughout this work, we will refer to an effective divisor satisfying the criteria in the conjecture above to be a ``Koll\'ar divisor".

Before beginning such an enumeration, it should be noted that there can in fact be many divisors $D$ of the form above for a single fibration structure in $X$. Thus, to count fibrations using this tool, the question of redundancy must be addressed. For a given fibration $\pi: X \to B$ there are in general, an infinite number of divisors $D \subset X$ satisfying the criteria above. For example, for a fibration $\pi: X \to \mathbb{P}^2$ not only will the pull-back of the hyperplane class, $H$, of the base $\mathbb{P}^2$, satisfy $D^2 \neq 0$ and $D^3 =0$, but also any multiple of it, $aH$ for $a \in \mathbb{Z}_{>0}$. This is not surprising since for any value of $a$, $D^2$ defines both a good volume form for $\mathbb{P}^2$ and the class of one or more fibers (i.e. $a^2$ fibers) of $\pi:X \to B$.

To eliminate this redundancy of counting, we will consider two divisors $D, D' \subset X$ to define \emph{generically the same fibration} if the fiber classes they define are proportional curves within $X$. That is,
\beq\label{proport_fibs}
D^2 \sim D'^2~~~\text{as curves in}~ X
\eeq
If this proportionality is satisfied, there are two immediate possibilities that are likely to arise: a) The fibers are proportional as in \eref{proport_fibs} and as in the $\pi: X \to \mathbb{P}^2$ example above, are associated to the same base and hence, the same fibration (in this case $D$ and $D'$ just count multiple copies of the same fundamental fiber class) or b) The two fibrations differ at non-generic points over the base. This case would be expected in cases where the two base geometries (associated to $D$ and $D'$) are birational. We will study such possibilities in detail in Sections \ref{the48_sec} and Appendix \ref{shrink}. Throughout this work, the criteria in \eref{proport_fibs} will be most useful to us to establish that when proportionality fails, the two possible fibrations are definitely distinct (and not even birational). 

Finally, we note that since triple intersection numbers of divisors in CY threefolds are generally easier to compute than double intersection numbers, we will frequently apply this test as
\beq\label{prop_fib_triple}
D \cdot D \cdot D_r = a D' \cdot D' \cdot D_r
\eeq
for some $a$ and every divisor $D_r$, $r=1, \ldots h^{1,1}$ in the basis.

With these results in hand, we turn now to a brief summary of our approach and key results.

\subsection{Enumeration of fibrations and key results}\label{set_up_sec}
The goal of this work is to systematically count genus one fibrations in the dataset of CICY threefolds. There are two distinct ways that we undertake this study:
\begin{enumerate}
\item By enumerating obvious fibrations (OGFs as defined in Section \ref{sec:Intro}) that are apparent from the given algebraic (in this case complete intersection) form of the CY geometry.
\item By utilizing the criteria in Section \ref{oguiso-wilson} to scan for possible base divisors $D$ and thereby to systematically enumerate all fibrations.
\end{enumerate}
Since all surveys in the literature to date have involved the first approach, we will be interested in undertaking both and comparing the totals where possible. In addition, we would like to probe other fibration structures (i.e. $K3$- or Abelian Surface fibrations). It should also be noted that at present the ``obvious" fibration approach is our only tool to count $K3$ fibrations or to consider compatible (i.e. nested) $K3$ and genus-one fibrations.

It is clear from the Koll\'ar-Oguiso-Wilson criteria laid out in Section \ref{oguiso-wilson} that a systematic search for genus-one fibrations must begin with a clear determination of all intersection numbers in the CY geometry as well as the structure of the K\"ahler and Mori cones. Despite the fact that the CICY dataset has existed for nearly 30 years, this information was still incomplete for the majority of manifolds in the list. In the following sections we compute the triple intersection numbers of all CICY threefolds and provide a description of the K\"ahler and Mori cones for the subset of ``K\"ahler Favorable" geometries whose K\"ahler/Mori cones descend in a simple way from an ambient space. For this subset of 4957 manifolds out of 7890, we are able to completely classify all genus one fibrations.

Thus our first results, laid out in Section \ref{top_sec} are,

\begin{itemize}
\item Algorithmic tools are developed to systematically replace CICY configuration matrices with new descriptions that provide an easy determination of their topological data (i.e. Hodge numbers, $c_2(X)$ and triple intersection numbers, $d_{rst}$, $r=1, \ldots h^{1,1}$). We construct this complete topological data for \emph{all CICY threefolds}.
\item For the $4957$ K\"ahler favorable geometries their K\"ahler and Mori cones are constructed explicitly. $4874$ of these geometries are K\"ahler favorable with respect to an ambient product of projective spaces and $83$ are K\"ahler favorable with respect to an ambient space defined as the product of two almost del Pezzo surfaces.
\end{itemize}
With these tools available, we then undertake the fibration surveys described above. In Section \ref{ogf_sec} we enumerate obvious fibration structures extending the tools developed in \cite{Gray:2014fla,Anderson:2015yzz,Anderson:2016ler,Anderson:2016cdu}. These are applied to all 7868 CICY threefolds which are not direct products. We find
\begin{itemize}
\item  $139,597$ obvious genus one fibrations.
\item $30,974$ obvious K3 fibrations.
\item $208,987$ distinct nestings of these fibrations.
\end{itemize}

In section \ref{OGF_vs_Kollar} we complete a scan for Koll\'ar divisors of the type described in Section \ref{oguiso-wilson} for the $4874$ K\"ahler favorable geometries descending from an ambient space of the form $\mathbb{P}^{n_1} \times \ldots \times \mathbb{P}^{n_m}$ and compare this to the OGF count for these manifolds. We find that here 
\begin{itemize}
\item The number of OGF fibrations exactly matches the exhaustive list of fibrations (obtained by counting Koll\'ar divisors). In these cases the (special) chosen algebraic form of the manifold has captured all relevant fibration structures.
\end{itemize}
Finally, there remains to consider the $83$ CICY configurations which are K\"ahler favorable with respect to an ambient space of the form $S \times S'$ where $S, S'$ are almost del Pezzo surfaces (i.e. $\mathbb{P}^1 \times \mathbb{P}^1$, $dP_r$ with $r=0, \ldots 7$ or the smooth rational elliptically fibered surface denoted as $dP_9$ in the physics literature). This class of geometries is studied in Sections \ref{the48_sec} and \ref{schoen_sec}. For these CY geometries
\begin{itemize}
\item For the $83$ CICYs defined as hypersurfaces in a product of almost del Pezzo surfaces, the criteria given in Section \ref{oguiso-wilson} produce vastly more fibrations than the OGF count.
\item More precisely, for the CYs defined as an anticanonical hypersurface in a product of two del Pezzo surfaces, we find $327,340$ fibrations, of which at most $1,289$ are OGFs.
\item Combining the counts of genus-one fibrations classified in all K\"ahler favorable geometries (with ambient spaces consisting of products of projective spaces and almost del Pezzo surfaces) we provide a complete classification of $377,559$ fibrations in total on $4,957$ manifolds.
\item For one manifold -- the threefold\footnote{Sometimes referred to as the ``Schoen" manifold (due to its study in \cite{schoen}) or the ``split bi-cubic" (from its original inclusion in \cite{Candelas:1987kf}).}  with Hodge numbers $(h^{1,1},h^{2,1})=(19,19)$ -- our survey yields an infinite number of genus-one fibrations.
\end{itemize}

Finally, in Sections \ref{conclusions_sec} we provide an overview of our conclusions and future applications of this work. The Appendices provide a collection of useful technical results. All of the data outlined above, including a new augmented CICY list (with complete topological data), and all the fibration data described is publicly available at \cite{website} and in part through an arXiv attachment associated to this work.

\section{Completing the topological data of the CICY 3-folds: intersection numbers and K\"ahler cones}\label{top_sec}

As described in the Introduction, in any attempt to systematically classify all genus one fibrations within a dataset of Calabi-Yau manifolds, it must first be possible to fully determine, for each manifold, $X$:
\begin{itemize}
\item The K\"ahler and Mori Cones of $X$.
\item The triple intersection numbers of all effective divisors on $X$.
\end{itemize}

In this section we attempt to characterize both of these structures as far as possible for the entire CICY threefold dataset using all available tools. We will begin with a systematic approach to determining the Picard groups of CICY threefolds.

\subsection{Splitting configuration matrices to produce favorable descriptions} \label{splitsec}
In the context of this work, when all divisors (equivalently the Picard group) of a Calabi-Yau three-fold $X$ descend from the simple ambient space $\cA$, we refer to it as a ``favorable" geometry \cite{Anderson:2008uw}. To determine explicitly when this occurs, consider the adjunction sequence and its dual:
 \bea\label{adjun}
  0 \ra TX \ra T\cA|_X \ra \cN|_{X} \ra 0, \\ \nonumber
    0 \ra \cN^*|_{X} \ra T\cA^*|_X \ra TX^* \ra 0.
 \eea
The latter induces the long exact sequence in cohomology,
 {\small
\bea\label{eq_generalizedkoszulwithdiviso}
      \parbox{0.1cm}{\xymatrix{
          &
           \hspace*{-1cm} \cdots \fto H^{1} (X, \cN|_{X}^*)   \xrightarrow{\quad\quad\alpha\quad\quad}    &
         {H^{1} (X, T\cA|_{X}^*)}  \xrightarrow{\quad\quad\quad}  &
      H^1(X, TX^*) \ar`[rd]`[l]`[dlll]`[d][dll] &
         \\ &
           \quad  H^{2} (X, \cN^*|_{X})   \xrightarrow{\quad\quad\beta\quad\quad}     &
               {H^{2} (X, T\cA|_{X}^*)}  \xrightarrow{\quad\quad\quad}  &
             H^2(X, TX^*)  \fto \cdots\,. &
        }}%^\big._\big.
    \eea}
It follows that the K\"ahler moduli of $X$ can be decomposed as $H^1(X, TX^*) \cong H^{1,1}(X) \cong {\rm coker }(\alpha) \oplus {\rm ker} (\beta)$.  These two contributions correspond to the descent of the K\"ahler moduli on $\cA$ to K\"ahler moduli on $X$ and K\"ahler forms that arise on $X$ only (i.e. non-toric divisors) . If the contribution from $\rm{ker}(\beta)$ is zero, the only divisors on $X$ are those descending from ${\cal A}$ (possibly with additional linear relations) and we say the geometry is ``favorable". In such a case we see that $h^{1,1}(X)  = {\rm  dim}(H^{1,1} (X)) \leq {\rm dim}({\rm Pic}(\cA))$.  The simplest case of a favorable  geometry is when $h^{2} (X, \cN|_{X}^*) = 0$ (or by Serre duality, when $h^{1} (X, \cN|_{X}) = 0$). Of the original $7890$ configuration matrices in the CICY three-fold dataset \cite{Candelas:1987kf}, there are 4896 favorable geometries (including 22 direct product geometries) and $2994$ unfavorable geometries.
 
For $2994$ manifolds then, there are non-toric divisors present from the point of view of the given configuration matrix and it is clear that the standard tools (see for example \cite{Hubsch:1992nu}) will not suffice to determine the required data for a fibration scan. We turn next to one approach to remedying this deficit.

\subsubsection{A review of CICY splitting/contraction}
To improve this situation, in this work, we make systematic use of a known approach to exchanging one configuration matrix with another that \emph{describes the same CY threefold}. This process, known as ``splitting" or ``contracting" a CICY has long been utilized in the context of this dataset of manifolds \cite{Candelas:1987kf}. In fact, the original generating algorithm of the CICY threefold dataset was designed to remove many such redundancies from the list.

The notion of splitting/contracting first arose naturally in the context of conifold transitions \cite{Candelas:1989ug}. For example the famous conifold of the quintic: 

\vspace{7pt}

~~~~~~~~~{\small $ \left[\begin{array}[c]{c}\mathbb{P}^4\end{array}
\left|\begin{array}[c]{c}5
\end{array}
\right.  \right]^{1,101}$}~~~ (Def) $\Leftrightarrow$ {\small $l_1 q_2 -l_2 q_1=0$} (sing. locus) $\Leftrightarrow$ {\small $ \left[\begin{array}[c]{c}\mathbb{P}^1\\\mathbb{P}^4\end{array}
\left|\begin{array}[c]{ccc}1 & 1 \\
1& 4 \\
\end{array}
\right.  \right]^{2,86}$} (Res)~~~.

\vspace{7pt}

\noindent Here the left and right configuration matrices form the deformation and resolution sides of the conifold, respectively. The two topologically distinct geometries share a common \emph{singular} locus in their moduli space -- in this case the nodal quintic (given in the center above, where $l_i$ and $q_i$, $i=1,2$ are linear and quartic polynomials in the coordinates of $\mathbb{P}^4$). See \cite{Candelas:1989ug} for a review. An example of a CICY topology changing transition such as this is called an ``effective splitting" of the initial manifold (in this case the quintic). However, there is another possibility in that the shared locus in moduli space between two configuration matrices need not be singular. For example, the singularities arise from the nodal quintic  above for the 16 points where $l_1=l_2=q_1=q_2=0$. On $\mathbb{P}^4$ there exists a common solution to the four equations, however, if the ambient space had been say, $\mathbb{P}^3$, no such solution would exist. When the shared locus in moduli space is \emph{smooth} the splitting operation on the configuration matrix is referred to as an \emph{ineffective splitting}. Because the manifolds described by the initial configuration matrix and its split then share a common smooth locus in moduli space, they are topologically equivalent.

In the remainder of this section, we will use this observation and the technique of ``ineffective splitting" to try to determine when it is possible to split an unfavorable configuration matrix of a CICY three-fold to a favorable one.  It is clear in principle that such ineffective splittings in general increase the number of rows/columns of the configuration matrix and as a result, will likely change the number of ``obvious" genus one fibrations available.

More precisely, a $\IP^n$-splitting of a CICY configuration matrix (corresponding to the manifold, $X$) can be written
as follows: 
\beq\label{gen_split}
X=\left[ {\cal A\,} |\, {\bf c} \; {\cal C} \right] \longrightarrow { X'}= \left[ \begin{array}{c|c c c c c}
{\mathbb P}^n & 1 & 1 & \ldots & 1 & {\bf 0} \\
{\cal A} & {\bf c}_1 & {\bf c}_2 & \ldots & \bf{c}_{n+1} & {\cal C} 
\end{array} \right] \; ,\quad {\bf c}=\sum_{{\alpha}=1}^{n+1} {\bf c}_{\alpha}\; .
\eeq
We begin with an initial CICY three-fold, $X$, defined above by a starting configuration matrix of the form $\left[ {\cal A}\, |\, {\bf c}\; {\cal C} \right]$ where ${\cal A}= \mathbb{P}^{n_1} \times \ldots \mathbb{P}^{n_m}$ and ${\bf c}$ and ${\cal C}$ form an $m \times K$ matrix of polynomial degrees for the $K$ equations defining the complete intersection hypersurface. The first column of this matrix, ${\bf c}$, has been explicitly separated from the remainder of the columns, denoted by ${\cal C}$, to facilitate the rest of our discussion.  Since $X$ is a three-fold, $\sum_{r=1}^m n_r -K=3$. We can ``split" $X$ by introducing the new configuration matrix ${X'}$ where the vector ${\bf c}$ has been partitioned as the sum of $n+1$ column vectors ${\bf c}_i$ (of dimension $m$) with nonnegative components, as indicated. Since ${X'}$ is still a three-fold, the new configuration matrix is $(m+1) \times (K+n)$ dimensional. 

While the process of going from $X$ to ${X'}$ is called ``splitting", the reverse process, in which  ${X'} \to X$, is called a ``contraction" \cite{Candelas:2007ac}. As described above, in some cases, a splitting  of the form \eref{gen_split} will not produce a new (i.e. topologically distinct) Calabi-Yau three-fold, but rather a new description of the same manifold.  As in the case of the quintic above, in either an effective or ineffective splitting, two manifolds $X$ and $X'$ related as in (\ref{gen_split}) share common loci in their complex structure
moduli space -- the so-called ``determinental variety". It is defined as follows. Take the subset of the defining relations of $X'$ corresponding to the first
$n + 1$ columns on the right hand side of (\ref{gen_split}).
\beq\label{gendet}
\left( \begin{array}{cc c c}
f^{1}_{1} & f^{1}_{2} & \ldots & f^{1}_{n+1} \\
f^{2}_{1} & f^{2}_{2} & \ldots & f^{2 }_{n+1} \\
\vdots & \vdots & \ddots & \vdots \\
f^{n+1}_{1} & f^{n+1}_{2}& \ldots & f^{n+1}_{n+1} 
\end{array} \right)\left( \begin{array}{c}
x_0 \\
x_1 \\
\vdots \\
x_{n}
\end{array} \right)=0 \ ,
\eeq
Here $f_\alpha^k$
is of degree ${\bf c_\alpha}$ for all $k$. The determinental variety (i.e. shared locus in moduli space) is found by taking the determinant of the matrix in \eref{gendet} and combining it with the remaining equations whose degree is governed by ${\cal C}$.

We can thus state more clearly the observation made above: if the two configurations $X$ and $X'$ can be smoothly deformed into each other and hence represent the same topological type of Calabi-Yau manifolds, the
splitting is  called  ``ineffective" \cite{Candelas:2007ac}.  Otherwise it is an ``effective" splitting.
Thus, the question of whether a given splitting is effective or ineffective is decided by whether or not the determinental
variety defined via \eref{gendet} is smooth. For all CICY three-fold splittings, the singular locus of the determinental variety  is a
zero-dimensional space.  That is, it can either be the empty set or a collection of points. It
turns out that the number of singular points is counted  by the difference in Euler characteristic between the original and the split configuration (the 16 singular points lead to $\Delta \chi=32$ in the example above).
This leads to the simple rule that two three-fold configurations, related by splitting as
in (\ref{gen_split}), are equivalent if and only if they have the same Euler characteristic.
In this case, the
splitting is ineffective. 

\subsubsection{Finding favorable splitting chains}

With these definitions in hand, we now turn to the question of when can a chain of ineffective splittings of a configuration matrix take a non-favorable configuration matrix to a favorable one? An important tool in this regard was provided by a small lemma in \cite{Anderson:2013qca} which we re-state here for completeness:
 \begin{lemma}\label{lemma1}
 Suppose that $X$ and $X'$ are two Calabi-Yau three-folds realized as complete intersections in products of projective spaces and related by a splitting  of the type described in \eref{gen_split}. Let ${\cal{L}}=\cO_{X}(a^{1},\ldots, a^{m})$ be a ``favorable" line bundle on $X$--that is, a line bundle corresponding to a divisor $D \subset X$ such that $D={D}_{\cal{A}}|_X$ is the restriction of a divisor ${D}_{\cal{A}}$ in 
the ambient space. Then the calculation (and dimension) of the cohomology of ${\hat {\cal{L}}}=\cO_{{X'}}(0\ldots ,0,a^{1},\ldots, a^{m})$ on $X'$ (defined by \eref{gen_split}) is identical to that of ${\cal{L}}$ on $X$ on the locus in complex structure moduli space shared by $X$ and ${X'}$.
\end{lemma}
See \cite{Anderson:2013qca} for a proof/discussion.

Returning to the adjunction sequence \eref{adjun} above, we see that a CICY configuration matrix will be potentially non-favorable whenever $h^1(X, {\cal N}_X) > 0$ (denoting ${\cal N}_X={\cal N}|_X$ as a sum of line bundles on $X$). Thus, in the process of splitting, we would like to know when it is possible to generate a new configuration matrix such that $h^1(X', {\cal N'}_{X'})$ goes to zero? For any CICY configuration matrix, ${\cal N}_X$ is simply a sum of line bundles and from the lemma above, it is clear that the line-bundle cohomology of any line bundle, $\cL \subset {\cal N}_X$, on $X$ does not change if we \emph{do not split} the elements (i.e. partition the multi-degree) of the column in $X$ associated to that component of the normal bundle (i.e. $\cL$). 

As a result, to find an ineffective split that changes an unfavorable to a favorable manifold, it is not necessary to split any part of ${\cal N}_X$ -- i.e. column of the configuration matrix -- in which the associated line bundle cohomology gives  $h^{1} (X, \cL)= 0$.  Instead, we will systematically  consider splitting only those columns for which the associated line bundle cohomology is non-vanishing and determine whether splitting reduces that number.

To make the somewhat opaque description above more clear, it is useful to illustrate this with an explicit CICY configuration matrix. One example of a non-favorable CICY is given by the following configuration matrix. 

 \bea\label{first_split_eg}
X= \left[
\begin{array}{c|ccc}
& \cL_1  & \cL_2 & \cL_3 \\\hdashline
\IP^2 & 2 & 0 & 1  \\
\IP^1 & 1 & 1 & 0  \\
\IP^3 & 0 & 1 & 3 
\end{array}
\right].
\eea
 where $(h^{1,1}(X), h^{1,2}(X)) = (6, 51)$ and $\chi (X)= -90$. Since $h^{1,1}(X)=6$, but only three K\"ahler forms descend from the ambient projective spaces, this configuration matrix is clearly unfavorable in the original CICY list.
 
To begin, it can be verified that only one line bundle in ${\cal N}_X$ has a non-zero $h^1$. That is, denoting ${\cal N}_X={\cal L}_1 \oplus {\cal L}_2 \oplus {\cal L}_3$ (with multi-degree given by the columns in \eref{first_split_eg} above), we find the cohomology dimensions: $h^{\bullet}(X, \cL_1)=(11,3,0,0)$, $h^{\bullet}(X, \cL_2)=(7,0,0,0)$, $h^{\bullet}(X, \cL_3)=(59,0,0,0)$.  Overall, the dimension of the cohomology of the normal bundle is $h^{\bullet}(X, \cN_{X})$ $=(77,3,0,0)$.  Since $h^1(X, \cL_1)\neq0$, in order to find a favorable description we must begin with a splitting which partitions the column associated to the line bundle $\cL_1$.

For this column there is only one split available which will non-trivially partition the entries and add a $\IP^1$ factor to the ambient space as:
  \bea
X'= \left[
\begin{array}{c|ccc}
&\cL_1'& \cL_4  & \hat\cL_{2,3} \\\hdashline
\IP^1 & 1 & 1 & {\bf 0}   \\
\cA & {\bf c_1} & {\bf c_2} & \cC
\end{array}
\right]= \left[
\begin{array}{c|cccc}
&\cL_1'& \cL_4  & \hat\cL_2 & \hat\cL_3 \\\hdashline
\IP^1 & 1 & 1 & 0 & 0   \\
\IP^2 & 1 &1 & 0 & 1  \\
\IP^1 & 1 & 0 & 1 & 0  \\
\IP^3 & 0 & 0 & 1 & 3 
\end{array}
\right]
\eea
For this new configuration matrix, $h^{\bullet}(X', \cL_1')=(9,2,0,0)$ and $h^{\bullet}(X', \cL_4)=(5,0,0,0)$, while due to the lemma above, the cohomology of $\hat\cL_{2,3}$ stays the same. It is easy to verify that this splitting is ineffective with $\chi (X')=-90$. Moreover, by performing this $\mathbb{P}^1$-split, the dimension of the first cohomology of the normal bundle decreases from $h^{1}(X, \cN_{X})=3$ to  $h^{1}(X', \cN'_{X'})=2$ while $h^{2}(X', \cN'_{X'})=0$. It is clear that this splitting has produced a potentially slightly more favorable configuration matrix and furthermore, that this process can be continued -- that is, there are still further splittings of the configuration available to us.

Starting again from the configuration $X'$, we can proceed to split the first column in $\cL_1' \in X'$ with a $\IP^2$ in a way that the new submatrix ${\bf c_i}$ has the maximal rank: 
 \bea\label{secondstep}
X''= \left[
\begin{array}{c|cccc}
& \cL_1'' &\cL_6 & \cL_5  & {\hat\cL_{2,3,4}} \\\hdashline
\IP^2 & 1 & 1 &  1 &{\bf 0}   \\
\cA & {\bf c'_1} & {\bf c'_2} &  {\bf c'_3} & \cC
\end{array}
\right]= \left[
\begin{array}{c|cccccc}
& \cL_1'' &\cL_6 & \cL_5 &  \hat\cL_4  & \hat\cL_2 & \hat\cL_3 \\\hdashline
\IP^2 & 1 & 1 & 1 & 0 & 0 & 0 \\
\IP^1 & 0 & 0  & 1 & 1 & 0 & 0   \\
\IP^2 & 1 & 0  & 0 &1 & 0 & 1  \\
\IP^1 & 0 & 1 & 0 & 0 & 1 & 0  \\
\IP^3 & 0 & 0 & 0 & 0 & 1 & 3 
\end{array}
\right]
\eea
with $h^{\bullet}(X'', \cL_1'')=(7,1,0,0)$, $h^{\bullet}(X'', \cL_{5,6}) =(5,0,0,0)$. Once again, the remaining normal bundle cohomology and the overall Euler number of the manifold is unchanged. At this step in the splitting chain, the dimension of the first cohomology of the normal bundle decreases from $h^{1}(X', \cN'_{X'})=2$ to  $h^{1}(X'', \cN''_{X''})=1$ while the second cohomology group is still zero.  

It is important to note at this stage, that even having identified a problematic element of the normal bundle (such as ${\cal L}_1$ above), not all splittings will cause the relevant cohomology, $h^1(X, {\cal L}_1)$ to decrease. In general, an analysis of the associated long exact sequences in cohomology demonstrates that the maximal change is possible when  the new submatrix, ${\bf c_i}$, is of maximal rank. For example, an alternative splitting to \eref{secondstep} is
$$
\hat X= \left[
\begin{array}{c|cccccc}
& \cL_1'' &\cL_6 & \cL_5 &  \hat\cL_4  & \hat\cL_2 & \hat\cL_3 \\\hdashline
\IP^2 & 1 & 1 & 1 & 0 & 0 & 0 \\
\IP^1 & 0 & 0  & 1 & 1 & 0 & 0   \\
\IP^2 & 0 & 1  & 0 &1 & 0 & 1  \\
\IP^1 & 0 & 1 & 0 & 0 & 1 & 0  \\
\IP^3 & 0 & 0 & 0 & 0 & 1 & 3 
\end{array}
\right]
$$
For this configuration, $h^{\bullet}(\hat X, \cL_{1}'') =(2,0,0,0)$  while  $h^{\bullet}(\hat X, \cL_{6}) =(9,2,0,0)$,  $h^{\bullet}(\hat X, \cL_{1}'') =(3,0,1,0)$. Unfortunately, $h^{1}(\hat X, \hat{\cN}_{\hat X})$ does not decease while $h^{2}(\hat X, \hat{\cN}_{\hat X})$ increases. Finally, it should be noted that \emph{even with maximal rank splittings of a column} in some non-generic cases, the cohomology may not decrease in the desired manner. We will return to this in a moment, but for now it is enough to observe that in general there are only a few choices of maximal rank splittings available and thus this process is suitable for an automated, algorithmic search for ineffective, favorable splittings.

To conclude the example at hand, we have one further step to take from the configuration, $X''$, in \eref{secondstep}. Once again, the final $\mathbb{P}^1$ splitting is performed on the first column in the configuration matrix $\cL_1''$ on $X''$ as
\bea
X'''= \left[
\begin{array}{c|ccc}
& \cL_1''' &\cL_7 & {\hat\cL_{2,3,4,5,6}} \\\hdashline
\IP^1 & 1  &  1 &{\bf 0}   \\
\cA & {\bf c''_1} & {\bf c''_2}  & \cC
\end{array}
\right]= \left[
\begin{array}{c|ccccccc}
& \cL_1''' & \cL_7 &\hat\cL_6 &\hat\cL_5 &  \hat\cL_4  & \hat\cL_2 & \hat\cL_3 \\\hdashline
\IP^1 & 1 & 1 & 0 & 0 & 0 & 0 & 0\\
\IP^2 & 0 & 1 & 1 & 1 & 0 & 0 & 0 \\
\IP^1 &0 & 0 & 0  & 1 & 1 & 0 & 0   \\
\IP^2 & 1 & 0 &0  & 0 &1 & 0 & 1  \\
\IP^1 &0 & 0 & 1 & 0 & 0 & 1 & 0  \\
\IP^3 & 0 &0 & 0 & 0 & 0 & 1 & 3 
\end{array}
\right]
\eea
with $h^{\bullet}(X''', \cL_1''')=(5,0,0,0)$, $h^{\bullet}(X'', \cL_{7}) =(5,0,0,0)$ and $\chi(X''')=-90$. Now at last, after a three-step chain of splittings, a configuration matrix has been obtained with  $H^{1} (X''', \cN_{X'''}) = 0$. Thus, the procedure outlined above has produced a new, equivalent description of the same CY manifold, but one for which we have complete control of the divisors/line bundles via restriction from a simple ambient space.

In summary, it is clear that for a given CICY configuration matrix, there are a finite number of such splitting chains that have the potential to lead to a new, favorable description of the manifold via ineffective splitting. In practice, a computer search can easily be implemented. The algorithm we employed consists of the following steps: 

\begin{enumerate}

\item Begin by computing the line-bundle cohomology for each component of the normal bundle (i.e. column of the matrix) and split (in any order) those with non-zero $h^1(X, \cL)$ cohomology. Due to the Lemma, other line-bundle cohomology groups will not change in the splitting process.
\item If the maximal size of the degree entries in the chosen column/line bundle $\cL=\cO_X(a^1, \dots, a^m)$ is $2$, split it with $\IP^1$ as:\footnote{In the original CICY list with unfavorable descriptions, $2$ is the largest degree/entry for the line-bundle involved in the splitting.} 
\bea
 \left[
\begin{array}{c|cc}
\IP^n & 2 & {\bf b}  \\
\cA & {\bf c} & {\bf \cC}
\end{array}
\right]=
 \left[
\begin{array}{c|ccc}
\IP^1 & 1  &  1 &{\bf 0}   \\
\IP^n & 1 & 1  & {\bf b}  \\
\cA &   {\bf c_1} & {\bf c_2} & \cC
\end{array}
\right]
\eea
and at the same time choose degree partitions such that the submatrix ${[\bf c_{i,j}]}$ is maximal rank.
If the largest degree entry in the chosen line bundle $\cL$ is $1$, then perform a $\IP^n$-split, where $n=\sum_{i}^{m} a^i -1$ and choose the submatrix ${\left[\bf c_{i,j}\right]}$ to be of maximal rank.
\item For each step of splitting, verify that the split is ineffective by computing the Euler number of the new configuration matrix. 
\item Repeat these procedures whenever $h^{1} (X, \cN_{X})$  decreases while $h^{2} (X, \cN_{X})$ is unchanged.   Finish the procedure when  $h^{1} (X, \cN_{X})=0$ and a favorable description of the manifold has been obtained.
\end{enumerate}

Implementing this search in the original CICY database \cite{Candelas:1987kf}, there are $2994$ unfavorable configuration matrices to be analyzed. A search as described above readily provides a new, favorable description of $2946$ of them.  For the remaining 48 configuration matrices an exhaustive search demonstrates that no chain of splittings/contractions will lead to a favorable description. The remaining $48$ configuration matrices will be dealt with separately in Section \ref{the48_sec} where we will demonstrate that this set of $48$ geometries in fact contains $15$ descriptions of the same CY threefold (the so-called ``Schoen manifold" with Hodge numbers $(19,19)$) and $33$ others. Of these latter manifolds, we find a further $9$ redundancies and observe that the remaining 24 distinct geometries can all be simply described as hypersurfaces defined in an ambient product of two del Pezzo surfaces. 

For now, we see that the simple process of splitting has allowed to generate a new version of the CICY list in which we have dramatically increased the number of favorable configurations to 7842 in total. For each of these new descriptions, we can employ existing tools \cite{Hubsch:1992nu,Anderson:2008ex} to fully specify the topological data of the manifold, including the triple intersection numbers, line bundle cohomology, etc. By combining these results with those from Section \ref{the48_sec} for the remaining $48$ manifolds we have produced a new version of the CICY list with all topological data fully specified. It is available at \cite{website} and in an attachment to the arXiv submission of this work.

\subsection{K\"ahler favorable manifolds}\label{kah_fav_sec}
As observed in Section \ref{set_up_sec}, a fibration scan crucially relies on the characterization of the K\"ahler and nef cones. Although the K\"ahler cone of a hypersurface in a Fano variety descends simply from the ambient space, in general, few tools exist to characterize the K\"ahler cones of complete intersection Calabi-Yau manifolds (even those defined in simple ambient spaces). In this section, we review two useful tools that together help us to determine the K\"ahler and Mori cones of 4957 out of 7890 configuration matrices in the list of CICY threefolds. This set has the simple property that their K\"ahler cones descend from the ambient space in which they are embedded. We will refer to such manifolds as \emph{``K\"ahler Favorable"} configurations. For these, we will provide a complete classification of all genus one fibrations in the CY geometries.

To begin, let us review two simple results about K\"ahler and Mori cones of CICYs. The first is the following result for the cone of curves (denoted $\overline{NE}(X)$) of CY hypersurfaces in Fano fourfolds proved in \cite{descend}:
 \begin{lemma}\label{lemma2}
(Koll\'ar) Let $Y$ be a smooth Fano variety with $\rm{dim}(Y) \geq 4$. Let $X \subset Y$ be a smooth divisor in the class $-K_Y$ (in fact $X$ can have arbitrary singularities). Then the natural inclusion
 \beq
 i_*: \overline{NE}(X) \to \overline{NE}(Y)
 \eeq
 is an isomorphism.
 \end{lemma}
Thus, a CY hypersurface in any \emph{smooth} Fano fourfold has a cone of algebraic curves that descends simply from its ambient space. Moreover, from the simple form of intersection numbers on a hypersurface, it follows that the (dual) K\"ahler cone also descends from the ambient space (for a careful set of arguments on the descent of the effective, nef and ample cones of divisors see \cite{oguiso_peternell,2016arXiv161100556L}). We will utilize this, and one additional result, to describe the K\"ahler cones of 83 CICYs described as hypersurfaces in a product of two almost del Pezzo surfaces in Sections \ref{the48_sec} and \ref{schoen_sec}.

For more general complete intersections $X \subset {\cal A}$, the first observation to be made is that every K\"ahler form on ${\cal A}$ restricts to a K\"ahler form on $X$. For the CICY threefolds defined in products of projective spaces considered here, the K\"ahler cone of $\mathbb{P}^{n_1} \times \ldots \mathbb{P}^{n_m}$ is simply the positive orthant (see \cite{Candelas:1987kf} for details). Thus, for all the favorable manifolds in the CICY list, it is clear that the K\"ahler cone of $X$ is at least as big as the positive orthant. In general, however, it could be larger. To illustrate this, consider the following example:
\bea\label{kahler_illustrate}
X= \left[
\begin{array}{c|ccc:c}
\IP^2 & 0 & 0 & 0 &3  \\\hdashline
\IP^1 & 1 & 0 & 0& 1  \\  
\IP^1 & 0 & 1 & 0 & 1  \\  
\IP^1 & 0 & 0 &0 & 1 \\
\IP^2 & 1 & 1 & 1 & 0
\end{array}
\right].
\eea
This manifold (with Hodge numbers $(h^{1,1},h^{2,1})=(5,59)$) is an anti-canonical hypersurface in $\mathbb{P}^2 \times dP_3$. It is also a manifest genus one fibration over $dP_3$ (with a $\mathbb{P}^2[3]$ fiber). Although it is favorable in the sense that its Picard group descends from the ambient product of $5$ projective spaces, it is \emph{not K\"ahler favorable} since the K\"ahler cone of $X$ is actually larger than the positive orthant! To see this, note that by Lemma \ref{lemma2} above the K\"ahler cone of $X$ is simply that of $\mathbb{P}^2 \times dP_3$. However, the K\"ahler cone of $dP_3$ is non-simplicial (with 5 generators \cite{Cvetic:2014gia}). Written in terms of a basis of the $5$ restricted hyperplanes ($D_i$, $i=1, \ldots 5$), the six generators of the K\"ahler cone of $X$ are
\beq\label{expan_cone}
\{D_1, D_2, D_3, D_4, D_5, D_2 +D_3 + D_4 -D_5 \}
\eeq
This last generator, $D_2 + D_3 + D_4 -D_5$, is manifestly not in the K\"ahler cone of the ambient space.

How, then, are we to determine when the K\"ahler cone of a CICY, $X$ is ``enhanced" in this way relative to the K\"ahler cone of the ambient space? It is clear that whenever the K\"ahler cone expands (as in the example above), the dual (i.e. Mori) cone must shrink. Thus, one simple way to determine when the K\"ahler cone of $X$ descends from that of ${\cal A}$ is to determine when \emph{the Mori cone remains the positive orthant}.

More precisely, consider the basis of curves dual to the K\"ahler forms $J_i$ (i.e. basis of $H^{1,1}$ restricted from the ambient space in a favorable CICY), defined via
\beq
\int_{C_i} J_j = \delta^i_j
\eeq
A general curve can be written in terms of this basis as $C=a^i C_i$. The claim for a favorable CICY then is \\

\emph{If $C_i$ are in the (closure of the) Mori cone for all $i$ then the K\"ahler cone is exactly the positive orthant}. \\

The expectation then is that if all $C_i$ (i.e. all the boundaries of the dual positive orthant) are in the Mori cone then it is clear that the Mori cone cannot be smaller than that orthant (as would be the case if the K\"ahler cone expanded as in \eref{expan_cone}). For the example above, it is clear that because of the presence of the last K\"ahler generator, $D_2+D_3+D_4-D_5$, the curve $C_5$ (dual to $J_5$) is not an element of the Mori cone. 

It remains then to try to determine when can we establish that $X$ contains effective curves in the class $[C_i]$? One simple (but certainly not exhaustive) approach\footnote{We would like to thank Andre Lukas for helpful discussions on this topic and for pointing out the utility of Gromov-Witten Invariants for this question. See also the recent work \cite{Buchbinder:2017azb} for alternative approaches to curve counting in CICY geometries.} is to use existing tools to determine the existence of curves of a given class and genus in $X$ -- namely to compute the Gromov-Witten Invariants of $X$.

In the case at hand  -- that of complete intersection Calabi-Yau manifolds in smooth toric ambient spaces -- techniques to compute the \emph{genus zero} Gromov-Witten Invariants are well established in the literature using mirror symmetry \cite{Candelas:1990rm,Hosono:1993qy,Hosono:1994ax}. In particular, the tools laid out in \cite{Hosono:1994ax} provide a simple algorithmic way to enumerate simple, algebraic curves in genus zero. In general, caution must be used in interpreting a Gromov-Witten invariant as an actual count of algebraic curves, however for the CICYs in consideration here (defined in simple, smooth toric ambient spaces) the results of a mirror symmetry computation lead to positive integers which are expected to give a physically relevant, enumerative count (see \cite{clemens,S.Katz:1986cba,Pandharipande:2011jz} for mathematical conjectures in this regard).

We employ the method of \cite{Hosono:1994ax} to determine the vector:
\beq
n(0, [C_i])=(\#, \ldots, \#)
\eeq
where $i=1, \ldots h^{1,1}(X)$ for every favorable CICY in the augmented CICY list attached to this arXiv submission. We find that for $4874$ out of $7820$ non-product, favorable CICY configuration matrices (in the new list), 
\beq
n(0, [C_i]) >0 ~~\forall i=1, \ldots h^{1,1}(X)
\eeq
Thus, for this subset of CICY manifolds, every dual curve $C_i$ in the positive orthant should in fact be effective and hence in the Mori cone. It follows from the logic above that the K\"ahler cones are in turn exactly the (dual) positive orthant. Since the K\"ahler cones for these manifolds descend exactly from the ambient product of projective spaces, they are K\"ahler favorable as defined above. Of course, a zero entry in the genus zero Gromov-Witten Invariant vector \emph{does not} necessarily imply that the Mori cone is smaller than the positive orthant. However, the condition above should be sufficient (though not in general necessary) and for these geometries we will provide a complete classification of genus one fibrations. We leave it to future work to thoroughly explore the full curve enumeration on CICYs and their correspondence with Gromov-Witten invariants. In addition, we would also hope to search for other tools that might fully determine the K\"ahler cones of the remaining CICYs.

Summarizing the two approaches outlined above, we find that 4957 out of 7890 CICYs have K\"ahler cones that descend from a simple ambient space -- either a product of projective spaces (with entirely non-vanishing $n(0, [C_i])$ as described above) or a product of two almost del Pezzo surfaces. Of this latter type, there are 83 geometries in the CICY list that can be written this way and we will analyze them in Section \ref{the48_sec}.

In the augmented CICY list attached to this arXiv submission, there are simple flags added to each entry to denote the status of the Picard group (``Favorable $\to$ True'' indicates the Picard group descends from the ambient space) and K\"ahler cone (``KahlerPos $\to$ True'' denotes a K\"ahler cone that descends from the ambient product of projective spaces). In addition, the configuration matrix, second Chern class and Hodge numbers are also provided. A sample entry in the new CICY list, with all data annotated, is given in Appendix \ref{app_data}.

%%%%%%%%%%%%%%%%%%%%
\section{A search for ``obvious" genus one fibrations}\label{ogf_sec}

\subsection{General comments on obvious Calabi-Yau fibrations}

As was described in the introduction, row and column permutations can be applied to the configuration matrix of a CICY without affecting the geometries it describes. Permuting rows simply corresponds to writing the ambient projective space factors in a different order, and permuting columns corresponds to relabeling the defining equations. Consider a case where such row and column permutations can be used to put a configuration matrix in the following block form.
\begin{eqnarray} \label{fib1}
X = \left[ \begin{array}{c|c:c}  {\cal A}_1 & 0 & {\cal F} \\  \hdashline {\cal A}_2 & {\cal B} & {\cal T} \end{array}\right]
\end{eqnarray}
Here ${\cal A}_1$ is a product of $m$ projective spaces and ${\cal A}_2$ is a product of $N-m$ projective spaces (where $N$ is the total number of such factors in the initial configuration). The blocks $0$ and ${\cal B}$ contain $n$ columns while ${\cal F}$ and ${\cal T}$ contain $K-n$. We will include cases where $n=0$.

A configuration which can be put in the form (\ref{fib1}) describes a fibration of the fiber $\left[ \begin{array}{c|cc} {\cal A}_1 &  {\cal F} \end{array} \right]$ over the base $\left[ \begin{array}{c|c} {\cal A}_2 & {\cal B} \end{array} \right]$ where the twisting of the fiber over the base is encoded by the matrix ${\cal T}$. To see this, consider the following line of reasoning. First, pick a solution to the first $n$ equations by choosing a point in ${\cal A}_2$ which satisfies the equations whose degrees are encoded by ${\cal B}$. This furnishes us with a point in the base. Take this set of coordinates in ${\cal A}_2$ and substitute it into the remaining $K-n$ equations, whose multi-degrees are determined by the matrices ${\cal F}$ and ${\cal T}$. This results in a particular set of equations, whose degrees are described by the configuration matrix $\left[ \begin{array}{c|cc} {\cal A}_1 & {\cal F} \end{array} \right]$, associated to that base point. As we change the base point the complex structure of this fiber over that base point will change. Thus we end up with a non-trivial fibration of this type over the base.

Note that the fiber $\left[ \begin{array}{c|cc} {\cal A}_1 &{\cal F} \end{array} \right]$ is a Calabi-Yau manifold. This is a simple consequence of the Calabi-Yau condition applied to the original configuration matrix (\ref{deg-constraint}), together with the presence of a completely zero block in the top left of (\ref{fib1}). For an initial configuration describing a Calabi-Yau p-fold, we can in general find Calabi-Yau q-fold fibers of this type for any $q<p$. For the case of $q=1$ fibrations that can be seen in this manner have been referred to as Obvious Genus One Fibrations (OGFs) \cite{Gray:2014fla,Anderson:2016ler}.

In fact, as has been noted before \cite{Gray:2014fla}, a given configuration will generically admit a multitude of different such fibrations. In other words, a given configuration matrix can often be put in the form (\ref{fib1}) in several different ways. For example, the following are all rearrangements of the same configuration matrix.
\begin{eqnarray} \label{eg1}
\left[ \begin{array}{c|cc:cccc} \mathbb{P}^2& 0&0&0&0&2&1 \\ \mathbb{P}^3& 0&0&1&1&1&1  \\\hdashline  \mathbb{P}^1 &1&0&1&0&0&0 \\ \mathbb{P}^2 & 1&2&0&0&0&0  \\\mathbb{P}^1 &0&1&0&1&0&0\end{array} \right]
\left[ \begin{array}{c|c:ccccc} \mathbb{P}^1& 0&1&1&0&0&0 \\ \mathbb{P}^2& 0&0&0&0&2&1  \\ \mathbb{P}^3 &0&0&1&1&1&1 \\\hdashline \mathbb{P}^1 &1&0&0&1&0&0 \\ \mathbb{P}^2 & 2&1&0&0&0&0\end{array} \right] 
\left[ \begin{array}{c|c:ccccc} \mathbb{P}^1& 0&0&1&1&0&0 \\ \mathbb{P}^2& 0&0&0&0&2&1  \\ \mathbb{P}^3 &0&1&0&1&1&1 \\\hdashline \mathbb{P}^1 &1&1&0&0&0&0 \\ \mathbb{P}^2 & 1&0&2&0&0&0\end{array} \right]  \\ \nonumber
\left[ \begin{array}{c|:cccccc} \mathbb{P}^2& 0&0&0&0&2&1 \\ \mathbb{P}^2& 1&0&2&0&0&0  \\ \mathbb{P}^3 &0&1&0&1&1&1 \\\hdashline \mathbb{P}^1 &1&1&0&0&0&0 \\ \mathbb{P}^1 & 0&0&1&1&0&0\end{array} \right] \left[ \begin{array}{c|:cccccc} \mathbb{P}^1& 1&1&0&0&0&0 \\ \mathbb{P}^1& 0&0&1&1&0&0  \\ \mathbb{P}^2 &0&0&0&0&2&1 \\ \mathbb{P}^3 &0&1&0&1&1&1 \\\hdashline \mathbb{P}^2 & 1&0&2&0&0&0\end{array} \right] \left[ \begin{array}{c|:cccccc} \mathbb{P}^1& 1&1&0&0&0&0 \\ \mathbb{P}^1& 0&0&1&1&0&0  \\ \mathbb{P}^2 &1&0&2&0&0&0 \\ \mathbb{P}^3 &0&1&0&1&1&1 \\\hdashline \mathbb{P}^2 & 0&0&0&0&2&1\end{array} \right] 
\end{eqnarray}
The block matrix form, as described in (\ref{fib1}) has been denoted here with dotted lines. Computing the dimension of the fibers in this case the reader will find that these constitute six different torus fibrations of the CY manifold. Similarly we can find two different K3 fibrations in this case.
\begin{eqnarray} \label{eg2}
\left[ \begin{array}{c|:cccccc} \mathbb{P}^2& 0&0&0&0&2&1 \\ \mathbb{P}^3& 0&0&1&1&1&1  \\  \mathbb{P}^1 &1&0&1&0&0&0 \\\mathbb{P}^2 &1&2&0&0&0&0 \\\hdashline \mathbb{P}^1 & 0&1&0&1&0&0\end{array} \right] \left[ \begin{array}{c|:cccccc} \mathbb{P}^2& 0&0&0&0&2&1 \\ \mathbb{P}^3& 0&0&1&1&1&1  \\  \mathbb{P}^1 &0&1&0&1&0&0 \\\mathbb{P}^2 &1&2&0&0&0&0 \\\hdashline \mathbb{P}^1 & 1&0&1&0&0&0\end{array} \right]
\end{eqnarray}

Note that trivial redundancies have been removed in enumerating the fibrations in the (\ref{eg1}) and (\ref{eg2}) above. For example, column permutations that do not mix the first $n$ and the last $K-n$ columns generate obviously identical fibrations and thus this redundancy has been removed. Similarly for row permutations that do not mix the fiber (first $m$) and base (last $N-m$) rows. In the results we present here, however, there are certain potential redundancies, which have been removed in the previous literature \cite{Gray:2014fla} which we will {\it not} be removing from our data. These are best illustrated with an example. Consider the bi-cubic.
\begin{eqnarray}  \label{bicubic}
X_{\textnormal{bicubic}} = \left[ \begin{array}{c|:c} \mathbb{P}^2 & 3 \\\hdashline \mathbb{P}^2 & 3 \end{array}\right]
\end{eqnarray}
In past work this manifold would have been said to admit a single obvious fibration, a torus described as a cubic in $\mathbb{P}^2$ fibered over a $\mathbb{P}^2$ base. Here we will count both fibrations of this type that appear in the matrix - that is we will consider the two fibrations which arise by considering each of the two $\mathbb{P}^2$ factors in the ambient space to be the base in turn. 

There are two main reasons for making this choice in our approach to redundancy removal, one physical and one mathematical. First, counting distinct but identical fibrations like these will enable us to enumerate fibrations in a manner which agrees with the mathematics literature. After all, there are two fibrations in our example (\ref{bicubic}), albeit ones that are symmetric in structure. In particular, counting fibrations that appear with symmetry like this will make it easier to compare to the number of fibrations that are obtained by applying Koll\'ar's criteria. Second, from a physical perspective, the fact that there are two distinct fibrations in the example (\ref{bicubic}) does have important physical consequences. One would not obtain two different F-theory models by compactifying on the two fibrations, of course, as the moduli space of the two F-theory geometries would be identical. Nevertheless, in considering dualities, the fact that there are two fibrations can be key. Picking a particular complex structure for the bi-cubic and performing a heterotic compactification, for example, one finds that the two fibrations present in (\ref{bicubic}) will lead to two very different F-theory duals \cite{Anderson:2016cdu} (see also \cite{Morrison:1996na} for related ideas in $6$-dimensional heterotic/F-theory duality). This is due to the fact that at a given point in complex structure moduli space, the two torus fibers will be twisted over their $\mathbb{P}^2$ bases in distinct ways. Given the above discussion, we will not remove distinct but topologically isomorphic fibrations from our scans over the CICYs.

Another issue that must be addressed in enumerating obvious fibrations of the type being discussed in this section is that of multiple fibers. Consider, for example, the following configuration matrix and associated obvious fibration:
\begin{eqnarray} \label{multfib1}
X_{\textnormal{multiple}} = \left[ \begin{array}{c|:cc} \mathbb{P}^2 &0 &3 \\ \mathbb{P}^1 & 2 & 0 \\ \hdashline \mathbb{P}^2 & 1 & 2 \end{array} \right]~~~.
\end{eqnarray}
This follows all of the rules to be considered an OGF but exhibits an obvious problem. The fiber in this case, as described by the configuration,
\begin{eqnarray} \label{multfib2}
X_{\textnormal{fiber}}= \left[ \begin{array}{c|cc} \mathbb{P}^2 &0 &3 \\ \mathbb{P}^1 & 2 & 0  \end{array} \right] \;,
 \end{eqnarray} 
is not a single genus one curve. Instead it describes two disjoint tori embedded in the $\mathbb{P}^2 \times \mathbb{P}^1$ ambient space. All such cases can be removed from consideration by imposing the additional condition that no fiber can be described by a configuration matrix that can be put in block diagonal form by row and column permutations. We shall impose this requirement in all of the fibrations, by Calabi-Yau of any dimension, that we discuss in the remainder of this paper.

As a last point in the general discussion of this section, we should note that the Calabi-Yau fibers of different dimensions discussed above can be nested within one another. For example, if we look at the first matrices in (\ref{eg1}) and (\ref{eg2}) above, we see that the torus fibration depicted in (\ref{eg1}) is actually also a torus fibration of the $K3$ fibration in (\ref{eg2}). Such nesting is rather common, with the vast majority of higher dimensional Calabi-Yau fibers also being fibered themselves. However, not every torus fibration need be nested in a $K3$ fibration in this manner. As an example of this, the final torus fibration presented in (\ref{eg1}) clearly does not lie nested within a $K3$ fibration as its base is simply $\mathbb{P}^2$.

\subsection{Enumeration of obvious Calabi-Yau fibrations}

\begin{figure}[!h]\centering
\includegraphics[width=0.7\textwidth]{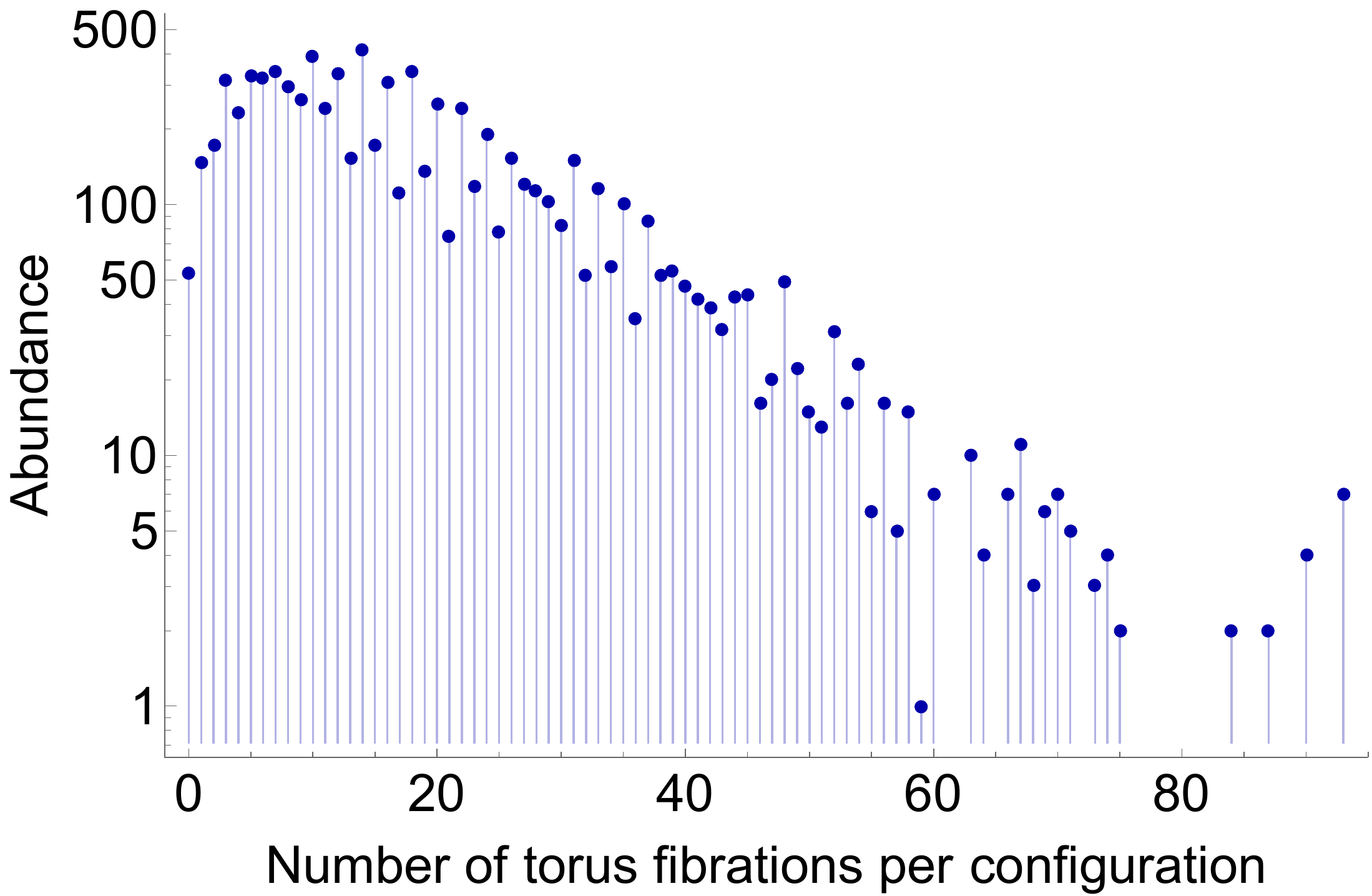}
\caption{{\it Distribution of obvious torus fibration abundance in the CICY threefold list (excluding product manifolds). The values lie in the range 0 - 93. We find $139,597$ fibrations in total and on average each CICY threefold configuration is elliptically fibered in $17.7$ different ways.}}
\label{ellfib}
\end{figure}

Classifying the obvious fibrations, as discussed in the previous subsection, results in the following numbers of inequivalent structures of this type. For torus fibers, the CICY threefolds, using the new favorable configurations mentioned in Section \ref{splitsec}, admit an average of $17.7$ fibrations per configuration matrix, for a total of $139,597$ such structures in the list. The maximum number of such torus fibrations admitted by any one configuration matrix is $93$. Note that these figures are somewhat larger than those given in \cite{Anderson:2016cdu}. This is for three reasons. First, we have favorable configurations describing more of the CICY manifolds and thus can find more torus fibrations. Second, as described in the proceeding subsection, we are not removing what was considered a redundancy in that work. That is, we are keeping symmetric fibrations that are nevertheless distinct. A plot of the number of configurations admitting a given number of obvious torus fibrations is presented in Fig. \ref{ellfib}.

\begin{figure}[!h]\centering
\includegraphics[width=0.72\textwidth]{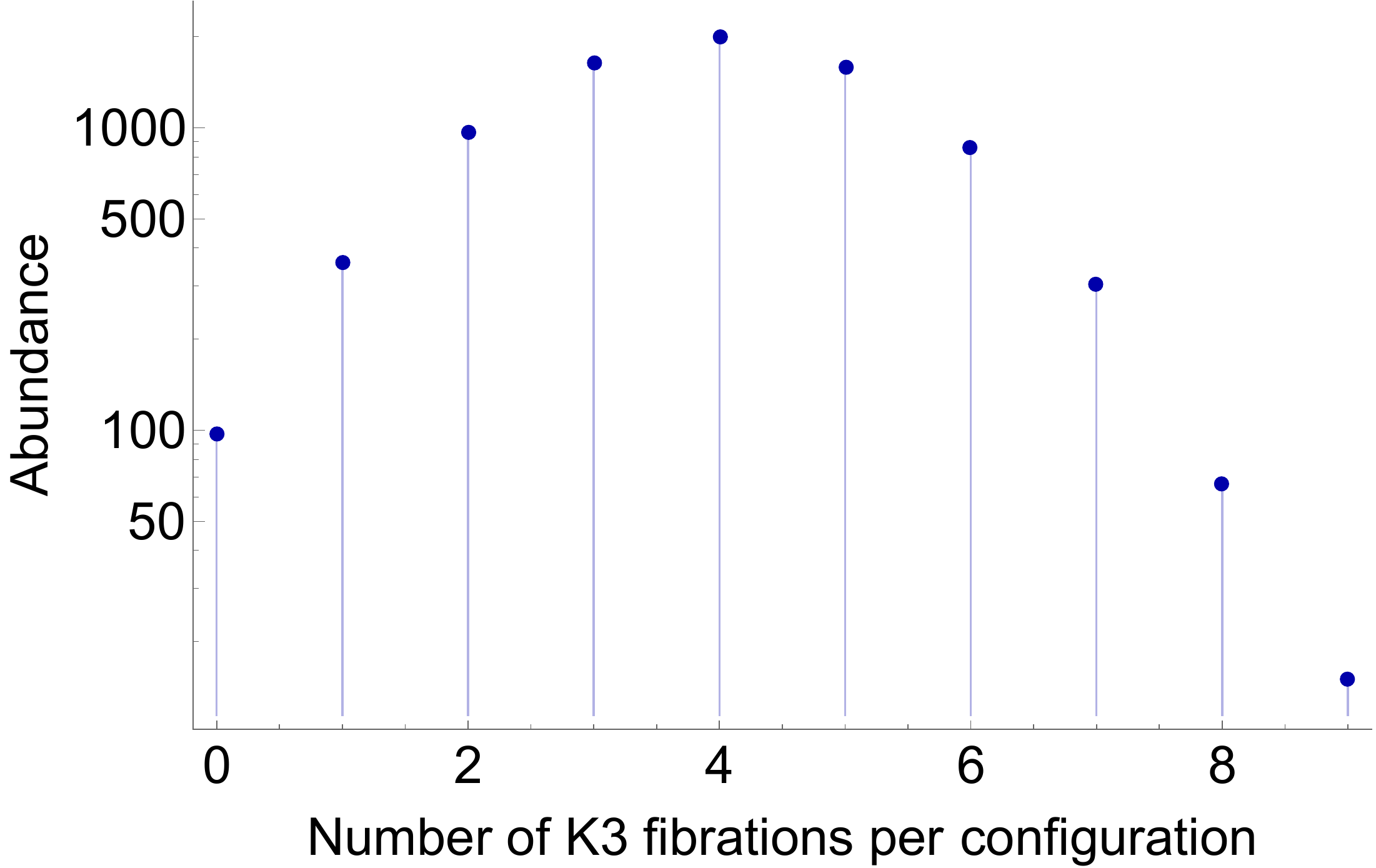}
\caption{{\it Distribution of obvious K3 fibration abundance in the CICY threefold list (excluding product manifolds). The values lie in the range 0 - 9. We find $30,974$ fibrations in total and on average each CICY threefold configuration is obviously K3 fibered in $3.9$ different ways.}}
\label{k3fib}
\end{figure}

For K3 fibrations, the threefolds in our data set admit an average of $3.9$ fibrations per configuration matrix, for a total of $30,974$ such structures in total. The maximum number of such $K3$ fibrations admitted by any one configuration matrix is $9$. A plot of the number of configurations admitting a given number of obvious K3 fibrations is presented in Fig. \ref{k3fib}.

\begin{figure}[!h]\centering
\includegraphics[width=0.85\textwidth]{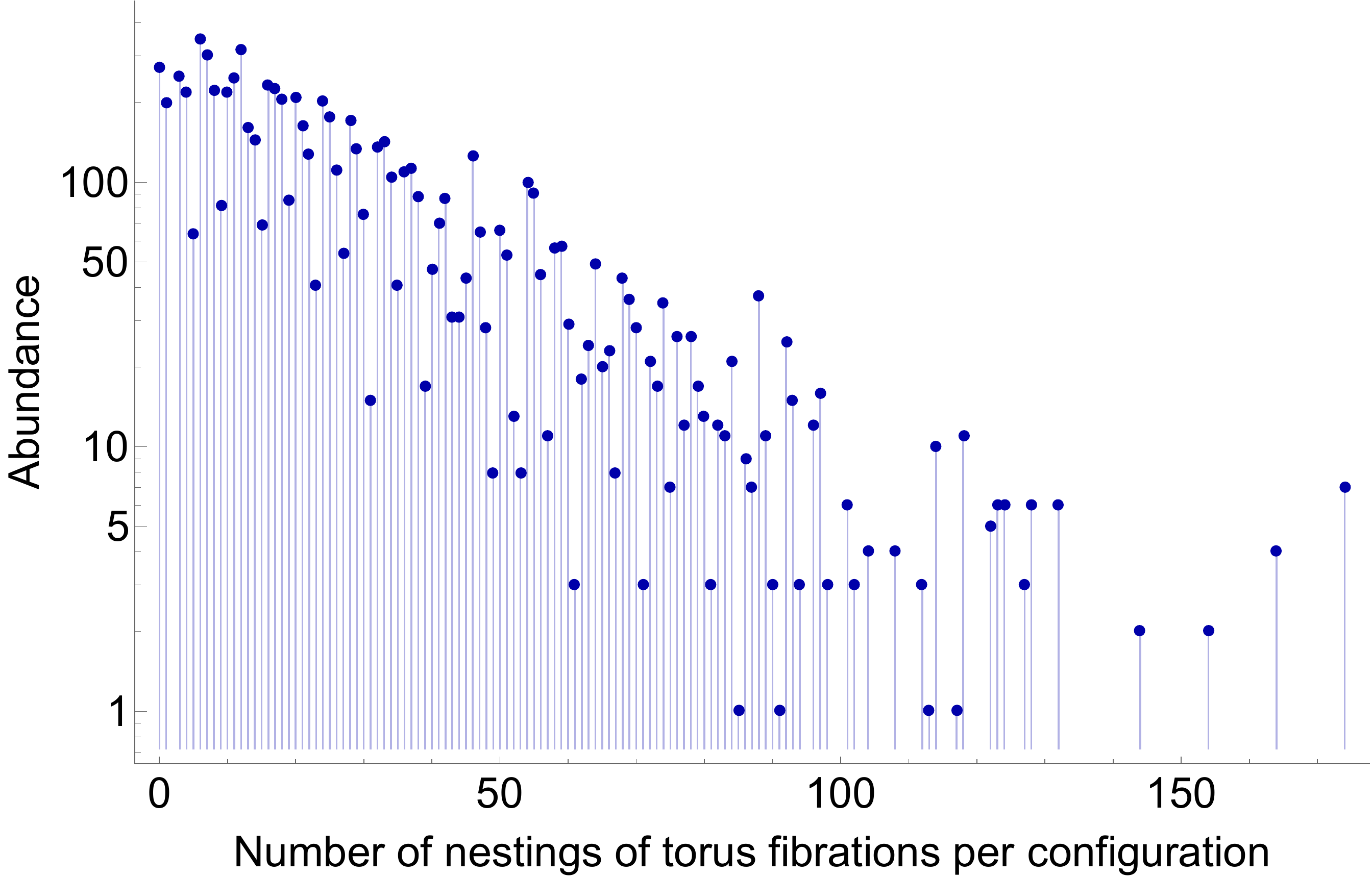}
\caption{{\it Distribution of the abundance of obvious torus fibrations  nested inside obvious K3 fibrations in the CICY threefold list (excluding product manifolds). The values lie in the range 0 - 174. We find $208,987$ such nested fibrations in total and on average each CICY threefold configuration admits $26.6$ different nested fibrations of this type.}}
\label{nestfib}
\end{figure}

Finally, we can ask about the nesting of the torus fibrations inside K3 fibrations. Counting each different obvious torus fibration with a multiplicity determined by how many obvious K3 fibrations it appears nested inside, we find that the average CICY threefold admits 26.6 such structures. Note that this is bigger than the average number of obvious torus fibrations given above as a given torus fibration can be nested inside multiple different K3 fibrations. The total number of such nested fibrations is $208,987$ with the largest example admitting $174$ such nested fibrations. A plot of the number of configurations admitting a given number of obvious torus fibrations nested inside obvious K3 fibrations is presented in Fig. \ref{nestfib}.

The full data describing these fibration structure can be found at \cite{website}. The data format is described in Appendix \ref{app_data}.

 \section{A comparison of obvious fibrations vs. all fibrations for K\"ahler favorable manifolds}\label{OGF_vs_Kollar}
 
As discussed in Section \ref{oguiso-wilson} (see the conjecture there), it has been established \cite{wil,ogu,kollar-criteria} that any effective divisor class $D$ of a Calabi-Yau threefold, $X$, leads to a genus-one fibration if and only if it obeys the following criteria:
\beq \label{kollar-app1} 
D \cdot C \geq 0 \quad \text{for all effective curves $C$ in $X$}\;; \quad\quad 
D^2 \neq 0 \;;  \quad\quad 
D^3 =0 \ . 
\eeq
One is thus led to classify the solutions to~\eqref{kollar-app1} for 
\beq\label{d_setup}
D=\sum\limits_{r=1}^{h^{1,1}(X)} a_r J_r \ , \quad \text{with}~a_r \in \IZ \ , 
\eeq 
where $\{J_r\}$ is a chosen basis of $H^{1,1}(X)$. 

In this section we will compare the results of a scan for Koll\'ar divisors of the form given above to the searches for OGFs described in Section \ref{ogf_sec}. It is of interest to see whether the total number of fibrations (as counted by the divisor criteria above) exceeds the number of ``obvious" fibrations that are visible from the algebraic form of the CICY configuration matrix. To make this comparison however, full control of the K\"ahler cone of $X$ is crucial. Thus, we will be able to make this comparison only for K\"ahler favorable manifolds as defined in Section \ref{kah_fav_sec}. For these, the K\"ahler cone of $X$ descends directly from the ambient product of projective spaces.

Given such a K\"ahler favorable CICY threefold $X$ embedded in $\cA=\prod_r \IP^{n_r}$, let us take the $J_r$ to be the harmonic $(1,1)$--form of the ambient $\IP^{n_r}$ pieces; we call $X \subset \cA$ favorable if the $J_r$'s form a basis\footnote{In general such a basis could be redundant but we will not need to consider such an eventuality here.} of $H^{1,1}(X)$. We begin by writing the conditions in \eref{kollar-app1} in terms of the explicit divisor in \eref{d_setup}. These take the following forms in terms of $a_r$, respectively, as
\beq\label{kollar-app2}
a_r \geq 0 ~\text{~for all}~r \;;\quad\quad
\sum_{s,t=1}^{h^{1,1}}d_{rst}a_s a_t  \neq 0 ~\text{~for some}~r \;;\quad\quad
\sum_{r,s,t=1}^{h^{1,1}} d_{rst} a_r a_s a_t = 0 \ , 
\eeq
where $d_{rst}:=\int_X J_r \wedge J_s \wedge J_t$ are the triple intersection numbers of $X$. We may then assume that a given solution to~\eqref{kollar-app2}, $a_r=a_r^{(0)}$, is ordered as $0= a_1 = \cdots a_{\nu} < a_{\nu +1}  \leq \cdots \leq a_{h^{1,1}}$, upon an appropriate permutation of the $\IP^{n_r}$'s in $\cA$. Here, $\nu$ is the number of $0$'s appearing in the solution. Because all the triple intersections $d_{rst}$ of a favorable CICY are nonnegative, it is then obvious that there is another solution, $a_r=a_r^{(1)}$:
 \begin{equation}
       a_r^{(1)} = 
        \begin{cases}
            0 & \text{if $r\leq\nu$} \\
            1 & \text{if $r>\nu$} \ , 
        \end{cases}
    \end{equation}
which must represent the same genus-one fibration as the original solution (using the conditions of equivalence outlined in Section \ref{oguiso-wilson}). Thus, for K\"ahler favorable CICYs, in searching for Koll\'ar divisors, we need only consider $D$ as in \eref{d_setup} with $a_r =0$ or $1$ in classifying the solutions to \eref{kollar-app1}. For each of the $4874$ CICYs that are K\"ahler favorable with respect to an ambient space $\mathbb{P}^{n_1} \times \ldots \times \mathbb{P}^{n_m}$, such a search for minimal Koll\'ar divisors (with $0\leq a_r \leq 1$) was carried out. 

Moreover, as described in Section \ref{ogf_sec}, a systematic scan for obvious genus-one fibrations (OGFs) has been completed for all the maximally favorable configuration matrices. In each case the classification proves to be finite and impressively, \emph{a one-to-one correspondence between the two classification results can immediately be found}. In each case we find $50,219$ fibrations in the set of $4874$ manifolds and an exhaustive comparison shows that each Koll\'ar divisor corresponds to an OGF (the converse is automatic). Thus, for K\"ahler favorable CICYs, the OGFs already provide a complete set of genus-one fibrations for these geometries. This result is not entirely surprising since the K\"ahler favorable form engineered in Section \ref{top_sec} has been chosen to provide a description in which the ambient projective space factors encode the maximal amount of information about the Picard group and K\"ahler cone of $X$. It should be noted that this correspondence \emph{does not persist for non-K\"ahler favorable CICY configurations}. For example, $2946$ out of $7890$ manifolds in the set described above have been described by new CICY configurations compared to the original CICY threefold data set. If the OGFs are counted for any non-favorable description of one of these geometries, many fibrations are missing. That is, the OGF count is found to be considerably less than the true count (based on \eref{kollar-app1}), as expected.

To conclude this comparison, it is worth making several remarks on the ubiquity of genus one fibrations within this dataset. Each of the $4874$ CICY configurations studied yields a finite number of fibration structures. Moreover, the maximal number of fibrations observed from any one threefold in this set is $39$ and the average number of fibrations is $10.3$. Finally, it should be noted that this search yields 53 configuration matrices which do not admit any fibration structure (either via \eref{kollar-app1} or the OGF criteria).
The CICY number of those 53 favorable configuration matrices are: 
\bea
&\{6771, 6803, 7036, 7208, 7242, 7530, 7563, 7566, 7571, 7578, 7588, 
7626, 7631, \nonumber \\ &7635, 7636, 7638, 
7644, 7647, 7648, 7679, 7717, 7721, 
7726, 7734, 7747, 7759, \nonumber \\ &7761, 7781,
 7799, 7806, 7809, 7816,
 7817, 
7819, 7822, 7823, 7840, 7842, \nonumber \\&7858, 7861, 7863, 7867, 7869, 7873, 7878, 7879,7882, 
7885-7890\}
\eea
(labels refer to the dataset in \cite{website}). Since this set are within the geometries for which we can scan exhaustively using the Koll\'ar criteria, we are certain that they are not genus-one fibered. This is intriguing since it demonstrates that the largest value of $h^{1,1}$ for a non-fibered CY manifold in the CICY threefold list is $h^{1,1}=4$ (note that for every other manifold in the CICY list, at least one OGF is present).

We will return to this point in Section \ref{conclusions_sec}, but for now it suffices to note that all existing fibration studies within CY threefolds indicate that genus one fibered geometries seem to become ubiquitous as $h^{1,1}$ increases. For the CICY threefold data set it is clear that this bound on $h^{1,1}$ in order for the manifold to be guaranteed a genus-one fibration is quite low indeed.
 
%%%%%%%%%%
%%%%%%%%%%
\section{Exceptional configurations}\label{the48_sec}
As described in Section \ref{splitsec}, the process of splitting/contraction yields a favorable description (in which a full basis for divisors is obtained via restriction from ambient projective space hyperplanes) for all but $48$ configurations in the CICY list. In this section and the next, we turn our attention to these $48$ seemingly non-favorable CICY threefold configurations. Fortunately, as we will see shortly, all $48$ configurations in fact have a simple structure that will allow us to not only determine their Picard groups, but also their K\"ahler and Mori cones and all their topological data, including the triple intersection numbers. We will thus analyze such topological properties and apply Koll\'ar's criteria \eref{kollar-app1} to exhaustively search for the genus-one fibration structures.
Unlike in the case of the favorable configurations studied in Section \ref{OGF_vs_Kollar}, here we will see that \emph{there exist many more fibration structures than are visible as OGFs}. We will enumerate these fully in the following sections.

To begin, it is worth considering the possible redundancy among the $48$ configurations. We state the results here and leave the proofs of the equivalences to Appendix \ref{surf_red}.  First, the set contains $15$ configurations with Hodge numbers $(h^{1,1},h^{2,1})=(19,19)$ which are equivalent to one another and which all describe the Schoen manifold. 
Since our fibration analysis shows qualitatively different features for the Schoen manifold, we will elaborate on the Schoen manifold in Section~\ref{schoen_sec}. 

It turns out that each of the remaining 33 configurations is favorably embedded as an anticanonical hypersurface in a product of two del Pezzo surfaces of the form,
\beq\label{amb}
\cA_{r,s} \equiv dP_r \times dP_s \quad \text{or} \quad \cA_s \equiv \IP^1 \times \IP^1 \times dP_s \ . 
\eeq
Here, $r = 0, \dots, 7$ and $s=5,6,7$, leading us to a total of $24$ geometries with $r \leq s$. This fact strongly suggests further redundancy and indeed some exists. 
Using equivalent descriptions of the ambient space surfaces and splittings/contractions, the $33$ configurations can be grouped into the $24$ distinct Calabi-Yau geometries as listed in Table~\ref{tb:dPxdP}. In addition, there exist $35$ favorable CICY configurations which are not K\"ahler favorable with respect to the ambient product of projective spaces, but which are still anticanonical hypersurfaces in ambient spaces of the form~\eqref{amb} with $r=0, \ldots, 7$ and $s=3, \ldots, 7$. All of these cases may be analyzed in the same fashion. Table~\ref{tb:dPxdP} thus has total of $68$ CICY configurations leading to $35$ distinct Calabi-Yau geometries. In this section, we will analyze topological properties of these $35$ anticanonical hypersurfaces, 
\beq
X_{r,s} \subset \mathcal A_{r,s} ~~\text{and}~~X_{s} \subset \mathcal A_s \ , 
\eeq
where $r=0,\ldots,7$ and $s=3 \ldots, 7$ with $r\leq s$, and will classify genus one fibrations therein.

\begin{table}
\begin{center}
\begin{tabular}{|c |c |}
\hline
Ambient Space & CICY Id Numbers \\ \hline\hline
$\IP^1 \times \IP^1 \times dP_3$ & (7709, 7731) \\ \hline
$\IP^1 \times \IP^1 \times dP_4$ & (7459) \\ \hline
$\IP^1 \times \IP^1 \times dP_5$ & 6826 (6829, 6925) \\ \hline
$\IP^1 \times \IP^1 \times dP_6$ & 5298  \\ \hline
$\IP^1 \times \IP^1 \times dP_7$ & 2565  \\ \hline
$dP_0 \times dP_3$ & (7800, 7810)   \\ \hline
$dP_0 \times dP_4$ & (7665)  \\ \hline
$dP_0 \times dP_5$ & 7232 (7233, 7298) \\ \hline
$dP_0 \times dP_6$ & 6021  \\ \hline
$dP_0 \times dP_7$ & 3405 (6748, 6796) \\ \hline
$dP_1 \times dP_3$ & (7716, 7743)  \\ \hline
$dP_1 \times dP_4$ & (7512)  \\ \hline
$dP_1 \times dP_5$ & 6795   \\ \hline
$dP_1 \times dP_6$ & 5282  \\ \hline
$dP_1 \times dP_7$ & 2548   \\ \hline
$dP_2 \times dP_3$ & (7546, 7603)  \\ \hline
$dP_2 \times dP_4$ & (7126)   \\ \hline
$dP_2 \times dP_5$ & 6291 (6292, 6369)  \\ \hline
$dP_2 \times dP_6$ & 4473 \\ \hline
$dP_2 \times dP_7$ & 1835  \\ \hline
$dP_3 \times dP_3$ & (7206, 7246, 7300)   \\ \hline
$dP_3 \times dP_4$ & (6533, 6619)  \\ \hline
$dP_3 \times dP_5$ & 5254, 5300 (5121, 5257, 5306, 5440) \\ \hline
$dP_3 \times dP_6$ & 3388, 3406\\ \hline
$dP_3 \times dP_7$ & 1257, 1268\\ \hline
$dP_4 \times dP_4$ & (5636)   \\ \hline
$dP_4 \times dP_5$ & 4077 (3936, 4081) \\ \hline
$dP_4 \times dP_6$ & 2199\\ \hline
$dP_4 \times dP_7$ & 708 \\ \hline
$dP_5 \times dP_5$ & 2564, 2566, 2638 (2568, 2641, 2837)\\ \hline
$dP_5 \times dP_6$ & 1266, 1267, 1289 \\ \hline
$dP_5 \times dP_7$ & 381, 382, 384 \\ \hline
$dP_6 \times dP_6$ & 536 \\ \hline
$dP_6 \times dP_7$ & 206 \\ \hline
$dP_7 \times dP_7$ & 95 \\ \hline

\end{tabular}\end{center}
\caption{{\it The $33$ exceptional configurations which are not favorable with respect to the ambient product of projective spaces and which are not the Schoen manifold. Each of them turns out to be an anticanonical divisor of either $dP_r \times dP_s$ or $\IP^1\times\IP^1\times dP_s$, for $r=0, \cdots, 7$ and $s=5,6,7$. The configurations are labelled by their standard CICY ID numbers~\cite{Candelas:1987kf,website}. In fact, there exist an additional $35$ configurations which describe an anticanonical divisor in ambient spaces of the same form with possibly some smaller values of $s$. These are favorable but not K\"ahler-favorable with respect to the ambient product of projective spaces, and are also listed here in parentheses.}} 
\label{tb:dPxdP}
\end{table}

%%%
\subsection{Favorable hypersurfaces in products of two del Pezzo surfaces}\label{dPdP}
\subsubsection{Topological data}
It is fortunate that the configurations in Table~\ref{tb:dPxdP} all take the form of an anticanonical hypersurface in an ambient space consisting of the direct product ${\cal S} \times {\cal S}'$ of two smooth K\"ahler Fano surfaces. Such geometries\footnote{Note that in fact \emph{all} anticanonical hypersurfaces in $dP_r \times dP_s$ with $r=0, \ldots 7$ and $s=r, \ldots, 7$ appear in the CICY list, however those not listed in Table \ref{tb:dPxdP} have K\"ahler favorable descriptions with respect to an ambient product of projective spaces and their fibration structures are considered in Section \ref{OGF_vs_Kollar}.} were systematically classified in \cite{Green:1987zp}. Since the ambient space takes such a simple form, we have once again found a favorable description of the geometry in which the divisors on $X$ descend simply from those on ${\cal A}={\cal S} \times {\cal S}'$. In addition, thanks to the Lemma \ref{lemma2} the K\"ahler and Mori cones of such CY threefolds $X$ can also be simply obtained via restriction. With these results in hand, the analysis is now as tractable as those studied in previous sections, although they are not K\"ahler favorable with respect to the ambient product of projective spaces.

Given the explicit embeddings, all of the relevant topological properties of the Calabi-Yau hypersurfaces can be obtained by choosing a description of the space $H^{1,1}$ in terms of the $(1,1)$-forms descended from those of the del Pezzo surface factors. 
This can be illustrated via concrete configuration matrix (e.g., number $4077$ in the CICY list \cite{Candelas:1987kf,website}):
\beq \label{conf-4077}
X \sim \left[ \begin{array}{c|cccc} 
\mathcal \IP^1 & 1 & 0&0&1 \\ 
\mathcal \IP^2 & 2 & 0&0&1 \\
\mathcal \IP^4 & 0 &2&2&1 
\end{array}\right]\; \ , 
\eeq
with Hodge numbers $(h^{1,1}(X), h^{2,1}(X))=(11, 31)$. This configuration matrix is highly non-favorable with respect to the ambient product of projective spaces as is obvious from $h^{1,1}(X)-h^{1,1}(\mathcal A) =11-3=8 >0$ with $\cA =\IP^1 \times \IP^2 \times \IP^4$. On the other hand, the threefold $X$ can also be thought of as an anticanonical divisor of the fourfold, $\mathcal A_{4,5} = dP_4 \times dP_5$, where the two del Pezzo surfaces are respectively given by the configurations,
\beq \label{conf-dP45}
dP_4\sim \left[ \begin{array}{c|c} 
\mathcal \IP^1 & 1 \\ 
\mathcal \IP^2 & 2\\
\end{array}\right]\; \ , \quad
dP_5\sim \left[ \begin{array}{c|cc} 
\mathcal \IP^4 & 2 & 2 \\ 
\end{array}\right]\; \ . 
\eeq
One can then easily see that the $h^{1,1}(\mathcal A_{4,5})=11$ K\"ahler forms of $\cA_{4,5}$ descend to the $h^{1,1}(X)=11$ independent K\"ahler forms of the Calabi-Yau hypersurface $X$ and hence that $X$ is favorably embedded in $\cA_{4,5}$. 

For any of the $35$ geometries in Table \ref{tb:dPxdP}, a simple description of the divisors can be obtained from the ambient product of surfaces. Let us set the notation for such a basis here. Recall that the del Pezzo surface $dP_r$ is constructed by blowing up a $\IP^2$ at $r$ generic points. The second homology group $H_2(dP_r, \IZ)$ is spanned by the hyperplane class $L$ in $\IP^2$ as well as the $r$ exceptional divisors $E_i$, $i=1, \cdots, r$, which intersect with one another as
\beq\label{double}
L\cdot L = 1, \quad L \cdot E_i = 0, \quad E_i \cdot E_j = -\delta_{ij} \ . 
\eeq
In this basis, the Mori cone generators of the del Pezzo surfaces $dP_r$ can be expressed as in Table~\ref{tb:dP-Mori} and the first Chern class of $dP_r$ is given by
\beq\label{c1}
c_1(dP_r) = 3L- \sum_{i=1}^r E_i \ . 
\eeq
See for example, \cite{Donagi:2004ia} for more details on the geometry of the surface $dP_r$. 
\begin{table}[h]
\begin{center}
\begin{tabular}{|>{\centering\arraybackslash}m{.2in} ||>{\centering\arraybackslash}m{3.8in}| | >{\centering\arraybackslash}m{.6in}|}
\hline
$r$ & Generators $(i<j<\cdots \leq r)$ & Number \\ \hline\hline
0& $L$ & 1 \\ \hline
1& $E_1, L-E_1$ & 2 \\ \hline
2& $E_i, L-E_i-E_j$ & 3 \\ \hline
3& $E_i, L-E_i-E_j$ & 6 \\ \hline
4& $E_i, L-E_i-E_j$ & 10 \\ \hline
5& $E_i, L-E_i-E_j, 2L-E_i-E_j-E_k-E_l-E_m$ & 16 \\ \hline
6& $E_i, L-E_i-E_j, 2L-E_i-E_j-E_k-E_l-E_m$ & 27 \\ \hline
7& 
$E_i, L-E_i-E_j, 2L-E_i-E_j-E_k-E_l-E_m, $ 
$3L-2E_i-E_j-E_k-E_l-E_m-E_n-E_o$& 56 \\ \hline
8& 
$E_i, ~L-E_i-E_j, ~2L-E_i-E_j-E_k-E_l-E_m, $ 
$3L-2E_i-E_j-E_k-E_l-E_m-E_n-E_o, $  
$4L-2(E_i+E_j+E_k)-\sum\limits_{a=1}^5 E_{m_a}, $
$5L-E_i - E_j - 2\sum\limits_{a=1}^6 E_{m_a} ,~ 6L-3E_i - 2 \sum\limits_{a=1}^7 E_{m_a} $ &  240 \\  \hline
9 & $f=3L-\sum\limits_{i=1}^9 E_i ,~\text{and~} \{y_a\} ~\text{such that}~y_a^2=-1, y_a \cdot f = 1$ &  $\infty$ \\ \hline
\end{tabular}\end{center}
\caption{{\it The Mori cone generators for the del Pezzo surfaces $dP_r$, $r=0, \cdots, 8$ and for the rational elliptic surface $dP_9$. The indices $i, j, \dots \in \{1, \dots, r\}$ appearing in each generator are distinct.}} 
\label{tb:dP-Mori}
\end{table}

Equipped with such topological information, the triple intersections $d_{mnp}^{(r,s)}$ of the Calabi-Yau hypersurfaces, $X_{r,s} \subset \cA_{r,s}$ can be straightforwardly computed as 
\beq\label{triple_la}
d_{m n p}^{(r,s)} \equiv \int_{X_{r,s}} J_m \wedge J_n \wedge J_p = \int_{\cA_{r,s}} J_m \wedge J_n \wedge J_p \wedge (3L - \sum_{i=1}^r E_i + 3L' - \sum_{i'=1}^s E'_{i'}) \ , 
\eeq
where $m$, $n$, $p$ are the indices labeling the $h^{1,1}(X)=r+s+2$ harmonic $(1,1)$-forms on $X_{r,s}$, 
\beq
L,~~E_{i=1, \dots, r}, ~~\text{and}~~L', ~~E'_{i'=1, \dots, s} \ , 
\eeq
descending from those on $dP_r$ and $dP_s$, respectively. 

It is clear that a similar approach will also yield information on surfaces of the form $\cA_s=\mathbb{P}^1 \times \mathbb{P}^1 \times dP_s$. Then 
\beq
H, ~~ \tilde H~~\text{and}~~L', ~~E'_{i'=1, \dots, s} \ , 
\eeq
label the divisors on $X_s \subset \cA_s$  descending from those on $\IP^1 \times \IP^1$ and $dP_s$. Furthermore, as in \eref{triple_la}, the triple intersections $d^{(s)}_{mnp}$ of $X_s$ can be obtained from
\beq
d^{(s)}_{mnp} \equiv \int_{X_s} J_m \wedge J_n \wedge J_p = \int_{\cA_{s}}  J_m \wedge J_n \wedge J_p \wedge (2H + 2\tilde H + 3L' - \sum_{i'=1}^s E'_{i'} ) \ , 
\eeq
where $m, n, p$ label the $h^{1,1}(X)=s+3$ $(1,1)$-forms on $X_s$. 

Finally, the Mori (and the K\"ahler) cones of $\cA_{r,s}$ and $\cA_s$ can also be straightforwardly obtained from those of the individual del Pezzo factors (see Table~\ref{tb:dP-Mori}). Via Lemma \ref{lemma2} this information can then be interpreted as the K\"ahler/Mori data of the Calabi-Yau threefold $X_{r,s}$ or $X_s$~\cite{descend}.

%%%
\subsubsection{Classification of genus-one fibrations}\label{case-distinction}
\label{dPdP-classify}
Recall that any divisors obeying the conditions~\eqref{kollar-app1} represent a genus-one fibration. In this subsection, we will classify such divisors for all the Calabi-Yau three-folds appearing in Table~\ref{tb:dPxdP}, which we label as 
\beq
X_{r,s} \subset \cA_{r,s} \text{~~and~~} X_s \subset \cA_s \ , 
\eeq
in terms of their four-fold ambient spaces, $\cA_{r,s}=dP_r \times dP_s$ and $\cA_s= \IP^1\times \IP^1 \times dP_s$. 

Let us start by analyzing $X_{r,s}$. Note first that divisors of $X_{r,s}$ can be parameterized as integer linear combinations,
\beq\label{div-gen}
\cD=a L - \sum_{i=1}^r a_i E_i + a' L' - \sum_{i'=1}^s a'_{i'} E'_{i'} \ , 
\eeq
of the basis divisors $L, E_1, \dots, E_r$ and $L', E_1', \dots, E_s'$.
The triple intersection of $\cD$ can then be expressed as
\bea
\int_{X_{r,s}} \cD^3 &=& \int_{\cA_{r,s}} (a L - \sum_{i=1}^r a_i E_i + a' L' - \sum_{i'=1}^s a'_{i'} E'_{i'} )^3 \wedge (3L - \sum_{i=1}^r E_i + 3L' - \sum_{i'=1}^s E'_{i'}) \\ \label{triple}
&=& 3 (a^2 - \sum_{i=1}^r a_i^2) (3a'  - \sum_{i'=1}^s a'_{i'}) + 3 (a'^2 - \sum_{i'=1}^s {a'_{i'}}^2) (3a  - \sum_{i=1}^r a_{i}) \ .
\eea
For the purpose of studying the geometries in Table~\ref{tb:dPxdP}, let us restrict our analysis to $0 \leq r, s \leq 7$ in particular. 

Note that the first of the Koll\'ar criteria~\eqref{kollar-app1} for the divisor $\cD$ of $X_{r,s}$ immediately leads to constraints on the divisors $D \equiv a L - \sum_{i=1}^r a_i E_i$ and $D' \equiv a' L' - \sum_{i'=1}^s a'_{i'} E'_{i'}$ of $dP_r$ and $dP_s$, respectively. For example, for $D$ considered as a divisor of $dP_r$, we should have $D \cdot C \geq 0$ for all curves $C$ in $dP_r$. 

We can, without loss of generality\footnote{Note that in general the cone of effective divisors  on $X$ (denoted $\rm{Eff}(X)$) is larger than that of the ambient space, $\rm{Eff}({\cal A})$, even for anticanonical hypersurfaces in Fano fourfolds. However, in the case at hand we are interested in divisors that are both effective and nef on $X$. That is the cone $\rm{Nef}(X) \cap  \rm{Eff}(X)$. For the hypersurfaces in consideration here this intersection is maximal and equal to $\rm{Nef}(X)$ which descends fully from the ambient space $A_{r,s}$ (see \cite{2016arXiv161100556L} for a review of these issues).}, consider the case when the effective divisor ${\cal D}$ on $X_{r,s}$ descends from an effective divisor on ${\cal A}_{r,s}$. Since the first Chern class $c_1(dP_r)= 3L- \sum_{i=1}^r E_i$ is ample and $D$ itself is to be an effective divisor of $dP_r$, we have $D \cdot (3L- \sum_{i=1}^r E_i) \geq 0 $ and $D \cdot D \geq 0$, which, respectively, lead to the inequalities,
\bea\label{simple}
3a- \sum_{i=1}^r a_i \geq 0 \ ,  \\
a^2 - \sum_{i=1}^r a_i^2 \geq 0 \ .  \label{hard}
\eea 
The situation where equality holds can then be described as follows (see Appendix~\ref{dio} for the derivation): the first inequality saturates only for the zero vector $(a, a_1, \cdots, a_r)=0$, while the second exhibits a more complicated solution set. This set contains not only the zero vector but also, depending on $r$, vectors of the form
\beq
(a, a_1, \cdots, a_r) = \alpha A_r^{(I)} \ , ~\text{with}~\alpha \in \IZ_{> 0} \ , ~I=1, \cdots, N_r \ , 
\eeq
where $A_r^{(I)}$ are the following vectors of length $r+1$:
\bea \nn
A_1^{(I)}&=&(1,1), \quad I=1  \ ; \\ \nn
A_2^{(I)}&=&(1,1,0),~\cdots, \quad I=1, 2  \ ;   \\ \nn
A_3^{(I)}&=&(1,1,0,0),~\cdots,  \quad I=1, 2, 3 \ ;  \\ \label{As}
A_4^{(I)}&=&(1,1,0,0,0), (2,1,1,1,1),~\cdots,  \quad 1\leq I \leq 5  \ ; \\ \nn
A_5^{(I)}&=&(1,1,0,0,0,0), (2,1,1,1,1,0),~\cdots,  \quad 1 \leq I \leq 10 \ ;  \\ \nn
A_6^{(I)}&=&(1,1,0,0,0,0,0), (2,1,1,1,1,0,0), (3,2,1,1,1,1,1),~\cdots, \quad 1 \leq I \leq 27 \ ;  \\ \nn
A_7^{(I)}&=&(1,1,0,0,0,0,0,0), (2,1,1,1,1,0,0,0), (3,2,1,1,1,1,1,0),\\  \nn
&&(4,2,2,2,1,1,1,1), (5,2,2,2,2,2,2,1),~\cdots,  \quad 1 \leq I \leq 126 \ .
\eea
In \eqref{As}, the ellipses represent all possible vectors obtained by permuting the $a_1, \cdots, a_r$ from any of the preceding vectors explicitly presented; E.g., with $r=2$, one obtains an additional vector $A_2^{(2)}=(1,0,1)$ by permuting $a_1$ and $a_2$ from the $A_2^{(1)}=(1,1,0)$ presented above. Note that the counting, $N_r$, of the vectors $A_r^{(I)}$ for each $r=0, \cdots, 7$ is given as
\beq
N_0=0,~~N_1=1,~~N_2=2,~~N_3=3,~~N_4=5,~~N_5=10,~~N_6=27,~~N_7=126 \ . 
\eeq
Similarly, on $dP_s$, we also have
\bea\label{simple'}
3a'- \sum_{i'=1}^s a'_{i'} \geq 0 \ ,  \\ \label{hard'}
a'^2 - \sum_{i'=1}^s {a'_{i'}}^2 \geq 0 \ , 
\eea 
where the first inequality saturates only for the zero vector $(a', a'_1, \cdots, a'_s)=0$, and the second, for the zero vector as well as for
\beq\label{equality-s}
(a', a'_1, \cdots, a'_s) = \alpha' A_s^{(I')} \ , ~\text{with}~\alpha' \in \IZ_{> 0} \ , ~I'=1, \cdots, N_s \ ,
\eeq
where $A_s^{(I')}$ are the same length-$(s+1)$ vectors as in \eqref{As}.

Given the constraints~\eqref{simple},~\eqref{hard},~\eqref{simple'}, and~\eqref{hard'}, together with the aforementioned equality conditions for them, the triple intersection~\eqref{triple} can only be set to zero in the following three cases. 

\paragraph{Case 1: $(a, a_1, \cdots, a_r) = 0$.} Each $(a', a'_1, \cdots, a'_s)\neq 0$ with $D'$ nef in $dP_s$ and $D'^2 \neq 0$ represents a genus-one fibration, where the base manifold is either $dP_s$ or its blow down. 
For a generic choice of $(a', a'_1, \cdots, a'_s)$ the base is $dP_s$ and the fibration is an OGF of the CICY configuration. For instance, in the configuration~\eqref{conf-4077} with $(r,s)=(4,5)$, an OGF with the base $dP_{s=5}\sim \left[ \begin{array}{c|cc} 
\mathcal \IP^4 & 2 & 2 \\ 
\end{array}\right]$ is immediately found. 

It should be noted that this same del Pezzo base can lead to other, related base geometries through the process of blowing down. Intuitively, any exceptional divisor in the del Pezzo base could be ``grouped" with the fiber rather than the base geometry (leading to a non-flat fiber over this locus). This phenomenon will be illustrated explicitly for a configuration matrix in Case 2 below. Such fibrations over the various blown-down bases can also be described by non-generic choices of $(a', a'_1, \cdots, a'_s)$, and these are not necessarily represented by an OGF. More details on these $dP_s$ base geometries, the Koll\'ar divisors (and how they relate to rational curves in the del Pezzo surface), as well as the enumeration of the distinct fibrations can be found in Appendix~\ref{enum}.

\paragraph{Case 2: $(a', a'_1, \cdots, a'_s) = 0$.} Each $(a, a_1, \cdots, a_r) \neq 0$ with $D$ nef in $dP_r$ and $D^2\neq 0$ represents a genus-one fibration, where the base is either $dP_r$ or its blow down. For a generic $(a, a_1, \cdots, a_r)$ the base is $dP_r$ and the fibration is an OFG. For instance, again in the configuration~\eqref{conf-4077} with $(r,s)=(4,5)$, an OGF with the base $dP_{r=4}\sim \left[ \begin{array}{c|c} 
\mathcal \IP^1 & 1 \\ 
\mathcal \IP^2 & 2\\
\end{array}\right]$ is immediately found. 

This geometry provides an explicit illustration of the possible birational relationship of bases within this case distinction. Consider the configuration matrix in \eref{conf-4077}, re-written to make manifest the $dP_4$ base:
\beq \label{conf-4077_again}
X \sim \left[ \begin{array}{c|c:ccc} 
\mathcal \IP^4 & 0 &2&2&1 \\ \hdashline
\mathcal \IP^1 & 1 & 0&0&1 \\ 
\mathcal \IP^2 & 2 & 0&0&1 \\
\end{array}\right]\; \ .
\eeq
This same configuration can also be re-grouped to make clear the fiber/base structure with a $\mathbb{P}^2$ base:
\beq \label{conf-4077_again2}
X \sim \left[ \begin{array}{c|:cccc} 
\mathcal \IP^4 & 0 &2&2&1 \\ 
\mathcal \IP^1 & 1 & 0&0&1 \\ \hdashline
\mathcal \IP^2 & 2 & 0&0&1 \\
\end{array}\right]\; \ . 
\eeq
Note that four exceptional divisors in the $dP_4$ base in \eref{conf-4077_again} are now part of the fiber in \eref{conf-4077_again2}. Consider the explicit form of the first defining equation associated to \eref{conf-4077_again2}
\beq
x_0 q_1(y) + x_1 q_2(y)=0
\eeq
where $x, y$ denote coordinates of $\mathbb{P}^1$ and $\mathbb{P}^2$ respectively. Over four points in the $\mathbb{P}^2$ base, the two quadratics $q_1$ and $q_2$ vanish, leading to a non-flat fiber over those points. In this case, the two bases are related by blowing up/down $4$ points in $\mathbb{P}^2$. In general, similar base relationships can arise for \emph{any choice of blow-downs for a del Pezzo base}, though not all may be visible as OGFs. In addition, these relationships can be seen via non-generic choices of the vectors $(a, a_1, \ldots, a_r)$ parameterizing Koll\'ar divisors. Enumeration of all the blown-down bases is worked out in Appendix~\ref{enum}.

\paragraph{Case 3: $(a, a_1, \cdots, a_r) \neq 0$ and $(a', a'_1, \cdots, a'_s) \neq 0$.} In this last case, $\cD^3=0$ is only achieved for $\cD=\cD_{r,s}(\alpha, I;\alpha',I')$ of the form~\eqref{div-gen} with 
\bea\label{kollar-rs}
(a, a_1, \cdots, a_r) &=& \alpha A_r^{(I)} \ , ~\alpha \in \IZ_{> 0} \ , ~I=1, \cdots, N_r \ , \\ \nn
(a', a'_1, \cdots, a'_s) &=& \alpha' A_s^{(I')} \ , ~\alpha' \in \IZ_{> 0} \ , ~I'=1, \cdots, N_s \ , 
\eea
where $A_r^{(I)}$ and $A_s^{(I')}$ are specified in \eqref{As}. Regarding the counting of genus-one fibrations, two important observations follow. Firstly, although there are infinitely many such divisors $\cD_{r,s}(\alpha, I; \alpha', I')$, one can show that different choices of $\alpha, \alpha' \in \IZ_{>0}$ for fixed $I$ and $I'$ lead to the same genus-one fibration. This can be observed by considering the redundancy criteria in Section \ref{oguiso-wilson} and \eref{proport_fibs}. In short, the fiber class $\cD^2$ can be compared up to scaling for each member of the family. By intersecting $\cD^2$ with the basis elements $D_m$ for $m=1, \cdots, h^{1,1}(X_{r,s})$ as in \eref{prop_fib_triple}, one immediately observes that
\beq\label{DSquare-prop}
\cD^2_{r,s}(\alpha, I; \alpha', I') = \alpha \alpha' \cD_{r,s}^2(1, I; 1, I') \ , 
\eeq
with $\cD_{r,s}^2(1, I; 1, I') \neq 0$. Therefore, the fiber classes of the two possible fibrations are proportional (see the discussion in Section \ref{oguiso-wilson}). Hence, the bases of these two fibrations should differ only at non-generic points -- for example, the two possible bases to the fibration are birational to each other (i.e. related by blowups in the base). On the other hand, it can also be shown that there are no curve classes which have a finite volume for one choice of $\alpha$ and $\alpha'$ but which shrink for another choice (see Appendix~\ref{shrink} for the details). This rules out possible disagreement of the fibrations. Secondly, having restricted ourselves to the divisors $\cD_{r,s}(1, I; 1, I')$, one can prove that the $N_r N_s$ such divisors all lead to distinct genus-one fibrations. The following sufficient condition turns out to distinguish all those fibrations: 
\bi
\item If two divisors $\cD_1$ and $\cD_2$ each obey the conditions~\eqref{kollar-app1} while $\cD_1+\cD_2$ does not, then $\cD_1$ and $\cD_2$ represent two distinct fibrations. 
\ei

\noindent The above completes the classification of genus-one fibrations for all the CICYs in Table~\ref{tb:dPxdP} except for the first five geometries. For these cases, the general divisors of $X_s \subset \cA_s$ are integer linear combinations of the form,
\beq\label{div-gen-P1P1dPs}
\cD= h H + \tilde h \tilde H  + a' L' - \sum_{i'=1}^s a'_{i'} E'_{i'} \ , 
\eeq
where $H$ and $\tilde H$ are the two hyperplane classes of the $\IP^1 \times \IP^1$, appropriately pulled back to the Calabi-Yau threefold. We can now go through exactly the same steps as we did for $X_{r,s}$. Firstly, the triple intersection of $\cD$ is given as
\bea
\int_{X_s} \cD^3 &=& \int_{\cA_s} (h H + \tilde h \tilde H  + a' L' - \sum_{i'=1}^s a'_{i'} E'_{i'} )^3 \wedge(2H + 2\tilde H + 3 L' - \sum_{i'=1}^s E'_{i'}) \\\label{DCube-P1P1dPs}
&=& h \tilde h (3a' - \sum_{i'=1}^s a'_{i'}) + 2 (h+\tilde h) (a'^2 - \sum_{i'=1}^s {a'_{i'}}^2) \ . 
\eea
As in the $X_{r,s}$ cases, we can begin by assuming that each factor in the two terms of \eqref{DCube-P1P1dPs} are non-negative: 
\bea
h \geq 0 \ , ~~\tilde h \geq 0 \ ,~~ h+ \tilde h\geq 0, \\ \label{simple''}
3a'- \sum_{i'=1}^s a'_{i'} \geq 0 \ ,  \\ \label{hard''}
a'^2 - \sum_{i'=1}^s {a'_{i'}}^2 \geq 0 \ , 
\eea
where the equality conditions for the last two inequalities have been described in the text around \eqref{equality-s}. It thus follows that the triple intersection of $\cD$ can only be set to zero in the following four cases. 

\paragraph{Case 1: $(h, \tilde h) =(0,0)$.} Each $(a', a'_1, \cdots, a'_s)\neq 0$ with $D'$ nef in $dP_s$ and $D'^2 \neq 0$ represents a genus-one fibration, where the base manifold is either $dP_s$ or its blow down.
Just as in the cases of $X_{r,s}$, for a generic $(a', a'_1, \cdots, a'_s)$ the base is $dP_s$ and the fibration is an OGF. Enumeration of all the blown-down bases is worked out in Appendix~\ref{enum}.

\paragraph{Case 2: $(a', a'_1, \cdots, a'_s)= 0$.} Each $(h, \tilde h) \neq 0$ corresponds to the genus-one fibration with the base $\IP^1 \times \IP^1$. Such a fibration is an OGF. 

\paragraph{Case 3: $h=0$, $\tilde h > 0$ and $(a', a'_1, \cdots, a'_s)\neq 0$.} In this case, $\cD^3=0$ is only achieved for $\cD=\cD_s(h=0, h';\alpha',I')$ of the form~\eqref{div-gen-P1P1dPs}, with 
\beq
(a', a'_1, \cdots, a'_s) = \alpha' A_s^{(I')} \ , ~\alpha' \in \IZ_{> 0} \ , ~I'=1, \cdots, N_s \ , 
\eeq
where $A_s^{(I')}$ are specified in \eqref{As}. Following exactly the same procedures as in the $X_{r,s}$ cases, we can confirm that $\cD_s(0,h'; \alpha', I')$ represent the same genus-one fibration as $\cD_s(0,1; 1, I')$.

\paragraph{Case 4: $h>0$ and $\tilde h = 0$ and $(a', a'_1, \cdots, a'_s)\neq0$.} In this case, $\cD^3=0$ is only achieved for $\cD=\cD_s(h, h'=0;\alpha',I')$ of the form~\eqref{div-gen-P1P1dPs}, again with 
\beq
(a', a'_1, \cdots, a'_s) = \alpha' A_s^{(I')} \ , ~\alpha' \in \IZ_{> 0} \ , ~I'=1, \cdots, N_s \ . 
\eeq
We can confirm that $\cD_s(h,0; \alpha', I')$ represent the same genus-one fibration as $\cD_s(1,0; 1, I')$. 
\vspace{2mm}

Finally, restricting ourselves to the $2N_s$ divisors, $\cD_s(0,1; 1, I')$ and $\cD_s(1,0;1, I')$, we can prove, based on the aforementioned sufficient criterion for distinguishing fibrations, that the $2N_s$ such divisors all lead to distinct genus-one fibrations. 

The counting of distinct genus-one fibrations for $X_{r,s} \subset \cA_{r,s}$ and $X_{s} \subset \cA_s$, with $r=0, \ldots, 7$ and $s=3,\ldots,7$, is summarized in Table~\ref{tb:stat}. We also provide in Table~\ref{tb:stat-red} another counting result that takes into account of some context-dependent potential redundancies (the permutations of the exceptional divisors of the del Pezzo factors, as well as the permutations of the two $\IP^1$ factors in the $X_{s}$ cases). Note, however, that such redundancies only arise from the topological view point. For the purpose of string dualities, a more relevant counting is the one given in Table~\ref{tb:stat}.

\begin{table}
\begin{center}
\begin{tabular}{|c ||cccc|}
\hline
Calabi-Yau & \multicolumn{4}{c|}{Number of Fibrations}   \\ 
Space & Case 1 & Case 2 & All Others & Tot.  \\ \hline\hline
$X_3$ & 18 & 1 & 6 & 25 \\ \hline
$X_4$ & 76 & 1 & 10 & 87 \\ \hline
$X_5$ & 393 & 1 & 20 & 414 \\ \hline
$X_6$ & 2764 & 1 & 54 &  2819 \\ \hline
$X_7$ & 27094 & 1 & 252 &  27347 \\ \hline
$X_{0,3}$ & 18 & 1 & 0 & 19 \\ \hline
$X_{0,4}$ & 76 & 1 & 0 & 77 \\ \hline
$X_{0,5}$ & 393 & 1 & 0 &  394   \\ \hline
$X_{0,6}$ & 2764 & 1 & 0 & 2765 \\ \hline
$X_{0,7}$ & 27094 & 1 & 0 & 27095 \\ \hline
$X_{1,3}$ & 18 & 2 & 3 & 23 \\ \hline
$X_{1,4}$ & 76 & 2 & 5 & 83\\ \hline
$X_{1,5}$ & 393 & 2 & 10 & 405  \\ \hline
$X_{1,6}$ & 2764 & 2 & 27  & 2793  \\ \hline
$X_{1,7}$ & 27094 & 2 & 126 & 27222 \\ \hline
$X_{2,3}$ & 18 & 5 & 6 & 29 \\ \hline
$X_{2,4}$ & 76 & 5 & 10 & 91 \\ \hline
$X_{2,5}$ & 393 & 5 & 20  & 418  \\ \hline
$X_{2,6}$ & 2764 & 5 & 54  & 2823  \\ \hline
$X_{2,7}$ & 27094 & 5 & 252  & 27351  \\ \hline
$X_{3,3}$ & 18 & 18 & 9 & 45 \\ \hline
$X_{3,4}$ & 76 & 18 & 15 & 109 \\ \hline
$X_{3,5}$ & 393 & 18 & 30  & 441  \\ \hline
$X_{3,6}$ & 2764 & 18 & 81  & 2863  \\ \hline
$X_{3,7}$ & 27094 & 18 & 378  & 27490  \\ \hline
$X_{4,4}$ & 76 & 76 & 25 & 177 \\ \hline
$X_{4,5}$ & 393 & 76 & 50  & 519  \\ \hline
$X_{4,6}$ & 2764 & 76 & 135  & 2975  \\ \hline
$X_{4,7}$ & 27094 & 76 & 630  & 27800   \\ \hline
$X_{5,5}$ & 393 & 393  & 100  & 886   \\ \hline
$X_{5,6}$ & 2764 & 393  & 270  & 3427   \\ \hline
$X_{5,7}$ & 27094  & 393  & 1260  & 28747  \\ \hline
$X_{6,6}$ & 2764  & 2764  & 729  & 6257   \\ \hline
$X_{6,7}$ & 27094  & 2764  & 3402  & 33260   \\ \hline
$X_{7,7}$ & 27094  & 27094  & 15876  & 70064  \\ \hline
\end{tabular}\end{center}
\caption{{\it The statistics of distinct genus-one fibrations for all the Calabi-Yau threefolds in Table~\ref{tb:dPxdP}, where the Calabi-Yau hypersurfaces are embedded in the Fano fourfolds as $X_{r,s} \subset dP_r \times dP_s$ and $X_{s} \subset \IP^1\times \IP^1\times dP_s$, with $r=0, \ldots, 7$ and $s=3,\ldots,7$. Case 1 and Case 2 denote those fibrations with the base surface being the second and the first ambient factor, respectively (incl. all their blow downs); this is consistent with the case distinctions in Section~\ref{case-distinction}.} }
\label{tb:stat}
\end{table}

\begin{table}
\begin{center}
\begin{tabular}{|c ||cccc|}
\hline
Calabi-Yau & \multicolumn{4}{c|}{Number of Fibrations}   \\ 
Space & Case 1 & Case 2 & All Others & Tot.  \\ \hline\hline
$X_3$ & 8  & 1 & 2 & 11 \\ \hline
$X_4$ & 13 & 1 & 4 & 18 \\ \hline
$X_5$ &  25 &  1 &  4 &  30 \\ \hline
$X_6$ &  51 &  1 &  6 &   58 \\ \hline
$X_7$ &  112 &  1 &  10 &   123 \\ \hline
$X_{0,3}$ & 8 & 1 & 0 & 9 \\ \hline
$X_{0,4}$ & 13 & 1 & 0 & 14 \\ \hline
$X_{0,5}$ &  25 &  1 &  0 &   26  \\ \hline
$X_{0,6}$ &  51 &  1 &  0 &  52 \\ \hline
$X_{0,7}$ &  112 &  1 &  0 &  113 \\ \hline
$X_{1,3}$ & 8 & 2 & 1 & 11 \\ \hline
$X_{1,4}$ & 13 & 2 & 2 & 17 \\ \hline
$X_{1,5}$ &  25 &  2 &  2&  29 \\ \hline
$X_{1,6}$ &  51 &  2 &  3 &  56 \\ \hline
$X_{1,7}$ &  112 &  2 &  5&  119\\ \hline
$X_{2,3}$ & 8 & 4 & 1 & 13 \\ \hline
$X_{2,4}$ & 13 & 4 & 2 & 19 \\ \hline
$X_{2,5}$ &  25 &  4 &  2 &  31 \\ \hline
$X_{2,6}$ &  51 &  4 &  3 &  58 \\ \hline
$X_{2,7}$ &  112 &  4 &  5 &  121 \\ \hline
$X_{3,3}$ & 8 & 8 & 1 & 9 \\ \hline
$X_{3,4}$ & 13 & 8 & 2 & 23 \\ \hline
$X_{3,5}$ &  25 &  8 &  2 &  35 \\ \hline
$X_{3,6}$ &  51 &  8 &  3 &  62 \\ \hline
$X_{3,7}$ &  112 &  8 &  5 &  125 \\ \hline
$X_{4,4}$ & 13 & 13 & 3 & 16 \\ \hline
$X_{4,5}$ &  25 &  13 &  4 &  42 \\ \hline
$X_{4,6}$ &  51 &  13 &  6 &  70 \\ \hline
$X_{4,7}$ &  112 &  13 &  10 &  135  \\ \hline
$X_{5,5}$ &  25 &  25 &  3 &  28  \\ \hline
$X_{5,6}$ &  51 &  25 &  6 &  82  \\ \hline
$X_{5,7}$ &  112 &  25 &  10 &  147 \\ \hline
$X_{6,6}$ &  51 &  51 &  6 &  57  \\ \hline
$X_{6,7}$ &  112 &  51 &  15 &  178  \\ \hline
$X_{7,7}$ &  112 &  112 &  15 &  127 \\ \hline
\end{tabular}\end{center}
\caption{{\it The results from Table \ref{tb:stat} for genus-one fibrations for all the Calabi-Yau threefolds in Table~\ref{tb:dPxdP}, grouped into families to reflect the symmetries of the del Pezzo ambient spaces. Case 1 and Case 2 denote those fibrations with the base surface being the second and the first ambient factor, respectively (incl. all their blow downs); this is consistent with the case distinctions in Section~\ref{case-distinction}. The families are defined by symmetries including: (a) for Case 1 and Case 2 the fibrations are only counted up to permutations of the exceptional divisors of the del Pezzo factors; (b) for All Others, up to the obvious permutation symmetries of the ambient space (i.e., permutations of the two $\IP^1$ factors in the $X_{s}$ cases and those of the two surface factors in the $X_{r,s}$ cases with $r=s$); (c) in the $X_{r,s}$ cases with $r=s$, the numbers for Case 1 and Case 2 do not add up for the Total counting as only one half of these is regarded as irredundant.} }{\label{tb:stat-red}}
\end{table}

%%%%%

\section{An infinite number of fibrations}\label{schoen_sec}
In this section we analyze the CICY with Hodge numbers $(19,19)$ which has long been known to have a number of unique and remarkable properties. This geometry proves to be the \emph{only K\"ahler favorable CICY studied in this work which admits an infinite number of genus one fibrations}\footnote{Note, as discussed in previous sections, we are limited here to the study of K\"ahler favorable CICY threefolds. There could of course exist other geometries, not in this class, which also admit an infinite numbers of fibrations.}. The existence of an infinite number of both genus one and $K3$-fibrations in this geometry has been observed in several contexts previously \cite{ogu,Aspinwall:1996mw}, however we will provide here a new and explicit parameterization of one such infinite family.
\subsection{The Schoen manifold}\label{non-dPdP}
%%%
\subsubsection{Topological data}
As described in Section~\ref{1.1}, of the $48$ non-favorable configurations studied in Section \ref{the48_sec}, there are $15$ with Hodge numbers $(h^{1,1},h^{2,1})=(19,19)$ and all can be proved to be equivalent to one another via splittings/contractions. Each is manifestly a fiber product of two generic rational elliptic surfaces (called $dP_9$ in the physics literature) identified over a common $\IP^1$. Thus, they are all equivalent to the Schoen manifold. 

One simple CICY configuration, similar in spirit to those studied in Section \ref{the48_sec} (i.e. a hypersurface in an ambient product of two surfaces) provides a particularly straightforward way to compute the topology -- including all of the triple intersection numbers -- of this manifold:
\beq\label{sch-20}
X_{\rm Schoen} \sim \left[ \begin{array}{c|ccc} 
\mathbb{P}^1 & 0 & 1 & 1 \\ 
\mathbb{P}^1 & 1 & 0 & 1 \\
\mathbb{P}^2 & 0 & 3  & 0\\ 
\mathbb{P}^2 & 3 & 0  & 0
\end{array}\right] \ .
\eeq
This configuration can be obtained by ineffectively splitting the configuration of the split bi-cubic (number 14 in the original CICY list \cite{Candelas:1987kf,website}),
\beq\label{Sch-eg}
\left[ \begin{array}{c|cc} 
\mathbb{P}^1 & 1 & 1  \\ 
\mathbb{P}^2 & 0 & 3  \\ 
\mathbb{P}^2 & 3 & 0 
\end{array}\right] \ . 
\eeq
The configuration~\eqref{sch-20} describes the Schoen manifold as an anticanonical divisor of $\cA_{9,9}=dP_9 \times dP_9$, with

\beq
dP_9 \sim 
\left[ \begin{array}{c|c} 
\mathbb{P}^1 & 1   \\ 
\mathbb{P}^2 & 3  
\end{array}\right]~.
\eeq

As in the other del Pezzo surface cases, the second homology group $H_2(dP_9, \IZ)$ of $dP_9$ is spanned by the hyperplane class $L$ in $\IP^2$ as well as the, in this case 9, exceptional divisors $E_{i=1, \cdots, 9}$. Their intersections and the first Chern class are as in \eqref{double} and~\eqref{c1}. The twenty ambient divisors of $\cA_{9,9}$ restrict to the anti-canonical hypersurface (i.e. $X_{\rm Schoen}$) with one linear relation that reduces the number of independent divisors to the expected $19$. Despite the fact that $h^{1,1}(\cA_{9,9})=20 = 1+ h^{1,1}(X_{\rm Schoen})$, $X_{\rm Schoen}$ is both favorable and (as we will see below) K\"ahler favorable, since its Picard group, K\"ahler, and Mori cones descend directly from $\cA_{9,9}$, albeit in this case with a non-trivial redundancy.

The linear relationship reducing the $20$ divisors of $\cA_{9,9}$ to the 19 dimensional Picard group of $X_{\rm Schoen}$ can be seen in several ways. The first of these is to consider imposing the third defining equation (given by the last column in \eref{sch-20}) first. From the well-known relation that
\beq\label{p1_rel}
\left[ \begin{array}{c|c} 
\mathbb{P}^1 & 1   \\ 
\mathbb{P}^1 & 1  
\end{array}\right] \sim \mathbb{P}^1 \ , 
\eeq
it is clear that two distinct divisors (the hyperplanes of the first two $\mathbb{P}^1$ factors of the ambient space) are made linearly equivalent by imposing the third defining relation. Alternatively, this same relation can be observed by considering the long exact sequence in cohomology associated to the dual of the adjunction sequence:
\beq
0 \to TX_{\rm Schoen} \to T\cA_{9,9}|_{X_{\rm Schoen}} \to \mathcal N|_{X_{\rm Schoen}} \to 0 \ , 
\eeq
where $\mathcal N|_{X_{\rm Schoen}}$ is the restriction of the normal bundle with the Chern class $c_1(\mathcal N)=3L - \sum_{i=1}^9 E_i + 3 L' - \sum_{i'=1}^9 E'_{i'}$. An explicit algebraic description of the following morphism (using the tools of \cite{Anderson:2013qca,Anderson:2008ex}) 
\beq
\phi: H^1(X_{\rm Schoen}, \mathcal N^*_{X_{\rm Schoen}}) \to H^1(X_{\rm Schoen}, T{\cal A}_{9,9}^*|_{X_{\rm Schoen}})
\eeq
demonstrates that $h^1(X_{\rm Schoen}, TX^*_{\rm Schoen}) =\rm{dim}(\rm{coker}(\phi))=19$ with the same linear relationship as in \eref{p1_rel} imposed. Choosing the obvious basis of divisors descended from the ambient product of $dP_9$'s:
\beq\label{20}
L, E_{i=1, \dots, 9}, \quad \text{and} \quad L', E'_{i'=1, \dots, 9} \ , 
\eeq
the linear relationship takes the following form:
\beq\label{red}
3L-\sum_{i=1}^9 E_i = 3L'-\sum_{i'=1}^9 E'_{i'} \ . 
\eeq
Having in mind of this redundancy, the Mori and the K\"ahler cones of $X_{\rm Schoen}$ can be immediately obtained by pulling back those of the ambient space $\cA_{9,9}$ (though the argument is distinct in this case from that used in Lemma \ref{lemma2} since the ambient space is not Fano. Instead the same conclusion -- that the K\"ahler and Mori cones descend from $\cA_{9,9}$ -- can be obtained by the results of ~\cite{gm}). The Mori cone generators of $dP_9$ are described in Table~\ref{tb:dP-Mori}. 

With this favorable basis of divisors in hand, the triple intersection numbers of $X_{\rm Schoen}$ can be derived in the usual way as
\beq\label{sch-inter}
d_{m n p}^{\rm Schoen} \equiv \int_{X_{\rm Schoen}} J_m \wedge J_n \wedge J_p = \int_{\cA_{9,9}} J_m \wedge J_n \wedge J_p \wedge (3L - \sum_{i=1}^9 E_i + 3L' - \sum_{i'=1}^9 E'_{i'}) \ . 
\eeq
Here, $m, n, p$ are the indices labeling the redundant ($20$-dimensional) basis of descended harmonic $(1,1)$-forms on $X_{\rm Schoen}$. Such a description of the intersection numbers is symmetric in the two $dP_9$ factors. Note that the intersection numbers with the indices ranging from $1$ to $19$ already form a complete dataset themselves, given the redundancy~\eqref{red} within the $20$ $(1,1)$-forms.

%%%
\subsubsection{A study of genus-one fibrations}
In this section we will set-up a systematic study of the fibrations of $X_{\rm Schoen}$ and demonstrate that unlike the other cases investigated in Section \ref{the48_sec}, the Schoen manifold manifestly admits an \emph{infinite number} of genus one fibrations. Remarkably, it is the only such manifold we have encountered in the CICY list. We will not attempt to classify all such infinite classes here, but will instead illustrate the phenomenon with one explicit infinite family.

In order to classify genus-one fibrations within the Schoen manifold, we will take a similar approach to the one in Subsection~\ref{dPdP-classify}, based on Koll\'ar's criteria. Even in the presence of the redundancy~\eqref{red}, the divisors of $X_{\rm Schoen}$ can be integrally parameterized as
\beq
\cD= a L - \sum_{i=1}^9 a_i E_i + a' L' - \sum_{i'=1}^9 a'_{i'} E'_{i'} \ ,
\eeq
whose triple intersection is given by 
\bea
\int_{X_{\rm Schoen}} \cD^3 &=& \int_{dP_9 \times dP_9} (a L - \sum_{i=1}^9 a_i E_i + a' L' - \sum_{i'=1}^s a'_{i'} E'_{i'} )^3 \wedge (3L - \sum_{i=1}^9 E_i + 3L' - \sum_{i'=1}^9 E'_{i'}) \\ \label{triple-Sch}
&=& 3 (a^2 - \sum_{i=1}^9 a_i^2) (3a'  - \sum_{i'=1}^9 a'_{i'}) + 3 (a'^2 - \sum_{i'=1}^9 {a'_{i'}}^2) (3a  - \sum_{i=1}^9 a_{i}) \ .
\eea

In a search for fibrations, one of the Koll\'ar criteria~\eqref{kollar-app1} says that $\cD \cdot {\cal C} \geq 0$ for all curves $\cal C$ in $X_{\rm Schoen}$. Given the description of the Mori and the K\"ahler cones of $X_{\rm Schoen}$ in the previous subsection, it follows that
\beq
(a L - \sum_{i=1}^9 a_i E_i) \cdot C \geq 0~~\text{and}~~(a' L' - \sum_{i'=1}^9 a'_{i'} E'_{i'} ) \cdot C' \geq 0 \ , 
\eeq
is required for all curves $C$ and $C'$ of the two $dP_9$ factors, respectively. 

For simplicity, we will from now on denote the two pieces in $\cD$ as
\bea\nn 
D&=&a L - \sum_{i=1}^9 a_i E_i \ , \\ \nn
D'&=&a' L' - \sum_{i'=1}^9 a'_{i'} E'_{i'} \ , 
\eea
and due to the complete symmetry in the two $dP_9$ pieces, it is sufficient to analyze this constraint for a single piece, say, $D$. 
Since the fiber class $f=3L- \sum_{i=1}^9 E_i$ is a Mori cone generator and $D$ must itself be an effective divisor of $dP_9$, we have $D \cdot f \geq 0 $ and $D \cdot D \geq 0$, which, respectively, lead to  
\bea\label{first}
3a-\sum_{i=1}^9 a_i &\geq& 0 \\ \label{second}
a^2 - \sum_{i=1}^9 a_i^2 &\geq& 0 \ . 
\eea
In order to saturate~\eqref{first}, one must have 
\beq\label{ineqs}
9a^2 = (\sum_{i=1}^9 a_i)^2 \leq 9(\sum a_i^2 ) \leq 9a^2 \ , 
\eeq 
where the first inequality comes from the Cauchy-Schwarz inequality and the second, from~\eqref{second}. Thus, both inequalities in~\eqref{ineqs} saturate, which can only happen when $(a, a_1, \cdots, a_9) \sim (3, 1, \cdots, 1)$, i.e., when $D$ is a positive multiple of $f$. However, since the fiber classes $f$ and $f'$ on the two $dP_9$ factors are identified on $X_{\rm Schoen}$, we may as well view this as $D=0$ by shifting $D'$ by $f'$. 

Thus, as in subsection~\eqref{dPdP-classify}, we are naturally lead to the following three cases. 

\paragraph{Case 1: $(a, a_1, \cdots, a_9) = 0$.} In this case, $(a', a'_1, \cdots, a'_9)\neq 0$ so that $\cD^2 \neq 0$, and each such divisor corresponds to the genus-one fibration with the base $dP_9$ (or its blow downs). Such a fibration is an OGF of the CICY configuration. For instance, in the configuration~\eqref{Sch-eg} of CICY 14, an OGF with the base $dP_{9}\sim \left[ \begin{array}{c|cc} 
\mathcal \IP^1 & 1 \\
\mathcal \IP^2 & 3 \\
\end{array}\right]$ is immediately found, where, for instance, the $\IP^2$ is that in the second row of the threefold configuration matrix.

\paragraph{Case 2: $(a', a'_1, \cdots, a'_9) = 0$.} In this case, $(a, a_1, \cdots, a_9) \neq 0$ so that $\cD^2 \neq 0$, and each divisor of this type corresponds to the genus-one fibration with the base being the other $dP_9$ (or its blow downs). Such a fibration is also an OGF. In the configuration~\eqref{Sch-eg}, this corresponds to the OGF with the base $dP_{9}\sim \left[ \begin{array}{c|c} 
\mathcal \IP^1 & 1 \\ 
\mathcal \IP^2 & 2\\
\end{array}\right]$, where the $\IP^2$ is now taken from the third row of the threefold configuration. 

\paragraph{Case 3: $(a, a_1, \cdots, a_r) \neq 0$ and $(a', a'_1, \cdots, a'_s) \neq 0$.} Note first that the cases with $D$ being a positive multiple of $f$ and those with $D'$ being a positive multiple of $f'$  for $D'$ have effectively been considered in the Case 2 and the Case 1 above, respectively. Therefore, we may strengthen the constraint~\eqref{first} so that $(a, a_1, \cdots, a_9)$ obey 
\beq\label{first-s}
3a-\sum_{i=1}^9 a_i > 0\ , 
\eeq
and $(a', a'_1, \cdots, a'_9)$ obey the analogously strengthened constraint. For the triple intersection of $\cD$ in \eqref{triple-Sch} to vanish, we must thus have the inequality~\eqref{second} saturated. 
Then, from the Riemann-Roch theorem, it follows that
\beq
0 < D \cdot f  = D\cdot D - 2 g + 2 = -2g + 2 \leq 2 \ , 
\eeq 
and hence, the last inequality has to saturate. Therefore, we only need to classify the solutions to
\bea\label{first-f}
3a-\sum_{i=1}^9 a_i = 2\ ,  \\ \label{second-f}
a^2 - \sum_{i=1}^9 a_i^2 = 0 \ , 
\eea
that describe certain rational curves of $dP_9$. Note that these two constraints for $D$ arose from $D \cdot C \geq 0$ for particular curves $C$ in $dP_9$ and as such they do not guarantee the inequality for every curve in $dP_9$ (i.e. the nef criterion). 

Once we require that $D$ should be nef, a somewhat lengthy argument (see Appendix~\ref{nefimplied} for details) demonstrates that the solutions can be characterized as 
\beq\label{mori_to_kollar}
D=y+z \ , 
\eeq
where $y, z\,(\neq f)$ are any two (of the \emph{infinite} number of) distinct Mori cone generators with $y \cdot z=1$. 

Given such a characterization, one can obtain a parametric family of infinitely many solutions for $D$ (as described in \eqref{inf-family-d}) by making use of specific parameterizations of the Mori cone generators of $dP_9$ (see Appendix~\ref{classify-dp9} for details). An analogous parameterization is obtained for $D'$ (as described in \eqref{inf-family-dprime}) and hence, one can parametrically describe an infinite family of divisors $\cD = D + D'$ of $X_{\rm Schoen}$, each of which represents a genus-one fibration. 
Here, for simplicity, we will present and analyze an infinite, one-parameter subfamily\footnote{This subfamily is obtained by turning off all but one parameter $k_8 \equiv k$ from the $14$-parameter family described in \eqref{inf-family-d}, \eqref{inf-family-dprime}, and \eqref{inf-family}. See Appendix~\ref{classify-dp9}.} with
\bea\label{DD'}
D&=& 2(2+k+k^2)f + 3E_1 - E_2 +E_3+E_4 +2k E_8 -2(1+k)E_9 \ ,  \\ \nn
D'&=& 4f' +3E'_1 - E'_2 +E'_3+E'_4 -2E'_9\ , 
\eea
where $f=3L-\sum_{i=1}^9 E_i$ and $f'=3L'-\sum_{i'=1}^9 E'_{i'}$ have been used for a simpler description and $k \in \mathbb{Z}$. It is straightforward to verify that $\cD$ is nef (via the criteria in \eref{kollar-app1}) and satisfies $\cD^3=0$ and $\cD^2 \neq 0$ as required.

Finally, it remains to verify that for each value of $k$ these are in fact \emph{inequivalent} Koll\'ar divisors. For this we will utilize the criteria laid out in \eref{proport_fibs} (i.e. that two Koll\'ar divisors are equivalent if they lead to proportional fiber classes in $X$) in Section \ref{oguiso-wilson}. Using the triple intersection numbers~\eqref{sch-inter} of $X_{\rm Schoen}$, the fiber class (up to a proportional constant) can be obtained as the following vector of length $19$, 
\bea\nn
\{{\cD}^2 \cdot J_m\}_{m=1}^{10} &=& \{48+3\kappa, 4+\kappa, 20+\kappa, 12+\kappa, 12+\kappa, 16+\kappa, 16+\kappa, 16+\kappa, 16+8k^2, 24+\kappa+8k\} \\ \nn
\{{\cD}^2 \cdot J_m\}_{m=11}^{19} &=& \{48, 4,20,12,12,16,16,16,16\} \ , 
\eea
where $\kappa=8k+8k^2$ and the $19$ $(1,1)$-forms $J_m$ are the first $19$ of those in \eqref{20} except the $E'_{9}$. It is therefore obvious that the fibers can never be the same for different values of $k$ and hence, all of the divisors $\cD$ described by \eqref{DD'} represent distinct genus-one fibrations. Thus, $\cD=D + D'$ with \eref{DD'} defines a true infinite family of genus one fibrations in $X_{\rm Schoen}$.

It may seem somewhat surprising that an infinite number of genus one fibrations can arise for any CICY threefold. In addition, the fact that this infinite structure appears to occur for only one geometry is also remarkable. However, the Schoen manifold has a number of special features that set the stage for this infinite structure. First, as remarked above, the K\"ahler and Mori cones of $X_{\rm Schoen}$ are infinitely generated. This alone was perplexing in the context of mirror symmetry and was the subject of a detailed study by Morrison and Grassi in \cite{gm}. There they observed that although the K\"ahler cone of $X_{\rm Schoen}$ is infinitely generated, the action of the automorphism group of $X_{\rm Schoen}$ on the K\"ahler cone has a rationally polyhedral fundamental domain (Morrison has conjectured\footnote{For a more precise definition of the Morrison-Kawamata Cone Conjecture see \cite{2016arXiv161100556L} for a recent review.} that this must hold for any CY manifold with such an infinitely generated K\"ahler cone \cite{morrison1993}). Like our constructions above, the automorphism action descends from the two ambient space copies of the rational elliptically fibered surface, $dP_9$, and the associated action on the K\"ahler/Mori cones of $dP_9$. Although it is beyond the scope of the present work to explore, it is reasonable to surmise that this automorphism action could collapse the infinite families parameterized above into a finite (physically distinct) set.

Finally, realizing the Schoen manifold as the blow up of the toroidal orbifold $T^6/\mathbb{Z}_2 \times \mathbb{Z}_2$ \cite{ogu} has lead to the observation that $X_{\rm Schoen}$ must contain an infinite number of both elliptic and $K3$ fibrations (see \cite{ogu} and \cite{piateckii_s} for details). Moreover, the physical consequences of this infinite fibration structure for string dualities (in particular for heterotic/Type IIA or heterotic/F-theory duality) were explored in \cite{Aspinwall:1996mw}, where an argument was made that such infinite families could be realized as U-duality symmetries \cite{Hull:1994ys} in the dual pairs. It would be interesting to find additional examples of such infinite fibration structures in other constructions of CY manifolds and to more explicitly explore their consequences for string dualities.

\vspace{0.1cm}

\section{Conclusions and future directions}\label{conclusions_sec}
As summarized in Section \ref{set_up_sec}, the goal of this work was to undertake a comprehensive survey of genus-one fibrations in the data set of CICY threefolds. We find that not only do more than $99\%$ of the 7890 configuration matrices in this dataset lead to a genus-one fibered geometry, but that this simple set of CY manifolds yields a vast number (indeed an infinite number) of distinct genus one fibrations. We have approached this survey using two tools: 1) A study of so-called ``obvious" fibrations manifestly realized in the algebraic form of the manifolds and 2) An exhaustive search for genus-one fibrations using the Koll\'ar criteria outlined in Section \ref{oguiso-wilson}. We explicitly provide examples of geometries for which these methods of enumerating fibrations agree (see Section \ref{OGF_vs_Kollar}) and those for which OGFs vastly undercount the possible fibration structures (see Sections \ref{the48_sec} and \ref{schoen_sec}).

The Koll\'ar criteria provides a means of classifying \emph{all genus-one fibrations} within the set of CICY threefolds. However, a complete classification is dependent on having explicit descriptions of the full K\"ahler and Mori cones of each manifold. In general, tools to determine this data had been lacking in the literature to date and we substantially expand these methodologies here. By means of splitting/contracting configuration matrices as outlined in Section \ref{top_sec} we have found new, favorable descriptions of $2946$ CICYs and moreover determined the full K\"ahler/Mori cone structure of $4957$ ``K\"ahler favorable" geometries. Thus, the survey completed here is a complete classification of genus-one fibrations for $4957$ out of the $7868$ CICYs that are not direct products. If new tools are developed to complete the K\"ahler/Mori data of the remaining CICY geometries, the techniques outlined here could be readily applied to the remainder of the CICY dataset. Such systematic scans for Koll\'ar divisors/fibrations in toric hypersurfaces could also be readily carried out. We leave such explorations for future work.  

The enhanced CICY list \cite{website} described above now contains the full topology of each manifold (the so-called ``Wall's theorem data" \cite{wall_thm} that can be used to distinguish these threefold geometries as real manifolds), and also, thanks to the OGF survey, many descriptions of them as genus one fibered geometries. It is our expectation that this data will provide a useful playground for the study of string compactifications, model building, and string dualities. For example, this set of CY geometries has already provided a fruitful arena for heterotic model building \cite{Anderson:2014hia,Anderson:2013xka,Anderson:2012yf,Anderson:2011ns,Anderson:2009mh,Anderson:2007nc,Anderson:2008uw, Buchbinder:2016jqr,Blesneag:2015pvz} (see also toric approaches in \cite{He:2009wi,He:2011rs,He:2013ofa}) and simple geometries in which to explore new approaches to moduli stabilization \cite{Anderson:2013qca,Anderson:2011ty,Anderson:2011cza,Anderson:2010ty,Anderson:2010mh,Anderson:2009nt,Anderson:2009sw}. The study of fibration structures completed here could yield valuable new approaches to the study of heterotic string and F-theory compactifications, as well as string dualities -- including providing novel backgrounds for $6$-dimensional compactifications of F-theory. For example, CICY threefolds have been observed to generically lead to higher rank Mordell-Weil groups \cite{Anderson:2016ler,Anderson:2016cdu} and have been used as compactification geometries for heterotic/F-theory duality for ${\cal N}=1$, 4-dimensional solutions. 

The results of the fibration survey completed here indicate that even existing data sets of CY geometries may contain many more genus-one fibrations than previously realized. Indeed, genus one fibrations have long provided intriguing structure that suggests a possible route to classifying all CY threefolds. The work of \cite{grassi, gross_finite} establishing that the set of all genus-one fibered CY manifolds is finite was motivated in no small part by the hope that this result could be used to bound the number of \emph{all CY geometries}. The essential idea is as follows: many studies of CY threefold geometry have indicated that as $h^{1,1}$ increases, the topology and triple intersection numbers of the threefold take on more specific forms \cite{kanazawa_wilson, wilson_top}. This fact, coupled to the ubiquitous presence of genus-one fibrations at large $h^{1,1}$ in known data sets of CY threefolds, has lead to speculation that perhaps all CY threefolds with large enough $h^{1,1}$ admit an elliptic fibration -- and hence their topology may be bounded by the classification of \cite{gross_finite} (see also \cite{Morrison:2012np,Morrison:2012js,Taylor:2012dr,Martini:2014iza,Taylor:2015isa} for a recent program of work explicitly enumerating such fibrations). In that spirit, the results of this systematic survey seem to provide more evidence for this conjecture (indeed every CICY threefold is fibered for $h^{1,1}>4$). We are hopeful that the approaches outlined here may be applied to further these ideas and study fibration structures in CY geometries more generally in the future.

\section*{Acknowledgements}
For the initial phase of this work, LA (and XG in part) was supported by NSF grant PHY-1417337 and JG (and SJL in part) was supported by NSF grant PHY-1417316. For the conclusion of this project, the work of LA and JG is supported by NSF grant PHY-1720321. This project is part of the working group activities of the 4-VA initiative ``A Synthesis of Two Approaches to String Phenomenology". The authors would like to thank Andre Lukas for extremely helpful discussions relating to the material in Section \ref{kah_fav_sec}.

\appendix

%%%%%%%%%%

%%%%%%%%%%
\section{Details of the Diophantine system for del Pezzo surfaces}
In this section we provide the details of the arguments leading to a complete fibration count for the $68$ CICY configurations analyzed in Section~\ref{the48_sec}, which correspond to an anticanonical hypersurface in an ambient product of two del Pezzo surfaces (see Table~\ref{tb:dPxdP} for the CICY ID numbers).

\subsection{Saturation of the inequalities}\label{dio}
Here, we will derive the equality conditions for the two inequalities~\eqref{simple} and~\eqref{hard},
\bea\label{simple-app}
3a- \sum_{i=1}^r a_i \geq 0 \ ,  \\ \label{hard-app}
a^2 - \sum_{i=1}^r a_i^2 \geq 0 \ ,
\eea 
that should hold for a divisor $D= aH-\sum\limits_{i=1}^r a_i E_i$ of $dP_r$ obeying $D \cdot C \geq 0$ for all curves $C$ in $dP_r$, where $r=0, \cdots, 7$. 
Without loss of generality, we may assume that $a_i$'s are ordered so that 
\beq
a_1 \geq \cdots \geq a_r \geq 0 \ , 
\eeq
where the last inequality can be seen from the fact that $E_r$, being a Mori cone generator, is effective. Further, when $r \geq 1$, $a\geq a_1$ since $L-E_1 - E_2$ (or $L-E_1$ in the $r=1$ case) is also effective as a Mori cone generator.  

Let us start with the inequality~\eqref{simple-app}. We will prove that $a$ vanishes (and hence so do $a_i$'s) when it saturates, i.e., when $3a=\sum_{i=1}^r a_i$. This can easily be seen for each $r=0, \cdots, 7$ as follows: 
\bi
\item $r=0$: Vacuously true.
\item $r=1$: $a_1= 3a \geq a \geq a_1$ (as $L-E_1$ is effective) ~$\Rightarrow$~ $a=0$.
\item $r=2$: $a_1+a_2 =3a \geq a \geq a_1+a_2$ (as $L-E_1-E_2$ is effective) ~$\Rightarrow$~ $a=0$. 
\item $r=3$: $\sum_{i=1}^3 a_i =3a \geq 2a \geq (a_1+a_2) + (a_1 + a_3)$ (as $L-E_i-E_j$ are effective) $\geq \sum_{i=1}^3 a_i$ ~$\Rightarrow$~ $a=0$. 
\item $r=4$: $\sum_{i=1}^4 a_i =3a \geq 2a \geq (a_1+a_2) + (a_3+a_4)$ (as $L-E_i-E_j$ are effective) $\geq \sum_{i=1}^4 a_i$ ~$\Rightarrow$~ $a=0$.
\item $r=5$: $\sum_{i=1}^5 a_i =3a \geq 2a \geq \sum_{i=1}^5 a_i$ (as $2L-\sum_{i=1}^5 E_i$ is effective) ~$\Rightarrow$~ $a=0$.
\item $r=6$: 
$\sum_{i=1}^6 a_i = 3a = \frac12 (2a + 2a + 2a) \geq \frac32 \sum_{i=1}^5 a_i$ (as $2L-\sum_{i=1}^5 E_i$ is effective) $\geq \sum_{i=1}^5 a_i + (\frac12 a_1 + \frac12 a_2 +\frac32 a_6)  \geq \sum_{i=1}^6 a_i$ ~$\Rightarrow$~ $a_1=0$ ~$\Rightarrow$~ $a=0$. 
\item $r=7$: 
$\sum_{i=1}^7 a_i = 3a = \frac12 (2a + 2a + 2a) \geq \frac32 \sum_{i=1}^5 a_i$ (as $2L-\sum_{i=1}^5 E_i$ is effective) $\geq \sum_{i=1}^5 a_i + (\frac12 a_1 + \frac22 a_6+ \frac22 a_7)\geq \sum_{i=1}^7 a_i$ ~$\Rightarrow$~ $a_1=0$~$\Rightarrow$~ $a=0$. 
\ei

Next, saturation of the second inequality~\eqref{hard-app} can be immediately analyzed on a computer, for each $r=1, \cdots, 7$ (when $r=0$, the inequality can not be saturated unless $a=0$). Practically, upon setting $a=1$, we can show that $g(a_i)\equiv \sum_{i=1}^r a_i^2 \leq 1$ and the maximum value $1$ can only be achieved when $a_i$'s take values from \eqref{As} (up to an appropriate rescaling of $a$). Note that the region for the values of the $r$ parameters $(1\geq) a_1 \geq \cdots \geq a_r (\geq 0)$ is further restricted by the Mori cone generators of Table~\ref{tb:dP-Mori} as follows,
\bi
\item $r=1$: $a_1 \leq 1$ 
\item $r=2, 3, 4$: $a_1 + a_2 \leq 1$
\item $r=5, 6$: $a_1+a_2 \leq 1$, $\sum_{i=1}^5 a_i \leq 2$
\item $r=7$: $a_1+a_2 \leq 1$, $\sum_{i=1}^5 a_i \leq 2$, $a_1+\sum_{i=2}^7 a_i \leq 3$ \ , 
\ei
and the maximum of $g$ may only occur at the critical points on the boundary of the bounded region defined as such. A quick computer analysis based on this leads to the aforementioned result.

\subsection{Shrinking curves}\label{shrink}
To a divisor $\cD$ of the Calabi-Yau threefold $X \subset \cA$ obeying the criteria~\eqref{kollar-app1}, there exists a corresponding genus-one fibration $\pi: X \to B$. Let us suppose that we are given another such divisor $\tilde \cD$ and the corresponding fibration $\tilde\pi: X \to \tilde B$ such that $c \cD^2 = \tilde c \tilde \cD^2$ for some $c, \tilde c\in \IZ_{>0}$. In general there is still a possibility of $\pi$ and $\tilde \pi$ being a different projection with $B$ and $\tilde B$ being only birational. 

In this subsection, we consider the divisors of a given Calabi-Yau threefold $X_{r,s} \subset \cA_{r,s}$, 
\beq
\cD(\alpha, \alpha')\equiv D_{r,s}(\alpha, I; \alpha', I') \ , 
\eeq 
described in \eqref{kollar-rs}, for fixed $I$ and $I'$. They obey \eqref{DSquare-prop} and may in principle represent different fibrations. The purpose of this subsection is to rule out such a possibility.

Let us begin by choosing one such divisor, $\cD=\cD(\alpha, \alpha')$, associated with the corresponding fibration $\pi:X_{r,s} \to B$. In the following, we will first characterize the Picard lattice of $B$. Note that any pulled-back divisors $\cD_B = D_B + D'_B$ from the base $B$ must obey $\cD^2 \cdot \cD_B = 0$ on $X_{r,s}$, where $D_B$ and $D'_B$ are the pieces pulled back from the $dP_r$ and $dP_s$ under their own projections, respectively. Similarly decomposing the divisor $\cD$ as $D+D'$, we thus have
\bea
0 &=& \int_{X_{r,s}} \cD^2 \wedge \cD_B = \int_{\cA_{r,s}} (D+D')^2 \wedge (D_B + D'_B) \wedge (3L - \sum_{i=1}^r E_i + 3L' - \sum_{i'=1}^s E'_{i'})\\\label{44}
&=& 4\alpha' \int_{dP_r} D\wedge D_B  + 4\alpha \int_{dP_s} D'\wedge D'_B \ , 
\eea
where in the last step we have used the properties of $D$ that $D^2=0$ and $D\cdot (3L - \sum_{i=1}^r E_i) = 2\alpha$ on $dP_r$, as well as the analogous properties of $D'$ on $dP_s$. Now, restricting to those divisors $\cD_B$ that are pull-backs of ample divisors of $B$, we must have the two integrals in \eqref{44} both non-negative since $D_B$ and $D'_B$ are effective curves of $dP_r$ and $dP_s$, in particular, and hence, each integral should vanish. In terms of the expansion coefficients in $D=a L - \sum_{i=1}^r a_i E_i$ and $D_B=b L - \sum_{i=1}^r b_i E_i$, we have
\beq
\int_{dP_r} D \wedge D_B = a b - \sum_{i=1}^r a_i b_i \geq ab -(\sum_{i=1}^r a_i^2)^{1/2}(\sum_{i=1}^r b_i^2)^{1/2} \geq 0 \ , 
\eeq
where we have used the Cauchy-Schwarz inequality, as well as the intersections $D\cdot D=0$ (by construction) and $D_B \cdot D_B \geq 0$ (as $D_B$ is ample) on $dP_r$. For the saturation of inequalities, we conclude that $D_B$ is proportional to $D$, and similarly, $D'_B$ to $D'$. Thus, ample cone of $B$, when pulled-back to $X_{r,s}$, should lie in the two-dimensional plane spanned by $D$ and $D'$. Therefore, the Picard lattice of $B$ also lies in that plane, and effective curves $C$ of $B$ can be expanded as $C=\lambda D + \lambda' D'$ for some rational numbers $\lambda$ and $\lambda'$. Note that we are not distinguishing the $(1,1)$-forms in $B$ from their pull-backs to $X_{r,s}$, having in mind of the obvious injection. 

For the rest of this subsection, we will show that there do not exist zero-volume effective curves of $B$ when K\"ahler forms are taken from the interior of the positive cone spanned by $D$ and $D'$. This, if proven, will guarantee that different choice of $\alpha$ and $\alpha'$ still represents the same genus-one fibration. For this purpose, let us consider a K\"ahler form $J$ and an effective curve $C$ of the form,
\beq
J=\mu D+\mu' D' \quad (\mu, \mu'>0)\quad\quad \text{and}\quad C=\lambda D + \lambda' D' \ , 
\eeq
in the base $B$ and demand 
\beq
{\rm Vol}(C)= J \cdot C =0\ .
\eeq
This, when pulled-back to $X_{r,s}$, implies the vanishing triple intersection,
\beq
J \cdot C \cdot e = 0 \ , 
\eeq
in the threefold $X_{r,s}$ where $e$ is an arbitrary divisor of $X_{r,s}$, and hence, leads to
\beq
J\cdot C \cdot e\cdot (3L - \sum_{i=1}^r E_i + 3L' - \sum_{i'=1}^s E'_{i'}) = 0 \ , 
\eeq
in the fourfold $\cA_{r,s}$. Upon some algebras, this gives 
\beq
\lambda : \lambda' = \mu : - \mu' \ , 
\eeq
and hence, 
\beq
C=c \mu D - c\mu' D' \ , 
\eeq
where $D=a L - \sum_{i=1}^r a_i E_i$ with $(a, a_1, \cdots, a_r) = \alpha A_r^{(I)}$ and similarly, $D'=a' L' - \sum_{i'=1}^s a'_{i'} E'_{i'}$ with $(a', a'_1, \cdots, a'_s)=\alpha' A_s^{(I')}$. Since $C$ is an effective divisor of $B$, so is its pull-back $\pi^*C$ in $X_{r,s}$. Denoting the line bundle on $\cA_{r,s}$ of degree $c \mu D - c\mu' D'$ by $\cL$, let us now consider the Koszul sequence,
\beq
0 \to \cL \otimes \cN^* \to \cL \to \cL|_{X_{r,s}} \to 0 \ , 
\eeq 
where $\cN=\cO(3L-\sum_{i=1}^r E_i + 3L' - \sum_{i'=1}^s E'_{i'})$. Using the explicit formulae for the line bundle cohomologies on del Pezzo surfaces (see e.g. Appendix of Ref.~\cite{Blumenhagen:2006wj}), one can then show that $H^0(\cA_{r,s}, \cL)=0 =H^1(\cA_{r,s}, \cL\otimes \cN^*)$, 
and hence, that $H^0(X_{r,s}, \cL|_{X_{r,s}})=0$. However, this contradicts the effectiveness of $\pi^*C$, and therefore, no zero-volume effective curves exist in $B$ for any K\"ahler forms. This completes the proof that the divisors $\cD(\alpha, \alpha')$ of $X_{r,s}$ represent the same fibration for any positive $\alpha$ and $\alpha'$. 

%%%%%%%%%%
\section{Details of the Diophantine system for $dP_9$}
In this section we provide further details on the characterization of the infinite family of genus one fibrations found for the Schoen manifold in Section \ref{schoen_sec}.

\subsection{A non-trivial fact}\label{nefimplied}
Here, we will show that nef divisors $D$ of rational elliptic surface $dP_9$ (i.e., $D\cdot C \geq 0$ for every curve $C$ in $dP_9$) that obey
\bea\label{df}
D\cdot f&=&2  \ , \\ \label{dd} 
D\cdot D&=&0 \ , 
\eea
for the fiber class $f$, can be classified as 
\beq\label{desform}
D=y+z \ , 
\eeq
where $y, z\,(\neq f)$ are two distinct Mori cone generators with $y \cdot z=1$. 

With the Mori cone generators described in Table~\ref{tb:dP-Mori}, $D$ can be expanded as 
\beq
D=c\, f+ \sum_ac_a\, y_a \ , 
\eeq
where $c$ and $c_a$'s are non-negative integers. \eqref{df} then demands that at most two of the $c_a$ can be non-zero and only leaves the two possibilities: (a) $D=c\, f + 2\,y$ for a Mori cone generator $y \,(\neq f)$; (b) $D=c\, f + y + z$ for distinct Mori cone generators $y, z\,(\neq f)$. In the former case, $D \cdot D = 4 c -4= 0$ and hence $c=1$. For $D$ to be nef, $D \cdot y' = 2 y \cdot y' + 1 \geq 0$ for all the Mori cone generators $y'$, which implies that $y \cdot y' \geq 0$ for all $y'$. Therefore, $y$ also has to be nef but this contradicts that $y$ is a Mori cone generator itself. In the latter case, $D\cdot D = -2+4c+2 y\cdot z = 0$ and hence $y \cdot z =1-2c$. On the other hand, since $y$ and $z$ are distinct irreducible effective curves, they must intersect non-negatively and hence $y\cdot z = 1$ and $c=0$. 

Having derived the desired form~\eqref{desform}, to complete the proof, let us now show that such a divisor $D$ is necessarily nef. Note first that the only non-nef irreducible effective curves of $dP_9$ are the rational $(-1)$-curves, i.e. the Mori cone generators $y_a$ described in Table~\ref{tb:dP-Mori} (see the argument of Donagi et al in Ref.~\cite{Donagi:1996yf}). Since $D \cdot D =0$, it is enough to show that $D$ is irreducible. For the rest of this subsection, therefore, we will show that  
\beq\label{lbcoh}
h^0(dP_9, \cO(y))=1=h^0(dP_9, \cO(z)) \ , \quad\quad h^0(dP_9, \cO(y+z))=2 \ ,
\eeq
which guarantee that $D = y+z$ is indeed irreducible. 

Note that the Mori cone generators $y_a$ with $y_a \cdot y_a = -1$ and $y_a \cdot f =1$ are sections of the elliptic fibration and vice versa. We may thus rely on the Leray spectral sequence to compute the line bundle cohomologies in \eqref{lbcoh}; see e.g. Appendix A of Ref.~\cite{Anderson:2015yzz} for a self-contained description of the Leray spectral sequence. 

Given the elliptic fibration, 
\beq
\pi: dP_9 \to \IP^1 \ , 
\eeq
it is straightforward to demonstrate that the push-forward functors act on the trivial bundle as
\beq
R^0\pi_* (\cO)=\cO_{\IP^1} \ , \quad\quad R^1\pi_* (\cO)=\cO_{\IP^1}(-1) \ . 
\eeq
Firstly, for a single section, $y$, the Koszul sequence,
\beq\label{sing}
0 \to \cO \to \cO(y) \to \cO(y\cdot y)|_{y} \to 0 \ , 
\eeq
can be push-forwarded to 
\bea
0 &\to& \cO_{\IP^1} \to R^0\pi_*\cO(y) \to \cO_{\IP^1}(-1) \\ 
&\to& \cO_{\IP^1}(-1) \to R^1\pi_*\cO(y) \to 0 \ .
\eea
This leads to 
\beq
R^0\pi_*\cO(y) = \cO_{\IP^1} \ , \quad\quad R^1\pi_*\cO(y)=0 \ , 
\eeq
and hence, $h^0(dP_9, \cO(y))=h^0(\IP^1, \cO_{\IP^1})=1$. 
Similarly, for two distinct sections $y$ and $z$ with $y \cdot z = 1$, we can twist the Koszul sequence~\eqref{sing} as
\beq
0 \to \cO(z) \to \cO(y+z) \to \cO((y+z)\cdot y)|_{y} \to 0 \ ,
\eeq
which can be push-forwarded to 
\bea
0 &\to& \cO_{\IP^1} \to R^0\pi_*\cO(y+z) \to \cO_{\IP^1} \\ 
&\to& 0 \to R^1\pi_*\cO(y+z) \to 0 \ .
\eea
This leads to 
\beq
R^0\pi_*\cO(y+z) = \cO_{\IP^1}^{\oplus 2} \ ,\quad \quad R^1\pi_*\cO(y+z)=0 \ ,
\eeq
and hence, $h^0(dP_9, \cO(y+z))=h^0(\IP^1, \cO_{\IP^1}^{\oplus 2})=2$.

\subsection{An infinite family of solutions}\label{classify-dp9}
From the specific form~\eqref{desform} of the general solutions, we are led to classify pairs of distinct sections,
\bea
y &=& aL-\sum_{i=1}^9a_iE_i \ , \\
z&=&b L-\sum_{i=1}^9b_iE_i \ ,  
\eea
with 
\beq\label{ydotz}
y \cdot z=a b - \sum_{i=1}^9 a_i b_i = 1 \ . 
\eeq
In this subsection, instead of attempting a complete classification, we will impose an additional (rather artificial) constraint $a=b$ and will solve the corresponding restricted Diophantine system. 

Note first that 
\beq
\sum_{i=1}^9 (a_i-b_i)^2 = \sum_{i=1}^9 a_i^2 + \sum_{i=1}^9b_i^2 - 2 \sum_{i=1}^9 a_i b_i = (a^2+1) +(b^2+1)-2(ab-1)=(a-b)^2+4 \ , 
\eeq
where $y\cdot y=z\cdot z=-1$ as well as \eqref{ydotz} have been used. Under the additional constraint that $a=b$, we thus have
\beq
\sum_{i=1}^9 (a_i-b_i)^2 = 4 \ , 
\eeq
which leads to the two possibilities: (a) $|a_i-b_i|$ is $2$ for one $i$ and $0$ for the others; (b) $|a_i - b_i|$ is $1$ for four $i$'s and $0$ for the others. On the other hand, since $a=b$, 
\beq
(\sum_{i=1}^9 a_i)- (\sum_{i=1}^9 b_i) = z \cdot f - y \cdot f = 0 \ , 
\eeq
and hence, case (a) is ruled out and case (b) is further restricted to the condition that $|a_i - b_i|$ is $1$ for two $i$'s, $-1$ for another two $i$'s, and $0$ for the others. Upon permutation of the exceptional divisors, we may assume that 
\beq
b_{1,2}=a_{1,2}+1 \ , \quad b_{3,4} = a_{3,4}-1 \ , \quad b_{i}=a_i \text{~for~} i=5, \dots, 9 \ . 
\eeq
Since $z$ is a section, we must have
\beq
-1 = z \cdot z = b^2 - \sum_{i=1}^9 b_i^2 = a^2 - \sum_{i=1}^9 a_i^2 + 2(a_3 + a_4- a_1 - a_2) -4= -1 +2(a_3 + a_4 -a_1-a_2) - 4 \ , 
\eeq
where in the last step $y \cdot y = -1$ has been used. Therefore, for any section $y = aL-\sum_{i=1}^9a_iE_i$ such that 
\beq\label{12=34}
a_1+a_2 +2= a_3 + a_4 \ , 
\eeq
we can construct another section,  
\beq\label{zintermsofa}
z= aL - (a_1+1)E_1 - (a_2+1) E_2 -(a_3-1)E_3 -(a_4-1)E_4 - a_5 E_5 - \cdots -a_9 E_9 \ , 
\eeq
and $D=y+z$ is the general solution to the restricted Diophantine system. 

Let us now use the following result for the classification of sections $y$ in terms of $8$ integer parameters $k_i$, $i=1, \dots, 8$ (see Ref.~\cite{morigens} for the details),
\bi
\item $a\equiv0$ (mod 3): $a=3d$, $a_i=d-k_i$ for $i=1, \dots, 8$, and $a_9=d+s-1$ 
\item $a\equiv1$ (mod 3): $a=3d+4+9s$, $a_i=d-k_1+1+3s$ for $i=1, \dots, 8$, and $a_9=d+4s+3$
\item $a\equiv2$ (mod 3): $a=3d+32+18s$, $a_i=d-k_i+10+6s$ for $i=1, \dots, 8$, and $a_9=d+7s+15$ 
\ei
where 
\beq\label{ds}
d=\sum_{i=1}^8 k_i^2 + \sum_{1\leq i<j \leq 8} k_i k_j - \sum_{i=1}^8 k_i \quad\quad \text{and}\quad\quad s=\sum_{i=1}^8 k_i \ .
\eeq
Here, we will choose $y$ from the first category with $a\equiv0$ (mod 3) (one can proceed in exactly the same way for the other two categories). The only constraint that $y$ is required to obey is \eqref{12=34}. Thus, $k_1 + k_2 -2= k_3+k_4$ and the general solution for $D$ is give as
\bea \nn
D&=&y+z \\ \nn 
& =& 2aL-(2a_1+1) E_1 - (2a_2 +1)E_2 -(2a_3-1)E_3 - (2a_4-1) E_4 -2a_5 E_5 -\cdots - 2a_9 E_9 \ , \\ 
&=& \alpha - \sum_{i=1}^9 \alpha_i E_i \ , 
\eea
where 
\beq\label{inf-family-d}
\alpha=6d \ ,\quad \alpha_{i=1,2}=2d-2k_i+1, \quad \alpha_{i=3,4}=2d-2k_i-1 \ , \quad \alpha_{i=5,\dots,8}=2d-2k_i \ , \quad \alpha_9 = 2d+2s-2 \ . 
\eeq
Here, $s$ and $d$ are again functions of $k_i$'s as in \eqref{ds} and $k_1=k_3+k_4-k_2+2$  is to be substituted so that the solutions~\eqref{inf-family-d} actually form a seven-parameter family of divisors with arbitrary integer parameters, $k_2, \cdots, k_8 \in \IZ$. 

Similarly, the restricted Diophantine system can also be solved for the other $dP_9$ factor and leads to the following parameterization for $D'=\alpha'L'-\sum_{i'=1}^9 \alpha'_{i'} E'_{i'}$ with 
\beq\label{inf-family-dprime}
\alpha'=6d' \ ,\quad \alpha'_{i'=1,2}=2d'-2l_{i'}+1, \quad \alpha'_{i'=3,4}=2d'-2l_{i'}-1 \ , \quad \alpha'_{i'=5,\dots,8}=2d'-2l_{i'} \ , \quad \alpha'_9 = 2d'+2s'-2 \ , 
\eeq
where $d'$ and $s'$ are defined as 
\beq\label{dsprime}
d'=\sum_{i'=1}^8 l_{i'}^2 + \sum_{1\leq i'<j' \leq 8} l_{i'} l_{j'} - \sum_{i'=1}^8 l_{i'} \quad\quad \text{and}\quad\quad s'=\sum_{i'=1}^8 l_{i'} \ .
\eeq
Here, as for the $k_i$'s, $l_1= l_3+l_4-l_2+2$ is to be substituted and $l_2, \cdots, l_8$ are the seven free integer parameters. Combining \eqref{inf-family-d} and~\eqref{inf-family-dprime}, we thus obtain an infinite family of divisors 
\beq\label{inf-family}
\cD=D+D'= \alpha - \sum_{i=1}^9 \alpha_i E_i +\alpha'-\sum_{i'=1}^9 \alpha'_{i'} E'_{i'} \ , 
\eeq 
meeting the criteria of Koll\'ar's, in terms of the $14$ free integer parameters, $k_2, \dots, k_8$ and $l_2, \dots, l_8$.

%%%%%%%%%%
\section{Enumerating the blow-downs of del Pezzo surfaces}\label{enum}
In this section, we provide our methodology for enumerating various blow downs of del Pezzo surfaces.  The enumeration result is used in counting Koll\'ar divisors in Section~\ref{the48_sec}. 

For an illustration, let us consider blowing down $dP_3$, which is the simplest del Pezzo surface with a non-simplicial K\"ahler cone. Note first that the (non-simplicial) K\"ahler cone of $dP_3$ is spanned by the following $5$ generators,~\footnote{In this Appendix, we phrase the blown down bases in terms of their K\"ahler cone generators. One can consider a generic divisor positively spanned by those generators and pull it back to the Calabi-Yau three-fold. That way, a divisor of the three-fold is reconstructed from the K\"ahler cone information.}
\beq
\cK = \left<L, L-E_1, L-E_2, L-E_3, 2L-E_1-E_2-E_3\right> \ , 
\eeq
which can be seen as the dual description of the Mori cone in Table~\ref{tb:dP-Mori} (see \cite{Cvetic:2014gia} for the complete list of K\"ahler cone generators for all del Pezzo surfaces). 
We will now consider possible chains of blow downs of the irreducible exceptional curves, $E_1$, $E_2$, $E_3$, $L-E_1-E_2$, $L-E_2-E_3$ and $L-E_3-E_1$. 

If the curve $E_1$ is blown down first, we may delete the generators of $\cK(dP_3)$ that do not intersect with $E_1$ and are led to the following three-dimensional boundary component, 
\beq
\cK_{E_1} = \left<L, L-E_2, L-E_3\right> \ , 
\eeq 
where the subscript denotes the blown down curve. The cone $\cK_{E_1}$ should thus be seen as the K\"ahler cone of $dP_2$. Next, all of its two-dimensional boundaries, 
\bea
\cK_{E_1, E_2} &=& \left<L, L-E_3\right> \ , \\ 
\cK_{E_1, E_3} &=& \left<L, L-E_2\right> \ , \\ 
\cK_{E_1, L-E_2-E_3}&=&\left<L-E_2, L-E_3\right> \ , 
\eea
can be approached by a further blow down. Here, the first two are the K\"ahler cones of $dP_1$ surfaces, which can be further blown down to $\IP^2$ with the K\"ahler cones,
\beq
\cK_{E_1, E_2, E_3} = \left<L\right>=\cK_{E_1, E_3, E_2} \ , 
\eeq
while the third is the K\"ahler cone of $\IP^1\times \IP^1$, which cannot be blown down any further. A similar analysis can be made for the cases where $E_2$ or $E_3$ is blown down first. Taking into account of all such blow downs, one ends up with the following descriptions of the base surfaces in terms of their K\"ahler cones,
\bea\label{dp2}
dP_2&:& \left<L, L-E_i, L-E_j\right> \ , \\ \label{dp1}
dP_1&:& \left<L, L-E_i\right> \ , \\  \label{p1p1}
\IP^1\times\IP^1&:&  \left<L-E_i, L-E_j\right> \ , \\  \label{p2}
\IP^2&:& \left<L\right> \ , 
\eea
where $i$ and $j$ are distinct indices from $\{1,2,3\}$. 

If the curve $L-E_1-E_2$ is first blown down on the other hand, a different pattern is found for the resulting K\"ahler cones. At the first step, we obtain a $dP_2$ surface with the K\"ahler cone, 
\beq
\cK_{L-E_1-E_2} = \left<L-E_1, L-E_2, 2L-E_1-E_2-E_3\right> \ , 
\eeq
of which boundaries can be approached as
\bea
\cK_{L-E_1-E_2, E_3} &=& \left<L-E_1, L-E_2\right> \ , \\
\cK_{L-E_1-E_2, L-E_2-E_3} &=& \left<L-E_2, 2L-E_1-E_2-E_3\right> \ , \\
\cK_{L-E_1-E_2, L-E_3-E_1} &=& \left<L-E_1, 2L-E_1-E_2-E_3\right> \ . 
\eea
Of these three, the first is the K\"ahler cone of $\IP^1 \times \IP^1$, which cannot be blown down any further,  while the other two are the K\"ahler cones of $dP_1$ surfaces, which can be further blown down to $\IP^2$ with the K\"ahler cones,  
\beq
\cK_{L-E_1-E_2, L-E_2-E_3, L-E_3-E_1} = \left<2H-E_1-E_2-E_3\right> =\cK_{L-E_1-E_2, L-E_3-E_1, L-E_2-E_3} \ . 
\eeq
A similar analysis can be made for the cases where $L-E_2-E_3$ or $L-E_3-E_1$ is blown down first and one ends up adding to the list in \eqref{dp2}--\eqref{p2} the following new types of base surfaces: 
\bea\label{dp2'}
dP_2&:&  \left<L-E_i, L-E_j, 2L-E_1-E_2-E_3\right>\ , \\ \label{dp1'}
dP_1&:& \left<L-E_i,2L-E_1-E_2-E_3\right> \ , \\ \label{p2'}
\IP^2&:& \left<2L-E_1-E_2-E_3\right> \ , 
\eea
where, once again, $i$ and $j$ are distinct indices from $\{1,2,3\}$. 

This completes the classification of the blown down bases. In summary, we have a total of $18$ different base surfaces: (a) the $dP_3$ with which we start; (b) $6$ types of $dP_2$ surfaces in \eqref{dp2} and~\eqref{dp2'}; (c) $6$ types of $dP_1$ surfaces in \eqref{dp1} and~\eqref{dp1'}; (d) $3$ types of $\IP^1\times \IP^1$ in \eqref{p1p1}; (e) $2$ types of $\IP^2$ in \eqref{p2} and \eqref{p2'}. 

Along the same line, we have classified all the blown down bases by starting from each of the del Pezzo base surfaces $dP_r$, $r=0, \cdots, 7$, that are relevant to this work. In Table~\ref{tb:bd-stat}, we summarize the counting of different types of blown down bases. We also provide in Table~\ref{tb:bd-stat-red} another counting result that takes into account of the permutation redundancies of the exceptional divisors of the del Pezzo factors, although, for the purpose of string dualities, a more relevant counting is the one given in Table~\ref{tb:bd-stat}.

\begin{table}[h]
\begin{center}
\begin{tabular}{|c||c|c|c|c|c|c|c|c|c||c|}
\hline
$r$ & $dP_0$ & $\IP^1\times \IP^1$ & $dP_1$ & $dP_2$ & $dP_3$ & $dP_4$ & $dP_5$ & $dP_6$ & $dP_7$ & {Tot.}  \\ \hline\hline 
0& 1 & - & - & - & - & - & - & - & -  & 1 \\ \hline
1& 1 & - & 1 & - & - & - & - & - & - & 2 \\ \hline
2& 1 & 1 & 2 & 1 & - & - & - & - & - & 5  \\ \hline
3& 2 & 3 & 6 & 6& 1 & - & - & - & - & 18 \\ \hline
4& 5 & 10 & 20 & 30 & 10 & 1 & - & - & - & 76 \\ \hline
5& 16 & 40 & 80 & 160 & 80 & 16 & 1 & - & - & 393 \\ \hline
6& 72 & 216 & 432 & 1080 & 720 & 216 & 27 & 1 & - & 2764 \\ \hline
7& 576 & 1974 & 4025 & 12104 & 3745 & 3857 & 756 & 56 & 1 & 27094 \\ \hline
\end{tabular}\end{center}
\caption{{\it The statistics of different blow downs of the $dP_r$ surfaces, $r=0, \cdots, 7$. The numbers in parentheses count the fibrations up to permutations of the exceptional divisors of the del Pezzo factors.}}
\label{tb:bd-stat}
\end{table}

\begin{table}[h]
\begin{center}
\begin{tabular}{|c||c|c|c|c|c|c|c|c|c||c|}
\hline
$r$ & $dP_0$ & $\IP^1\times \IP^1$ & $dP_1$ & $dP_2$ & $dP_3$ & $dP_4$ & $dP_5$ & $dP_6$ & $dP_7$ & {Tot.}  \\ \hline\hline 
0& 1 & - & - & - & - & - & - & - & -  & 1 \\ \hline
1& 1 & - & 1 & - & - & - & - & - & - & 2 \\ \hline
2& 1 & 1 & 1 & 1 & - & - & - & - & - & 4  \\ \hline
3& 2 & 1 & 2 & 2& 1 & - & - & - & - & 8 \\ \hline
4& 2 & 2 & 3 & 3 & 2 &1 & - & - & - & 13 \\ \hline
5& 3 & 3 & 5 & 6& 4 & 3 & 1 & - & - & 25 \\ \hline
6& 5 & 6 & 9 & 12 & 9 & 6 & 3 & 1 & - & 51 \\ \hline
7& 10 & 12 & 18 & 24 & 18 & 16 & 9 & 4 & 1 & 112 \\ \hline
\end{tabular}\end{center}
\caption{{\it A redundancy removed version of the statistics of different blow downs of the $dP_r$ surfaces, $r=0, \cdots, 7$; Here, the blow downs are counted up to permutations of the exceptional divisors of the del Pezzo factors.}}
\label{tb:bd-stat-red}
\end{table}

\section{A class of CICY redundancies via base surface splittings/contractions}\label{surf_red}
%%%%%

In this section, we give a brief overview of a simple method for identifying equivalent CICY threefold configurations. 
Suppose that a given configuration matrix can be written, upon row and column permutations, in the following block form,

\beq \label{conf-gen}
X \sim \left[ \begin{array}{c|cc} 
\mathcal A_1 & \bold 0 & \mathcal F \\ 
\mathcal A_2 & \mathcal B & \mathcal T \\
\mathcal A_3 & \mathcal C & \bold 0  
\end{array}\right]\; \ , 
\eeq
where the $\bold 0$'s are zero matrices of appropriate sizes and the upper-right corner, 
\beq
F \sim  \left[ \begin{array}{c|c} \mathcal A_1& \mathcal F \end{array}\right] \ ,
\eeq 
represents a curve of complex dimension one. Given that $X$ is a Calabi-Yau threefold, one observes the obvious genus-one fibration (OGF) structure~\cite{Gray:2014fla}, in which $F$ necessarily represents an elliptic curve fibered over the surface base,
\beq \label{baseconf-gen}
B \sim \left[ \begin{array}{c|c} 
\mathcal A_2 & \mathcal B  \\
\mathcal A_3 & \mathcal C   
\end{array}\right]\; \ , 
\eeq
with the twist described by $\mathcal T$. 

Let us now consider varying the configuration~\eqref{conf-gen} via a chain of splittings and contractions~\cite{Candelas:1987kf}, 
so that $\mathcal A_3$ may lose and/or gain some projective space factors together with the associated rows of the configuration matrix, 
while the columns involving $\mathcal F$ and $\mathcal T$ are kept intact. In the end, the chain will relate the two configurations,

\beq \label{equiv-gen}
X\sim \left[ \begin{array}{c|cc} 
\mathcal A_1 & \bold 0 & \mathcal F \\ 
\mathcal A_2 & \mathcal B & \mathcal T \\
\mathcal A_3 & \mathcal C & \bold 0  
\end{array}\right]\;\quad \leftrightsquigarrow \quad   X'\sim \left[ \begin{array}{c|cc} 
\mathcal A_1 & \bold 0 & \mathcal F \\ 
\mathcal A_2 & \mathcal B' & \mathcal T \\
\mathcal A_3' & \mathcal C' & \bold 0  
\end{array}\right]\;  \ . 
\eeq
The observation to made here is that all such splittings and contractions are {\it ineffective}~\cite{Candelas:1987kf}; in particular, the two configurations in \eqref{equiv-gen} are equivalent. 

To see this, note first that it is sufficient to show the equivalence for a single splitting/contraction transition and we may restrict to the case where $\mathcal A_3$ is a single projective space. For example, suppose that $\mathcal A_3 = \IP^1$ and ${\rm dim}_{\IC} \mathcal A_2=3$ (the generalization to the cases with $\mathcal A_3=\IP^n$ and ${\rm dim}_{\IC} \mathcal A_2=m$ is straightforward) and consider the $\IP^1$ contraction, 
\beq \label{P1contraction}
X\sim \left[ \begin{array}{c|ccc} 
\mathcal A_1 & \bold 0 & \bold 0  & \mathcal F \\ 
\mathcal A_2 & u & v & \mathcal T \\
\mathcal \IP^1 & 1 & 1  & \bold 0  
\end{array}\right]\;\quad \rightsquigarrow \quad   X'\sim \left[ \begin{array}{c|cc} 
\mathcal A_1 & \bold 0 & \mathcal F \\ 
\mathcal A_2 & u+v & \mathcal T 
\end{array}\right]\;  \ . 
\eeq
$X$ is guaranteed to be smooth, given a generic complex structure, and we denote the two defining equations of $X$ associated to the first two columns of its configuration by
\bea
\left( \begin{array}{cc} 
P & Q \\
R & S   
\end{array}\right)
\left( \begin{array}{c} 
a_0 \\
a_1   
\end{array}\right) 
=
\left( \begin{array}{c} 
0 \\
0   
\end{array}\right) \ , 
\eea
where $P$, $Q$, $R$, and $S$ are generic polynomials in $\mathcal A_2$ with ${\rm deg} P= {\rm deg} Q=u$, ${\rm deg} R= {\rm deg} S=v$ and $(a_0: a_1)$ are the homogeneous coordinates of $\IP^1$. The first defining equation of $X'$ can then be written as
\beq
PS - QR = 0 \ . 
\eeq
It is then immediately seen that as long as at least one of $P$, $Q$, $R$, and $S$ is non-vanishing so that 
\beq
{\rm rk}\left( \begin{array}{cc} 
P & Q \\
R & S   
\end{array}\right)=1 \ ,
\eeq
there exists a local isomorphism between $X$ and $X'$, and hence, the only source of singularities in $X'$ is the locus where $P=Q=R=S=0$. For a generic complex structure of $X$, however, there is no solution to such a simultaneous vanishing since ${\rm dim}_{\IC}\mathcal A_2 = 3$. Therefore, the $\IP^1$ contraction,~\eqref{P1contraction}, does not involve singularities and is necessarily ineffective. 

Note that such a redundancy for the configurations of CICY threefolds,~\eqref{equiv-gen}, arises essentially from those of the base surfaces, 
\beq \label{base-equiv-gen}
B\sim \left[ \begin{array}{c|c} 
\mathcal A_2 & \mathcal B \\
\mathcal A_3 & \mathcal C   
\end{array}\right]\;\quad  \leftrightsquigarrow \quad B'\sim \left[ \begin{array}{c|c} 
\mathcal A_2 & \mathcal B'  \\
\mathcal A_3' & \mathcal C'  
\end{array}\right]\;  \ ,
\eeq
and can easily be spotted by comparing the latter. 
We will thus call the CICY threefold redundancies of this type a {\it surface redundancy}.

%%%%%
\subsection{Illustration}\label{1.1}
An example of this CICY configuration matrix redundancy can be illustrated with the following example of two configurations,  

\begin{eqnarray} \label{eg1ab} 
\left[ \begin{array}{c|cc} 
\mathbb{P}^1 & 1 & 1  \\ 
\mathbb{P}^2 & 0 & 3  \\ 
\mathbb{P}^2 & 3 & 0 
\end{array}\right]\;
~\text{and}~
\left[ \begin{array}{c|cccc} 
\mathbb{P}^1 & 1 & 1 & 0 & 0 \\ 
\mathbb{P}^2 & 3 & 0 & 0 & 0 \\ 
\mathbb{P}^4 & 0 & 1 & 2& 2
\end{array}\right]\; , 
\end{eqnarray}
labelled respectively as $\#14$ and $\#16$ in the original CICY threefold dataset~\cite{Candelas:1987kf}. 
We first permute the rows and columns of each configuration appropriately so that the two configurations are written respectively as  
\begin{eqnarray} \label{eg1-perm} 
\left[ \begin{array}{c|c|c} 
\mathbb{P}^2 & 0 & 3  \\ \hline 
\mathbb{P}^1 & 1 & 1  \\ \hline
\mathbb{P}^2 & 3 & 0 
\end{array}\right]\;
~\text{and}~
\left[ \begin{array}{c|ccc|c} 
\mathbb{P}^2  & 0 & 0 & 0 & 3\\ \hline
\mathbb{P}^1  & 1 & 0 & 0 & 1\\  \hline
\mathbb{P}^4  & 1 & 2& 2 & 0
\end{array}\right]\; .
\end{eqnarray}
Such permutations only correspond to an appropriate relabeling of the defining equations as well as the ambient homogeneous coordinates and hence never affect the geometry. 
Here, the horizontal and the vertical lines have been added to manifestly distinguish the six blocks as in \eqref{equiv-gen}. Note that the two configurations \eqref{eg1-perm} are exactly of the form \eqref{equiv-gen} and the following is a relevant chain of base surface equivalences,
\begin{eqnarray} \nn
\left[ \begin{array}{c|cc} 
\mathbb{P}^1_{\bold x} & 1  \\ 
\mathbb{P}^2_{\bold y} & 3   
\end{array}\right]\;
&\underset{sp.}{\overset{\bold a}{\leftrightsquigarrow}}&
\left[ \begin{array}{c|cccc} 
\mathbb{P}^1_{\bold x} & 1 & 0 \\ 
\mathbb{P}^2_{\bold y} & 2 & 1 \\ 
\mathbb{P}^1_{\bold a} & 1 & 1 
\end{array}\right]\;
\underset{sp.}{\overset{\bold b}{\leftrightsquigarrow}}
\left[ \begin{array}{c|cccc} 
\mathbb{P}^1_{\bold x} & 1 & 0 & 0\\ 
\mathbb{P}^2_{\bold y} & 1 & 1 & 1\\ 
\mathbb{P}^1_{\bold a} & 1 & 0 & 1 \\
\mathbb{P}^1_{\bold b} & 1 & 1 & 0
\end{array}\right]\;
\underset{cont.}{\overset{\bold y}{\leftrightsquigarrow}}
\left[ \begin{array}{c|cccc} 
\mathbb{P}^1_{\bold x} & 1  \\ 
\mathbb{P}^1_{\bold a} & 2  \\ 
\mathbb{P}^1_{\bold b} & 2 
\end{array}\right]\; \\ \label{chain}
&\underset{sp.}{\overset{\bold w}{\leftrightsquigarrow}}&
\left[ \begin{array}{c|ccccc} 
\mathbb{P}^1_{\bold x} & 1 & 0 & 0 & 0 & 0 \\ 
\mathbb{P}^1_{\bold a} & 0 & 1 & 1 & 0 & 0  \\ 
\mathbb{P}^1_{\bold b} & 0 & 0 & 0 & 1 & 1 \\ 
\mathbb{P}^4_{\bold w} & 1 & 1 & 1 & 1 & 1  
\end{array}\right]\; 
\underset{cont.}{\overset{\bold b}{\leftrightsquigarrow}}
\left[ \begin{array}{c|cccc} 
\mathbb{P}^1_{\bold x} & 1 & 0 & 0 & 0  \\ 
\mathbb{P}^1_{\bold a} & 0 & 1 & 1 & 0   \\ 
\mathbb{P}^4_{\bold w} & 1 & 1 & 1 & 2  
\end{array}\right]\; \\ \nn
&\underset{cont.}{\overset{\bold a}{\leftrightsquigarrow}}&
\left[ \begin{array}{c|cccc} 
\mathbb{P}^1_{\bold x} & 1 & 0 & 0  \\ 
\mathbb{P}^4_{\bold w} & 1 & 2 & 2  
\end{array}\right]\; .
\end{eqnarray}
Here, the bold subscripts for the projective spaces label their homogeneous coordinates and the symbol ``$sp.$'' (resp., ``$cont.$'') below each arrow indicates that the configuration on the right is obtained by splitting (resp., by contracting) the one on the left along the projective space with homogeneous coordinates denoted above the arrow. This then leads to the equivalence between the two CICY threefold configurations~\eqref{eg1ab} according to the arguments in the previous subsection. 

In fact, there are total of $15$ configurations in the original dataset that are shown to be equivalent exactly for the families of surfaces described above (numbers $14$;\, $16, \dots, 21$;\, $23, \dots, 30$). They all represent the Schoen manifold, in particular. See Section \ref{schoen_sec} for further details on this geometry.

%%%%%
\subsubsection{The complete network of base surface splitting/contraction redundancies}
It is possible to search for such redundancies in the entire dataset of $7890$ CICY threefolds. In turns out that this class of redundancies arise only with the base surfaces, $dP_3$, $dP_5$ and $dP_9.$\footnote{We believe that this is an artifact of the particular algorithm based on which the original dataset of CICY threefolds was generated in \cite{Candelas:1987kf}.} The statistics for each of the three base types can be summarized as follows.
\bi 
\item $dP_3$: There arise $16$ pairs involving the surface equivalence, 
\beq 
\left[ \begin{array}{c|cc} 
\mathcal \IP^2 & 1 & 1 \\
\mathcal \IP^2 & 1 & 1   
\end{array}\right]\;  \leftrightsquigarrow  \left[ \begin{array}{c|cccc} 
\mathcal \IP^2 & 1 & 0 & 1 & 0   \\
\mathcal \IP^2 & 0 & 1 & 0 & 1 \\ 
\mathcal \IP^1 & 0 & 0 & 1 & 1 \\
\mathcal \IP^1 & 1 & 1 & 0 & 0 
\end{array}\right]\;  \ , 
\eeq
as well as $15$ pairs involving
\beq 
\left[ \begin{array}{c|c} 
\mathcal \IP^1 & 1  \\
\mathcal \IP^1 & 1 \\
\mathcal \IP^1 & 1 \\
\end{array}\right]\;  \leftrightsquigarrow  \left[ \begin{array}{c|cccc} 
\mathcal \IP^1 & 1 & 0 & 0   \\
\mathcal \IP^1 & 0 & 1 & 0 \\ 
\mathcal \IP^1 & 0 & 0 & 1 \\
\mathcal \IP^2 & 1 & 1 & 1 
\end{array}\right]\;  \ .
\eeq
Thus, total of $62$ configurations are grouped into $31$ classes. The pair of CICY configurations in each class are equivalent. 
\item $dP_5$: Total of $666$ configurations are grouped into $119$ classes via the surface equivalence, 
\beq 
\left[ \begin{array}{c|c} 
\mathcal \IP^1 & 1  \\
\mathcal \IP^1 & 1 \\
\mathcal \IP^1 & 2 \\
\end{array}\right]\;  \leftrightsquigarrow  \left[ \begin{array}{c|cc} 
\mathcal \IP^1 & 1 & 0   \\
\mathcal \IP^1 & 0 & 1 \\ 
\mathcal \IP^2 & 1 & 2  
\end{array}\right]\;  \ .
\eeq
The maximal number of equivalent threefold configurations in a class is $24$.
\item $dP_9$: The $15$ configurations discussed in the previous subsection are all that involves a $dP_9$ surface redundancy; they are grouped into a single class that describes the Schoen manifold. 
\ei
 
 \section{A guide to the new CICY data set and fibration data}\label{app_data}
The new favorable CICY configurations, and their obvious fibrations, are publicly available and can be found here: \url{www1.phys.vt.edu/cicydata}. The website includes two files. The first is a new version of the CICY list, with as many configuration matrices replaced by favorable examples as possible. The second contains the obvious fibrations of the CICYs.

\vspace{0.1cm}

\begin{figure}[!h]\centering
\includegraphics[width=1.0\textwidth]{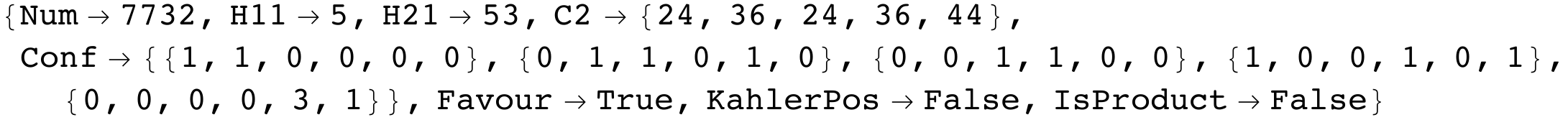}
\caption{\it An example entry in the new maximally favorable CICY list.}
\label{seg1}
\end{figure}

An example of the format of the entries of the CICY list is given in Figure \ref{seg1}.
The first entry gives the number labeling the CICY, which are compatible with the numbers in the original CICY list \cite{Candelas:1987kf} obtainable \href{http://www-thphys.physics.ox.ac.uk/projects/CalabiYau/cicylist/}{here}. The next two entries specify the Hodge data of the threefold. The fourth entry gives the second Chern class of the manifold. The format here is as follows. Express the second Chern class in the form
\begin{eqnarray}
c_2(TX) = C^{ij} J_i \wedge J_j
\end{eqnarray}
where the $J_i$ are the K\"ahler forms of the projective space factors restricted to the Calabi-Yau and the $C$'s are some numerical coefficients. One can then contract the $C$'s with the intersection form for the manifold to get the numbers given (no information is lost in performing this operation). The fifth entry in the list is simply the configuration matrix. In this case we have the following.
\begin{eqnarray} \label{elmattychops}
		X_{7732} = \left[ \begin{array}{c|cccccc} \mathbb{P}^1 & 1&1&0&0&0&0 \\\mathbb{P}^2&0&1&1&0&1&0 \\ \mathbb{P}^1&0&0&1&1&0&0 \\ \mathbb{P}^2 &1&0&0&1&0&1 \\ \mathbb{P}^3&0&0&0&0&3&1\end{array} \right]
		\end{eqnarray}
The sixth entry says whether the description provided is favorable, the seventh says whether the naive ambient space K\"ahler cone descends to give that of the Calabi-Yau and the final entry indicates whether the configuration matrix describes a direct product manifold.

\vspace{0.1cm}

The obvious fibrations of the maximally favorable CICY list described above  are provided in a second download file available at \cite{website}. In this instance a list of 7868 cases is provided, one for each CICY threefold excluding manifolds that are direct products (i.e. those for which ``$\textnormal{IsProduct} \to \textnormal{True}$" in the above CICY list). Each list entry has four components. The first entry is simply the relevant CICY number. The second entry lists the obvious genus one fibrations of the configuration, the third entry lists the obvious K3 fibrations of the configuration and the final entry describes how the K3 and torus fibrations are nested.

The list of torus fibrations for each configuration contains one entry per fibration. An example for CICY 7732 is the following (this is the first genus one fibration listed).
		\begin{eqnarray} \label{elfiby}
		\{ \{5\},\{5,6\}\}
		\end{eqnarray}
This means that the fiber in this example is described by the $5^{th}$ row and $5^{th}$ and $6^{th}$ columns, with reference back to the configuration provided in the maximally favorable CICY list. Referring back to the configuration matrix (\ref{elmattychops}), we see that the fibration (\ref{elfiby}) can be presented in our usual format as follows.
\begin{eqnarray} \label{elfiby2}
X_{7732}' = \left[ \begin{array}{c|cccc:cc} \mathbb{P}^3&0&0&0&0&3&1 \\\hdashline \mathbb{P}^1 & 1&1&0&0&0&0 \\\mathbb{P}^2&0&1&1&0&1&0 \\ \mathbb{P}^1&0&0&1&1&0&0 \\ \mathbb{P}^2 &1&0&0&1&0&1 \end{array} \right]
\end{eqnarray}
This abbreviated formatting is used to keep the fibration list to a manageable size.

The list of K3 fibrations in the third entry for each configuration follows the same format. So for example, for CICY 7732, the first entry in the K3 fibration list is the following.
		\begin{eqnarray}
		\{ \{1,2,4,5\},\{1,2,3,4,5,6\}\}
		\end{eqnarray}		
This corresponds to the following K3 fibration in our usual notation (again referring back to (\ref{elmattychops})).
		\begin{eqnarray} \label{k3fiby2}
X_{7732}'' = \left[ \begin{array}{c|:cccccc} \mathbb{P}^1 & 1&1&0&0&0&0 \\  \mathbb{P}^2&0&1&1&0&1&0 \\ \mathbb{P}^2 &1&0&0&1&0&1 \\   \mathbb{P}^3&0&0&0&0&3&1\\ \hdashline \mathbb{P}^1&0&0&1&1&0&0   \end{array} \right]
\end{eqnarray}
		
As mentioned above, the fourth and final entry in the list provided for each configuration matrix describes how these fibrations are nested. Each case in this fourth entry is a list of two numbers. The first number specifies a K3 fibration from the previous list and the second a genus one fibration from the second entry for that configuration matrix. If a given pair exists, then those two fibrations are compatible (that is, the genus one fibers are also fibers of the K3 of the K3 fibration). For example, in the case of CICY 7732, the first case we have is simply the following.
		\begin{eqnarray}
		 \{1,1\}
		\end{eqnarray}	
This just states that the first genus one fibration (\ref{elfiby2}) is nested in the first K3 fibration (\ref{k3fiby2}) in a compatible way. This can be confirmed by performing row and column permutations on those two matrices that are compatible with both fibration structures. That is, we only consider permutations that do not mix the matrix blocks of the form (\ref{fib1}). In the case at hand, we can obtain the following in this manner.
		\begin{eqnarray}
X_{7732}''' = \left[ \begin{array}{c|:cccc:cc}   \mathbb{P}^3&0&0&0&0&3&1\\ \hdashline \mathbb{P}^1 & 1&1&0&0&0&0 \\  \mathbb{P}^2&0&1&1&0&1&0 \\ \mathbb{P}^2 &1&0&0&1&0&1\\ \hdashline \mathbb{P}^1&0&0&1&1&0&0   \end{array} \right]~~.
\end{eqnarray}

%\clearpage

\end{document}